%% file: main.tex
\documentclass{lmcs}

\usepackage{amsmath}
\makeatletter
\newcommand{\allignLabel}[1]{\refstepcounter{equation}(\theequation)\def\tmplab{#1}\ltx@label\tmplab}
\makeatother
\usepackage{anyfontsize}
\usepackage{amsthm}
\usepackage{amssymb}
\usepackage{proof}
\usepackage{hyperref}
\usepackage{amssymb}
\usepackage{mathrsfs}
\usepackage{stmaryrd}
\usepackage{dsfont}
\usepackage{cleveref}
\usepackage[all]{hypcap}

\usepackage{tikz}

\usepackage{regexpatch}

\input{tikzit.tikzdefs}
\usepackage{tikzit}
\input{tikzit.tikzstyles}

\newcommand{\typ}{\mathtt{type}}

\newcommand{\alt}{~\mid~}

\newcommand{\tact}[1][c]{\mathtt{act}_\TT(#1)}
\newcommand{\zact}[1][c]{\mathtt{act}_\SC(#1)}

\newcommand{\xact}[1][c]{\mathtt{act}_\XX(#1)}

\newcommand{\xret}[1]{\mathtt{ret}_\XX~#1}
\newcommand{\sret}[1]{\mathtt{ret}_\SC~#1}
\newcommand{\zret}[1]{\mathtt{ret}_\SC~#1}
\newcommand{\tret}[1]{\mathtt{ret}_\TT~#1}
\newcommand{\xdo}[3]{\mathtt{do}_\XX~#1\leftarrow #2;~#3}

\newcommand{\zdo}[3]{\mathtt{do}_\SC~#1\leftarrow #2;~#3}
\newcommand{\tdo}[3]{\mathtt{do}_\TT~#1\leftarrow #2;~#3}

\newcommand{\secref}[1]{\S \ref{#1}}

\newcommand{\sem}[1]{\left\llbracket #1\right\rrbracket}
\newcommand{\pv}[2]{\langle #1,#2 \rangle}
\newcommand{\naturalto}{\ensuremath{\Rightarrow}}
\newcommand{\Id}{\ensuremath{\mathrm{id}}}
\newcommand{\iid}{\ensuremath{\mathrm{id}}}

\newcommand{\rret}[1]{\mathtt{ret}~#1}
\newcommand{\ddo}[3]{\mathtt{do}~#1\leftarrow #2;~#3}

\newcommand{\VV}{\ensuremath{\mathcal V}}

\DeclareFontFamily{U}{min}{}
\DeclareFontShape{U}{min}{m}{n}{<-> udmj30}{}

\begin{document}

\title[Central Submonads and Notions of Computation]{Central Submonads and
Notions of Computation: Soundness, Completeness and Internal Languages}

\author{Titouan Carette}[a]

\author{Louis Lemonnier}[b]

\author{Vladimir Zamdzhiev}[c]

\address{Institut Polytechnique de Paris, Ecole Polytechnique, LIX, 91128
Palaiseau Cedex, France}

\address{University of Edinburgh, United Kingdom}

\address{Université Paris-Saclay, CNRS, ENS Paris-Saclay,
Inria, Laboratoire Méthodes Formelles, 91190, Gif-sur-Yvette, France}


\begin{abstract}
	Monads in category theory are algebraic structures that can be used to
	model computational effects in programming languages. We show how the
	notion of ``\emph{centre}", and more generally ``\emph{centrality}",
	\textit{i.e.} the property for an effect to commute with all other
	effects, may be formulated for strong monads acting on symmetric monoidal
	categories. We identify three equivalent conditions which characterise the
	existence of the centre of a strong monad (some of which relate it to the
	premonoidal centre of Power and Robinson) and we show that every strong
	monad on many well-known naturally occurring categories does admit a
	centre, thereby showing that this new notion is ubiquitous. More generally,
	we study \emph{central submonads}, which are necessarily commutative, just
	like the centre of a strong monad. We provide a computational
	interpretation by formulating equational theories of lambda calculi with
	central submonads, we describe categorical models for these theories, and
	prove soundness, completeness, and internal language results for our
	semantics.
\end{abstract}

\maketitle

\noindent \textbf{Publication history.}
The present article is a journal version of our paper \cite{nous23monads} that
was published in the proceedings of LICS'23. Compared to the conference version
of the paper, the present article adds many proofs that were omitted in the
conference version.  We also provide a more detailed background section, as
well as a discussion on the link with Lawvere theories (see
\secref{sub:lawvere}), and an updated list of examples, some of which come from
recent works.

\section{Introduction}

The importance of monads in programming semantics has been demonstrated in
seminal work by Moggi \cite{moggi-lics,moggi}. The main idea is that monads
allow us to introduce computational effects (e.g. state, input/output,
recursion, probability, continuations) into pure type systems in a controlled
way. The mathematical development surrounding monads has been very successful
and it directly influenced modern programming language design through the
introduction of monads as a programming abstraction into languages such as
Haskell, Scala and others (see \cite{benton-monads}). Inspired by this, we
follow in the same spirit: we start with a mathematical question about monads,
we provide the answer to it and we present a computational interpretation. The
mathematical question that we ask is simple and it is inspired by the theory of
monoids and groups:
\begin{center}
	\label{eq:question}
	\emph{Is there a suitable notion of ``centre'' that may be formulated for
	monads and what is a ``central'' submonad?}
\end{center}

We show that, just as every monoid $M$ (on $\Set$) has a centre, which is a
\emph{commutative} submonoid of $M$, so does every (canonically strong) monad
$\TT$ on $\Set$ and the centre of $\TT$ is a \emph{commutative} submonad of
$\TT$ (\secref{sub:sets}). A central\footnote{Given a group $G$, a
\emph{central subgroup} is a subgroup of the centre of $G$, equivalently, a
subgroup whose elements commute with every element of $G$.} submonad of $\TT$
is simply a submonad of the centre of $\TT$ (Definition \ref{def:central-sub})
and the analogy to the case of monoids and groups is completely preserved. Note
that our construction has nothing to do with the folklore characterisation of
monads as monoid objects in a functor category, wherein the notion of
commutativity is unclear. The relevant analogy with monoids in $\Set$ is fully
explained in Example \ref{ex:free-monad}. Generalising away from the category
$\Set$, the answer is a little bit more complicated: not every monoid object
$M$ on a symmetric monoidal category $\CC$ has a centre, and neither does every
strong monad on $\CC$ (\secref{sub:bad-monad}). However, we show that under
some reasonable assumptions, the centre does exist (Theorem
\ref{th:centralisability}) and we have not found any naturally occurring monads
in the literature that are not centralisable (\emph{i.e.} monads other than the
artificially constructed one we used as a counter-example). Furthermore, we
show that for many categories of interest, all strong monads on them are
centralisable (\secref{sub:examples}) and we demonstrate that the notion of
centre is ubiquitous. The centre of a strong monad satisfies interesting
universal properties (Theorem \ref{th:centralisability}) which may be
equivalently formulated in terms of our novel notion of \emph{central cone} or
via the \emph{premonoidal centre} of Power and Robinson
\cite{power-premonoidal}. The notion of a central submonad is more general and
it may be defined without using the centre. When the centre exists, a central
submonad may be equivalently defined as a strong submonad of the centre
(Theorem \ref{th:centrality}).

The computational significance of these ideas is easy to understand: given an
effect, modelled by a strong monad, such that perhaps not \emph{every} pair of
effectful operations commute (\emph{i.e.} the order of monadic sequencing matters),
identify only those effectful operations which do commute with any other
possible effectful operation. The effectful operations that satisfy this
property are called \emph{central}. When the monad is centralisable, the
collection of \emph{all} central operations determine the centre of the monad
(which is a commutative submonad). \emph{Any} collection of central operations
that may be organised into a strong submonad determines a central submonad
(which also is commutative). We argue that central submonads have greater
computational significance compared to the centre of a strong monad
(\secref{sub:theories}) for two main reasons: (1) central submonads are
strictly more general; (2) central submonads have a simpler and considerably
more practical axiomatisation via an equational theory, whereas the centre of a
monad requires an axiomatisation using a more complicated logical theory. We
cement our categorical semantics by proving soundness, completeness and
internal language results (see \cite{maietti2005relating} for a
convincing argument why internal language results are important and why
soundness and completeness \emph{alone} might not be sufficient).

\paragraph{Related Work}
A notion of commutants for enriched algebraic theories has been defined in
\cite{commutants} from which the author derives a notion of centre of an
enriched algebraic theory. In the case of enriched monads, in other words,
strong monads arising from enriched algebraic theories, their notion of
commutant extends to monad morphisms. While not explicitly stated in the paper,
applying the commutant construction on the identity monad morphism from a monad
to itself provides a notion of centre of a monad that appears to coincide with
ours. However, enriched algebraic theories correspond to $\mathcal{J}$-ary
$\VVV$-enriched monads (see \cite{commutants} for a definition of
$\mathcal{J}$-ary monads w.r.t. a system of arities $\mathcal{J}$) on a
symmetric monoidal \emph{closed} category $\VVV$ (equivalently $\JJ$-ary strong
monads on $\VVV$). In this article, we show that monoidal closure of $\VVV$ is not
necessary to define the centre and neither is the $\mathcal J$-ary assumption
on the monad. Other related work \cite{garner2016commutativity} considers a
very general notion of commutativity in terms of certain kinds of duoidal
categories. As a special case of their treatment, the authors are able to
recover the commutativity of bistrong monads and with some additional effort
(not outlined in the paper), it is possible to construct the centre of a
bistrong monad acting on a monoidal \emph{biclosed} category. Our construction
of the centre appears to coincide with theirs in the special case of strong
monads defined on symmetric monoidal \emph{closed} categories, but as discussed
above, our method does not require any kind of closure of the category.
Therefore, compared to both works \cite{garner2016commutativity,commutants}, as
far as symmetric monoidal (not necessarily closed) categories are concerned,
our methods can be used to construct the centre for a larger class of strong
monads and we establish our main results, together with our universal
characterisation of the centre, under these assumptions. Furthermore, we also
place a heavy emphasis on \emph{central} submonads in this article and these kinds
of monads are not discussed in either of these works and neither is there a
computational interpretation (which is our main result in
\secref{sec:computational}).

Another related work is \cite{power-premonoidal}, which introduces
premonoidal categories. We have established important links between our
development and the premonoidal centre (Theorem \ref{th:centralisability}).
While premonoidal categories have been influential in our understanding of
effectful computation, it was less clear (to us) how to formulate an appropriate
computational interpretation of the premonoidal centre for higher-order
languages. We show that under some mild assumptions (which are easily
satisfied see \secref{sec:examples}), the premonoidal centre of the Kleisli
category of a strong monad induces an adjunction into the base category
(Theorem \ref{th:centralisability}) and this allows us to formulate a suitable
computational interpretation by using monads, which are already well-understood
\cite{moggi,moggi-lics} and well-integrated into many programming languages
\cite{benton-monads}.

Staton and Levy introduce the novel notion of \emph{premulticategories}
\cite{premulticategories} in order to axiomatise impure/effectful computation
in programming languages. The notion of centrality plays an important role in
the development of the theory there as well. However, they do not focus, as we
do, on providing suitable programming abstractions that identify both central
and non-central computations (e.g. by separating them into different types
like us) and from what we can tell from our reading, there are no universal
properties stated for the collection of central morphisms. Also, our results
provide a computational interpretation in terms of monads, which are standard
and well-understood, so it is easier to incorporate them into existing
languages.

Central morphisms in the context of computational effects have been studied
among other sorts of \emph{varieties}~\cite{fuhrmann1999lambda} of morphisms:
thunkable, copyable, and discardable. The author links their notion of central
morphisms with the ones from the premonoidal centre in Power and Robinson
\cite{power-premonoidal}, and also proves under some conditions that those
varieties form a subcategory with similar properties to the original category.
However, they do not claim that a central submonad or a centre can be
constructed out of those central morphisms. 

The work in \cite{fuhrmann1999lambda} has an impact in \cite{dylan2022galois},
where a Galois connection is established between call-by-value
and call-by-name. In that paper, the order in which operations
are done matters, and central computations are mentioned. Again,
the central computations are not linked to submonads in that work.

\section{Background}
\label{sec:background-monads}

We recall some background on strong and commutative monads and their 
premonoidal structure. We recall also that monads may be seen as a generalisation of monoids in 
category theory.

\subsection{Strong and Commutative Monads}

We begin by recalling the definition of a monad.

\begin{defi}[Monad]\label{def:monad}
	A \emph{monad} over a category $\CC$ is an endofunctor $\TT \colon \CC \to
	\CC$ equipped with two natural transformations $\eta \colon \mathrm{id} \Rightarrow
	\TT$ and $\mu \colon \TT^2 \Rightarrow \TT,$ such that the following diagrams
  \[\input{monad-def.tikz}\qquad\qquad\input{monad-def2.tikz}\]
 commute. We call $\eta$ the \emph{unit} of $\TT$ and we say that $\mu$ is the
 \emph{multiplication} of $\TT$.
\end{defi}

Next, we recall the definition of a \emph{strong} monad. As we already
explained in the introduction, these monads are more computationally relevant
(compared to non-strong ones) for most use cases. The additional structure,
called the \emph{monadic strength}, ensures the monad interacts appropriately
with the monoidal structure of the base category.

\begin{defi}[Strong Monad]\label{def:strong-monad}
 A \emph{strong monad} over a monoidal category $(\CC,\otimes,I,\alpha,\lambda, \rho)$
 is a monad $(\TT,\eta,\mu)$ equipped with a natural transformation
 $\tau_{X,Y}:X\otimes\TT Y\to\TT(X\otimes Y),$ called \emph{left strength},
 such that the following diagrams commute:
  \[\stikz[1]{./figures/strength-def1.tikz}\]
  \[\stikz[1]{./figures/strength-def2.tikz}\]
\end{defi}

We now recall the definition of a \emph{commutative} monad which is of central
importance for our work. Compared to a strong monad, a commutative monad enjoys even
stronger coherence properties with respect to the monoidal structure of the
base category (see also \secref{sub:premonoidal}).

\begin{defi}[Commutative Monad]
 \label{def:commutative-monad}
	Let $(\TT, \eta, \mu, \tau)$ be a strong monad on a \emph{symmetric} monoidal
	category $(\CC, \otimes, I, \gamma)$. The \emph{right strength} $\tau'_{X,Y}
	\colon \TT X \otimes Y \to \TT(X \otimes Y)$ of $\TT$ is given by the
	assignment $\tau'_{X,Y} \eqdef
	\TT(\gamma_{Y,X})\circ\tau_{Y,X}\circ\gamma_{\TT X,Y}$. Then, $\TT$ is said
	to be \emph{commutative} if the following diagram:
	\begin{equation}
		\label{eq:commutative-monad}
		\stikz[1]{./figures/commutative-monad-def.tikz}
	\end{equation}
	commutes.
\end{defi}

\begin{rem}
	In the literature, the left and right strengths are sometimes called
	``strength" and ``costrength", respectively. Additionally, the two paths in
	the diagram above \eqref{eq:commutative-monad} are ``double strenghs'', and
	we write $\mathrm{dst}_{X,Y}$ for the morphism $\mu_{X \otimes Y} \circ \TT
	\tau'_{X,Y} \circ \tau_{\TT X, Y}$ and $\mathrm{dst}'_{X,Y}$ for $\mu_{X
	\otimes Y} \circ \TT \tau_{X,Y} \circ \tau'_{X, \TT Y}$.
\end{rem}

\begin{defi}[Morphism of Strong Monads \cite{jacobs-coalgebra}]\label{def:morph-monad}
	Given two strong monads $(\TT, \eta^\TT, \mu^\TT, \tau^\TT)$ and
	$(\PP,\eta^\PP, \mu^\PP, \tau^\PP)$ over a category $\CC$, a \emph{morphism
	of strong monads} is a natural transformation $\iota : \TT \Rightarrow \PP$
	that makes the following diagrams commute:
 \[\stikz[0.9]{figures/map-of-monads-def.tikz}\]
\end{defi}

Strong monads over a (symmetric) monoidal category $\CC$
and strong monad morphisms between them form a category which we denote by
writing $\mathbf{StrMnd}(\CC).$
In the situation of Definition \ref{def:morph-monad}, if $\iota$ is a
monomorphism in $\mathbf{StrMnd}(\CC)$, then $\TT$ is said
to be a \emph{strong submonad} of $\PP$ and $\iota$ is said to be a
\emph{strong submonad morphism}.

\begin{defi}[Kleisli category]\label{def:kleisli}
	Given a monad $(\TT,\eta,\mu)$ over a category $\CC$, the \emph{Kleisli
	category} $\CC_\TT$ of $\TT$ is the category whose objects are the same as
	those of $\CC$, but whose morphisms are given by $\CC_\TT(X,Y)=\CC(X,\TT Y)$.
	Composition in $\CC_\TT$ is given by $g\odot f \eqdef \mu_Z\circ\TT g\circ f$
	where $f:X\to\TT Y$ and $g:Y\to\TT Z$. The identity at $X$ is given by the
	monadic unit $\eta_X \colon X \to \TT X.$
\end{defi}

\begin{prop}[\cite{jacobs-coalgebra}]
 \label{prop:embed}
 If $\iota : \TT \naturalto \PP$ is a submonad morphism, then
 the functor $\II:\CC_\TT\to \CC_\PP,$ defined by $\II(X) = X$ on objects and
 $\II(f:X\to \TT Y) = \iota_Y\circ f \colon X \to \PP Y$ on morphisms,
 is an embedding of categories.
\end{prop}

The functor $\II$ above is the canonical embedding of $\CC_\TT$ into $\CC_\PP$
induced by the submonad morphism $\iota \colon \TT \naturalto \PP .$

\subsection{Semantics of the $\lambda$-calculus with effects}
\label{sub:sem-lambda-effects}

We present a brief summary of the work of Moggi \cite{moggi, moggi-lics}
on computational effects. This work has been very influential and it inspired the
introduction of monads in programming practice, e.g. monads in Haskell.

The grammar and typing rules for Moggi's metalanguage are summarised in
Figure~\ref{fig:moggi-grammars}. Compared to the simply-typed
$\lambda$-calculus, a type construction $\TT(-)$ is added. Given a type $A$,
the type $\TT A$ is called a \emph{monadic} type. Terms of type $\TT A$ can be thought of as
effectful computations of type $A$ in the metalanguage. The $\mathtt{ret}$
constructor can be seen as an introduction rule for monadic types. A useful intuition
for it is that a \emph{pure} or \emph{non-effectful} computation can be
seen as a monadic computation (with trivial effect). The $\mathtt{do}$ operation
performs the sequencing of monadic computations.

The equational theory for monadic types, added on top of the equational theory
for the simply-typed $\lambda$-calculus, is summarised in
Figure~\ref{fig:moggi-monad-rules}. As often done by many authors, we
implicitly identify terms that are $\alpha$-equivalent. The rules for
$\beta$-equivalence and $\eta$-equivalence are explicitly specified.

\begin{figure}[!h]
 \noindent $\text{(Types)}\quad A, B ~~::= 1 \alt A \to B\alt A\times B
 \alt \TT A$ \\
 $\quad$\\
 $\text{(Terms)} \quad M,N ~~ ::= x \alt * \alt \lambda x^A.M \alt MN \alt \pv M N$ \\
 $\text{ }\qquad\alt \pi_i M \alt \rret M \alt \ddo x M N $
 $\quad$\\
 \[\begin{array}{c}
  \infer{
   \Gamma,x \colon A\vdash x \colon A}{}
  \qquad
  \infer{
   \Gamma\vdash MN \colon B
  }{
   \Gamma\vdash M \colon A\to B
   &
   \Gamma\vdash N \colon A}
  \\[1.5ex]
  \infer{
   \Gamma\vdash * \colon 1}{}
  \qquad
  \infer{\Gamma\vdash\lambda x^A.M \colon A\to B}{\Gamma,x \colon A\vdash M \colon B}
  \qquad
  \infer{
   \Gamma\vdash\pi_i M \colon A_i}{\Gamma\vdash M \colon A_1\times A_2}
  \\[1.5ex]
  \infer{
   \Gamma\vdash \pv{M}{N} \colon A\times B
  }{
   \Gamma\vdash M \colon A
   &
   \Gamma\vdash N \colon B
  }
  \qquad
  \infer{
   \Gamma\vdash \rret M \colon \TT A}{\Gamma\vdash M \colon A}
		\qquad
  \infer{
   \Gamma\vdash\ddo x M N \colon \TT B
  }{
   \Gamma\vdash M \colon \TT A
   &
   \Gamma, x \colon A\vdash N \colon \TT B
  }
 \end{array}\]
 \caption{Grammars and typing rules.}
 \label{fig:moggi-grammars}
\end{figure}

\begin{figure}[!h]
	\resizebox{\hsize}{!}{
		$
		\begin{array}{c}
			\infer[(refl)]{
				\Gamma\vdash M=M \colon A
			}{
				\Gamma\vdash M \colon A
			}
			\qquad
			\infer[(symm)]{
				\Gamma\vdash M=N \colon A
			}{
				\Gamma\vdash N=M \colon A
			}
			\\[1.5ex]
			\infer[(trans)]{
				\Gamma\vdash M=P \colon A
			}{
				\Gamma\vdash M=N \colon A
				&
				\Gamma\vdash N=P \colon A
			}
			\\[1.5ex]
			\infer[(1.\eta)]{
				\Gamma,x \colon 1 \vdash * = x \colon A
			}{}
			\qquad
			\infer[(subst)]{
				\Gamma\vdash N[M/x] = P[M/x] \colon B
			}{
				\Gamma\vdash M \colon A
				&
				\Gamma, x \colon A \vdash N = P \colon B
			}
			\\[1.5ex]
			\infer[(\pv{}{} .eq)]{
				\Gamma\vdash \pv M N = \pv{M'}{N'} \colon A\times B
			}{
				\Gamma\vdash M=M' \colon A
				&
				\Gamma\vdash N=N' \colon B
			}
			\qquad
			\infer[(\times.\beta)]{
				\Gamma\vdash\pi_i\pv{M_1}{M_2} = M_i \colon A_i
			}{
				\Gamma\vdash M_1 \colon A_1
				&
				\Gamma\vdash M_2 \colon A_2
			}
			\\[1.5ex]
			\infer[(\times.\eta)]{
				\Gamma\vdash \pv{\pi_1 M}{\pi_2 M} = M \colon A\times B
			}{
				\Gamma\vdash M \colon A\times B
			}
			\\[1.5ex]
			\infer[(app.eq)]{
				\Gamma\vdash MN=M'N' \colon B
			}{
				\Gamma\vdash M=M' \colon A\to B
				&
				\Gamma\vdash N=N' \colon A
			}
			\\[1.5ex]
			\infer[(\lambda.eq)]{
				\Gamma\vdash \lambda x^A.M = \lambda x^A.N \colon A\to B
			}{
				\Gamma, x \colon A \vdash M = N \colon B
			}
			\qquad
			\infer[(\lambda.\beta)]{
				\Gamma\vdash (\lambda x^A. M) N = M[N/x] \colon B
			}{
				\Gamma,x \colon A \vdash M \colon B
				&
				\Gamma\vdash N \colon A
			}
			\\[1.5ex]
			\infer[(\lambda.\eta)]{
				\Gamma\vdash \lambda x^A. Mx = M \colon A\to B
			}{
				\Gamma\vdash M \colon A\to B
			}
			\qquad
			\infer[(weak)]{
				\Gamma,x \colon A \vdash M=N \colon B
			}{
				\Gamma\vdash M=N \colon B
			}
			\\[1.5ex]
			\infer[(ret.eq)]{
				\Gamma\vdash \rret M = \rret N \colon \TT A
			}{
				\Gamma\vdash M = N \colon A
			}
			\qquad
			\infer[(do.eq)]{
				\Gamma\vdash \ddo x M N = \ddo{x}{M'}{N'} \colon \TT B
			}{
				\Gamma\vdash M=M' \colon \TT A
				&
				\Gamma, x \colon A\vdash N=N' \colon \TT B
			}
			\\[1.5ex]
			\infer[(\TT.\beta)]{
				\Gamma\vdash \ddo{x}{\rret M}{N} = N[M/x] \colon \TT B
			}{
				\Gamma\vdash M \colon A
				&
				\Gamma,x \colon A\vdash N \colon \TT B
			}
			\qquad
			\infer[(\TT.\eta)]{
				\Gamma\vdash \ddo{x}{M}{\rret x} = M \colon \TT A
			}{
				\Gamma\vdash M \colon \TT A
			}
		\end{array}
		$
		}
		\caption{Equational rules.}
		\label{fig:moggi-monad-rules}
	\end{figure}

\paragraph{Denotational semantics.}
The denotational semantics of Moggi's metalanguage is formulated in a cartesian
closed category $\CC$ equipped with a strong monad $\TT$. Pure computations
are interpreted as morphisms in the category $\CC$, while monadic
computations, \emph{e.g.} $\Gamma \vdash M \colon \TT A$, are interpreted as
morphisms $\sem\Gamma \to \TT \sem A$, thus living in the Kleisli category of
$\TT$. The interpretation of the $\mathtt{ret}$ term is given by the unit of the monad
$\TT$, and the interpretation of the $\mathtt{do}$ term is defined using the
composition of the Kleisli category.

\subsection{Premonoidal Structure of Strong Monads}
\label{sub:premonoidal}

Let $\TT$ be a strong monad on a symmetric monoidal category $(\CC, I,
\otimes)$. Then, its Kleisli category $\CC_\TT$ does \emph{not} necessarily
have a canonical monoidal structure. However, it does have a canonical
\emph{premonoidal structure} as shown by Power and Robinson
\cite{power-premonoidal}. In fact, they show that this premonoidal structure is
monoidal iff the monad $\TT$ is commutative. Next, we briefly recall the
premonoidal structure of $\CC_\TT$ as outlined by them.

For every two objects $X$ and $Y$ of $\CC_\TT$, their tensor product $X \otimes
Y$ is also an object of $\CC_\TT$, but the monoidal product $\otimes$ of $\CC$
does not necessarily induce a bifunctor on $\CC_\TT$. However, by using
the left and right strengths of $\TT$, we can define two families of functors
as follows:
\begin{itemize}
 \item for any object $X$, a functor $(-\otimes_l X) \colon \CC_\TT \to
  \CC_\TT$ whose action on objects sends $Y$ to $Y\otimes X$, and sends
  $f:Y\to \TT Z$ to $\tau'_{Z,X}\circ(f\otimes X):Y\otimes X\to\TT(Z\otimes X)$;
 \item for any object $X$, a functor $(X\otimes_r -) \colon \CC_\TT \to
  \CC_\TT$ whose action on objects sends $Y$ to $X\otimes Y$, and sends
  $f:Y\to \TT Z$ to $\tau_{X,Z}\circ(X\otimes f):X\otimes Y\to\TT(X\otimes Z)$.
\end{itemize}
This categorical data satisfies the axioms and coherence properties of
\emph{premonoidal categories} as explained in \cite{power-premonoidal}, but
which we omit here because it is not essential for the development of our
results. What is important is to note that in a premonoidal category, $f \otimes_l X'$ and
$X\otimes_r g$ do not always commute. This leads us to the following definition,
which plays a crucial role in the theory of premonoidal categories and
has important links to our development.

\begin{defi}[Premonoidal Centre \cite{power-premonoidal}]\label{def:central-morph}
	Given a strong monad $(\TT, \eta, \mu, \tau)$ on a symmetric monoidal
	category $(\CC, I, \otimes)$, we say that a morphism $f:X\to Y$ in $\CC_\TT$
	is \emph{central} if for any morphism $f':X'\to Y'$ in $\CC_\TT$, the diagram
 \[ \input{def-central-morphism.tikz} \]
 commutes in $\CC_\TT$.
 The \emph{premonoidal centre} of $\CC_\TT$ is the
 subcategory $Z(\CC_\TT)$ which has the same objects as those of $\CC_\TT$ and
 whose morphisms are the central morphisms of $\CC_\TT$.
\end{defi}

In \cite{power-premonoidal}, the authors prove that $Z(\CC_\TT)$, is a
symmetric \emph{monoidal} subcategory of $\CC_\TT$. In particular, this means
that Kleisli composition and the tensor functors $(- \otimes_l X)$ and $(X
\otimes_r -)$ preserve central morphisms. However, it does not necessarily hold
that the subcategory $Z(\CC_\TT)$ is the Kleisli category for a monad over
$\CC$. Nevertheless, in this situation, the left adjoint of the Kleisli
adjunction $\JJ \colon \CC \to \CC_\TT$ always corestricts to $Z(\CC_\TT)$. We
write $\hat \JJ \colon \CC \to Z(\CC_\TT)$ to indicate this corestriction
(which need not be a left adjoint).

\begin{rem}
 In \cite{power-premonoidal}, the subcategory $Z(\CC_\TT)$ is called the
 centre of $\CC_\TT$. However, we refer to it as the \emph{premonoidal
 centre} of a premonoidal category to avoid confusion with the new
 notion of the centre of a monad that we introduce next. In the sequel, we show
 that the two notions are very strongly related to each other (Theorem
 \ref{th:centralisability}).
\end{rem}

\section{The Centre of a Strong Monad}
\label{sec:centralisable}

We begin by showing that any (necessarily strong) monad on $\Set$ has a
centre (\secref{sub:sets}) and we later show how to define the centre of a
strong monad on an arbitrary symmetric monoidal category
(\secref{sub:centre-general}). Unlike the former, the latter submonad does not
always exist, but it does exist under mild assumptions and we show that the
notion is ubiquitous.

\subsection{The Centre of a Monad on $\Set$}
\label{sub:sets}

The results we present next are a special case of our more general development
from \secref{sub:centre-general}, but we choose to devote special attention to
monads on $\Set$ for illustrative purposes.

\begin{defi}[Centre]\label{def:centre-set}
	Given a strong monad $(\TT, \eta, \mu, \tau)$ on $\Set$
	with right strength $\tau'$,
	we say that the \emph{centre} of $\TT$ at $X$, written $\ZZ X$, is the set
	\begin{align*}
		\ZZ X \eqdef \left \{ t \in \TT X \ |\ \forall Y \in \Ob(\Set). \forall s \in \TT Y.
		\mu(\TT\tau'(\tau(t,s))) = \mu(\TT\tau(\tau'(t,s))) \right \} .
	\end{align*}
	We write $\iota_X \colon \ZZ X \subseteq \TT X$ for the indicated subset inclusion.
\end{defi}

In other words, the centre of $\TT$ at $X$ is the subset of $\TT X$ which
contains all monadic elements for which \eqref{eq:commutative-monad} holds when
the set $X$ is fixed and the set $Y$ ranges over all sets.


Notice that $\ZZ X \supseteq \eta_X(X),$ \emph{i.e.} the centre of $\TT$ at $X$
always contains all monadic elements which are in the image of the monadic
unit. This follows easily from the axioms of strong monads. In fact, the
assignment $\ZZ(-)$ extends to a \emph{commutative submonad} of $\TT$. In
particular, the assignment $\ZZ(-)$ extends to a functor $\ZZ \colon \Set \to \Set$
when we define \( \ZZ f \eqdef \TT f|_{\ZZ X} \colon \ZZ X \to \ZZ Y, \) for any
function $f: X \to Y,$ where $\TT f|_{\ZZ X}$ indicates the restriction of $\TT
f \colon \TT X \to \TT Y$ to the subset $\ZZ X.$ Moreover, for any two sets $X$ and
$Y$, the monadic unit $\eta_X \colon X \to \TT X$, the monadic multiplication $\mu_X
: \TT^2 X \to \TT X$, and the monadic strength $\tau_{X,Y} \colon X \times \TT Y \to
\TT(X \times Y)$ (co)restrict respectively to functions $\eta_X^\ZZ \colon X \to \ZZ
X$, $\mu_X^\ZZ \colon \ZZ^2 X \to \ZZ X$ and $\tau_{X,Y}^\ZZ \colon X \times \ZZ Y \to
\ZZ(X \times Y)$. That the above four classes of functions (co)restrict as
indicated follows from our more general treatment presented in the next
section. It then follows, as a special case of Theorem
\ref{th:centralisability}, that the data we just described constitutes a
commutative submonad of $\TT$.

\begin{thm}
	\label{thm:centre-set}
	The assignment $\ZZ(-)$ can be extended to a \emph{commutative submonad} $(\ZZ, \eta^\ZZ, \mu^\ZZ, \tau^\ZZ)$ of $\TT$
	with the inclusions $\iota_X \colon \ZZ X \subseteq \TT X$ being the submonad morphism.
	Furthermore, there is a canonical isomorphism of categories $\Set_\ZZ \cong
	Z(\Set_\TT)\footnote{Theorem \ref{th:centralisability} states precisely in what sense this isomorphism is canonical.}.$
\end{thm}

The final statement of Theorem \ref{thm:centre-set}
shows that the Kleisli category of $\ZZ$ is canonically
isomorphic to the premonoidal centre of the Kleisli category of $\TT$. Because
of this, we are justified in saying that $\ZZ$ is not just \emph{a} commutative
submonad of $\TT$, but rather it is \emph{the centre} of $\TT,$ which
is necessarily commutative (just like the centre of a monoid is a commutative
submonoid). In \secref{sub:specific-examples} we provide concrete examples of
monads on $\Set$ and their centres and we see that the construction
of the centre aligns nicely with our intuition.

\subsection{The General Construction of the Centre}
\label{sub:centre-general}

Throughout the remainder of
the section, we assume we are given a symmetric monoidal category
$(\CC,\otimes,I,\alpha,\lambda, \rho, \gamma)$ and a strong monad $(\TT, \eta,
\mu, \tau)$ on it with right strength $\tau'$.

In $\Set$, the centre is defined pointwise through subsets of $\TT X$ which
only contain elements that satisfy the coherence condition for a commutative
monad. However, $\CC$ is an arbitrary symmetric monoidal category, so we
cannot easily form subojects in the required way. This leads us to the
definition of a \emph{central cone} which allows us to overcome this problem.

\begin{defi}[Central Cone]\label{def:central-cone}
	Let $X$ be an object of $\CC$. A \emph{central cone} of $\TT$ at $X$ is given
	by a pair $(Z, \iota)$ of an object $Z$ and a morphism $\iota \colon Z \to
	\TT X,$ such that for any object $Y,$ the diagram
	\[\input{def-central-cone.tikz}\]
	commutes.
	If $(Z, \iota)$ and $(Z', \iota')$ are two central cones of $\TT$ at $X$,
	then a \emph{morphism of central cones} $\varphi \colon (Z', \iota') \to (Z,
	\iota)$ is a morphism $\varphi \colon Z' \to Z,$ such that $\iota \circ \varphi =
	\iota'.$ Thus central cones of $\TT$ at $X$ form a category. A
	\emph{terminal central cone} of $\TT$ at $X$ is a central cone $(Z, \iota)$
	for $\TT$ at $X$, such that for any central cone $(Z', \iota')$ of $\TT$ at
	$X$, there exists a unique morphism of central cones $\varphi \colon (Z', \iota')
	\to (Z, \iota).$ In other words, it is the terminal object in the category of
	central cones of $\TT$ at $X$.
\end{defi}

In particular, Definition \ref{def:centre-set} gives a terminal central cone
for the special case of monads on $\Set.$ The names ``central morphism'' (in
the premonoidal sense, see \secref{sub:premonoidal}) and ``central cone''
(above) also hint that there should be a relationship between them. In fact, the
two notions are equivalent.

\begin{prop}\label{prop:central}
	Let $f:X\to\TT Y$ be a morphism in $\CC$. The pair $(X,f)$ is a central cone
	of $\TT$ at $Y$ iff $f$ is central in $\CC_\TT$ in the premonoidal sense
	(Definition~\ref{def:central-morph}).
\end{prop}
\begin{proof}
	Let $(X,f)$ be a central cone and let $f' \colon X'\to\TT Y'$ be a morphism.
	The following diagram: \[\scalebox{0.8}{\input{central-cone-to-morph.tikz}} \]
	commutes because: (1) $\CC$ is monoidal; (2) $\tau'$ is natural; (3) $\tau$
	is natural; and (4) the pair $(X, f)$ is a central cone. Therefore, the
	morphism $f$ is central in the premonoidal sense.\\ For the other direction,
	if $f$ is central in $\CC_\TT$, the following diagram:
	\[\input{central-morph-to-cone.tikz} \] commutes because: (1) $\tau$ is natural;
	(2) $f$ is a central morphism; all remaining subdiagrams commute trivially.
	This shows the pair $(X,f)$ is a central cone.
\end{proof}

From now on, we rely heavily on the fact that central cones and central
morphisms are equivalent notions, and we use Proposition \ref{prop:central}
implicitly in the sequel. On the other hand, \emph{terminal} central cones are
crucial for our development, but it is unclear how to introduce a similar
notion of ``terminal central morphism'' that is useful. For this reason, we
prefer to work with (terminal) central cones.

It is easy to see that if a terminal central cone for $\TT$ at $X$ exists, then
it is unique up to a unique isomorphism of central cones. Also, one can easily
prove that if $(Z, \iota)$ is a terminal central cone, then $\iota$ is a
monomorphism. The main definition of this subsection follows next and gives
the foundation for constructing the centre of a strong monad.

\begin{defi}[Centralisable Monad]
	\label{def:centralisable}
	We say that the monad $\TT$ is \emph{centralisable} if, for any object $X$, a
	terminal central cone of $\TT$ at $X$ exists. In this situation, we write
	$(\ZZ X, \iota_X)$ for the terminal central cone of $\TT$ at $X$.
\end{defi}

In fact, for a centralisable monad $\TT$, its terminal central cones induce a
commutative submonad $\ZZ$ of $\TT$, as the next theorem shows, and its proof
reveals constructively how the monad structure arises from them.

\begin{thm}\label{th:submonad-from-cone}
	If the monad $\TT$ is centralisable, then the assignment $\ZZ(-)$ extends to a
	commutative monad $(\ZZ, \eta^\ZZ, \mu^\ZZ, \tau^\ZZ)$ on $\CC$. Moreover,
	$\ZZ$ is a commutative submonad of $\TT$ and the morphisms $\iota_X \colon \ZZ X
	\to \TT X$ constitute a monomorphism of strong monads $\iota \colon \ZZ
	\Rightarrow \TT$.
\end{thm}

The proof of this theorem relies on several lemmas that are formulated next.

\begin{lem}\label{lem:precompose-central}
	If $(X,f \colon X\to \TT Y)$ is a central cone of $\TT$ at $Y,$ then for any
	$g \colon Z\to X$, it follows that $(Z,f\circ g)$ is a central cone of $\TT$
	at $Y$.
\end{lem}
\begin{proof}
	This is obtained by precomposing the definition of central cone by $g\otimes \id$. The diagram
	\[ \input{central-precomposing.tikz} \]
	commutes directly from the definition of central cone for $f$.
\end{proof}

\begin{lem}\label{lem:postcompose-central}
	If $(X,f \colon X\to \TT Y)$ is a central cone of $\TT$ at $Y$ then for any
	$g \colon Y\to Z$, it follows that $(X,\TT g\circ f)$ is a central cone of
	$\TT$ at $Z$.
\end{lem}
\begin{proof}
	The naturality of $\tau$ and $\mu$ allow us to push the application of
	$g$ to the last postcomposition, in order to use the central property
	of $f$. In more details, the following diagram:
	\[\scalebox{0.8}{\input{central-postcomposing.tikz}}\] commutes, because:
	(1) $f$ is a central cone, (2) $\tau'$ is natural, (3) $\tau$ is
	natural, (4) $\mu$ is natural (5) $\tau$ is natural, (6) $\tau'$ is
	natural, (7) $\mu$ is natural.
\end{proof}

\begin{lem}\label{lem:monic}
	If $(Z,\iota)$ is a terminal central cone of $\TT$ at $X$, then $\iota$ is a monomorphism.
\end{lem}
\begin{proof}
	Let us consider $f,g:Y\to Z$ such that $\iota\circ f=\iota\circ g$; this
	morphism is a central cone at $X$ (Lemma~\ref{lem:precompose-central}), and
	since $(Z,\iota)$ is a terminal central cone, it factors uniquely through
	$\iota$. Thus $f = g$ and therefore $\iota$ is monic.
\end{proof}

\begin{proof}[Proof of Theorem~\ref{th:submonad-from-cone}]
	First let us describe the functorial structure of $\ZZ$. Recall that $\ZZ$
	maps every object $X$ to its terminal central cone at $X$. Let $f:X\to
	Y$ be a morphism. We know that $\TT f\circ\iota_X:\ZZ X\to \TT Y$ is a
	central cone according to Lemma~\ref{lem:postcompose-central}.
	Therefore, we define $\ZZ f$ as the unique map such that the following
	diagram commutes:
	\[\input{z-functor.tikz}\]

	It follows directly that $\ZZ$ maps the identity to the identity, and that
	$\iota$ is natural. The assignment $\ZZ(-)$ also preserves composition, which follows from
	the commutative diagram
	\[\input{z-functor-comp.tikz}\]
	This proves that $\ZZ$ is a functor. Next, we describe its monad structure
	and after that we show that it is commutative. \\ The monadic unit
	$\eta_X$ is central, because it is the identity morphism in
	$Z(\CC_\TT)$, thus it factors through $\iota_X$ to define $\eta^\ZZ_X$.
	\[\input{central-unit.tikz}\]
	Next, observe that, by definition, $\mu_X\circ\TT\iota_X\circ\iota_{\ZZ X} =
	\iota_X\odot\iota_{\ZZ X}$, where $(- \odot -)$ indicates Kleisli
	composition. Since $\iota$ is central and Kleisli composition preserves
	central morphisms (see Definition~\ref{def:central-morph}, central morphisms
	form a subcategory of the Kleisli category), it follows that this morphism
	factors through $\iota_X$ and we use this to define $\mu^\ZZ_X$ as in the
	diagram below.
	\[\input{central-mult.tikz}\]
	Again, by definition, $\tau_{A,B}\circ(A\otimes\iota_B) = A\otimes_r\iota_B$.
	Central morphisms are preserved by the premonoidal products (see
	\ref{sub:premonoidal}) and therefore, this morphism factors through
	$\iota_{A\otimes B}$ which we use to define $\tau^\ZZ_{A,B}$ as in the
	diagram below.
	\[\input{central-strength.tikz}\]
	Note that the last three diagrams are exactly those of a morphism of strong
	monads (see Definition \ref{def:morph-monad}). Using the fact that $\iota$ is
	monic (see Lemma~\ref{lem:monic}), the following commutative diagram shows
	that $\eta^\ZZ$ is natural.
	\[\input{eta-z-natural.tikz}\]
	This diagram commutes, because:
	(1) definition of $\eta^\ZZ$; (2) $\iota$ is natural; (3) $\eta$ is natural;
	and (4) definition of $\eta^\ZZ$. Thus, we have proven that for any $f \colon
	X\to Y$, $\iota_Y\circ\ZZ f\circ\eta^\ZZ_X = \iota_Y\circ\eta^\ZZ_Y\circ f$.
	Since $\iota$ is monic, it follows $\ZZ f\circ\eta^\ZZ_X =\eta^\ZZ_Y\circ f$
	which proves that $\eta^\ZZ$ is natural. We prove the commutativity of the remaining
	diagrams using similar reasoning.

	The following commutative diagram shows that $\mu^\ZZ$ is natural.
	\[\input{mu-z-natural.tikz}\]
	This diagram commutes, because:
	(1) definition of $\mu^\ZZ$; (2) $\iota$ is natural; (3) $\mu$ is natural;
	(4) $\iota$ is natural; and (5) definition of $\mu^\ZZ$. The
	following two commutative diagrams show that $\tau^\ZZ$ is natural in both components.
	\[\input{tau-z-natural.tikz}\]
	This diagram commutes, because:
	(1) definition of $\tau^\ZZ$; (2) $\iota$ is natural; (3) $\tau$ is natural;
	(4) $\iota$ is natural; and (5) definition of $\tau^\ZZ$.
	\[\input{tau-z-natural-1.tikz}\]
	This diagram commutes, because:
	(1) definition of $\tau^\ZZ$; (2) $\iota$ is natural; (3) $\tau$ is natural;
	(4) $\iota$ is natural; and (5) definition of $\tau^\ZZ$. The following
	commutative diagrams prove that $\ZZ$ is a monad.
	\[\scalebox{0.8}{\input{z-monad-1.tikz}}\]
	These diagrams commute, because:
	(1) and (2) involve the definition of $\mu^\ZZ$ and the naturality of $\iota$
	and $\mu^\ZZ$; (3) is by definition of monad; (4) definition of $\mu^\ZZ$ and
	(5) also; (6) and (7) involve the definition of $\eta^\ZZ$ and the
	naturality of $\iota$ and $\eta^\ZZ$; (8) is by definition of monad; (9)
	definition of $\mu^\ZZ$ and (10) also.
	
	We can prove that $\ZZ$ is strong with very similar diagrams. The commutative diagram:
	\begin{equation}\label{eq:proof-commutative}
		\scalebox{0.7}{\input{z-monad-commutative.tikz}}
	\end{equation}
	proves that $\ZZ$ is a commutative monad. It commutes, because: (1) $\tau'^\ZZ$ is natural;
	(2) definition of $\tau^\ZZ$; (3) $\tau^\ZZ$ is natural; (4) $\CC$ is
	monoidal; (5) definition of $\tau'^\ZZ$; (6) $\iota$ is natural; (7)
	definition of $\mu^\ZZ$; (8) definition of $\tau^\ZZ$; (9) $\iota$ is
	central; (10) definition of $\tau'^\ZZ$; (11) $\iota$ is natural; and (12)
	definition of $\mu^\ZZ$.
\end{proof}

Theorem~\ref{th:submonad-from-cone} shows that centralisable monads always
induce a canonical commutative submonad. Next, we justify why this submonad
should be seen as the centre of $\TT.$ Note that since $\ZZ$ is a submonad of
$\TT$, we know that $\CC_\ZZ$ canonically embeds into $\CC_\TT$ (see
Proposition \ref{prop:embed}). The next theorem shows that this embedding
factors through the premonoidal centre of $\CC_\TT$, and moreover, the two
categories are isomorphic.

\begin{thm}\label{th:iso-of-categories}
	In the situation of Theorem \ref{th:submonad-from-cone}, the canonical
	embedding functor $\II \colon \CC_\ZZ\to\CC_\TT$ corestricts to an isomorphism of
	categories $\CC_\ZZ\cong Z(\CC_\TT)$.
\end{thm}
\begin{proof}
	That 
	$\II$ corestricts as indicated follows easily: for any morphism $f \colon X \to
	\ZZ Y$, we have that $\II f = \iota_Y \circ f$ which is central by Lemma
	\ref{lem:precompose-central}. Let us write $\hat \II$ for the corestriction
	of $\II$ to $Z(\CC_\TT)$. Next, to prove that $\hat \II \colon \CC_\ZZ \to
	Z(\CC_\TT)$ is an isomorphism, we define the inverse functor
	$G:Z(\CC_\TT)\to\CC_\ZZ$.

	On objects, we have $G(X) \eqdef X.$ To define its mapping on morphisms,
	observe that if $f \colon X\to\TT Y$ is a central morphism (in the
	premonoidal sense), then $(X,f)$ is a central cone of $\TT$ at $Y$
	(Proposition \ref{prop:central}) and therefore there exists a unique morphism
	$f^\ZZ \colon X\to\ZZ Y$ such that $\iota_Y \circ f^\ZZ = f$; we define $Gf
	\eqdef f^\ZZ$. The proof that $G$ is a functor is direct considering that
	any $f^\ZZ$ is a morphism of central cones and that all components of $\iota$
	are monomorphisms.

	To show that $\hat \II$ and $G$ are mutual inverses, let $f:X\to \TT Y$ be a
	morphism of $Z(\CC_\TT)$, \emph{i.e.} a central morphism. Then, $\hat \II
	Gf=\iota_Y\circ f^\ZZ = f$ by definition of morphism of central cones (see
	Definition~\ref{def:central-cone}). For the other direction, let $g:X\to \ZZ
	Y$ be a morphism in $\CC$. Then, $\iota_Y\circ G\hat \II g=\iota_Y\circ
	(\iota_Y\circ g)^\ZZ = \iota_Y \circ g$ by Definition~\ref{def:central-cone}
	and thus $G\hat \II g=g$ since $\iota_Y$ is a monomorphism (Lemma
	\ref{lem:monic}).
\end{proof}

It should now be clear that Theorem \ref{th:submonad-from-cone} and Theorem
\ref{th:iso-of-categories} show that we are justified in naming the submonad
$\ZZ$ as \emph{the} centre of $\TT$. The existence of terminal central cones
is not only sufficient to construct the centre (as we just showed), but it also
is necessary and we show this next. Furthermore, we provide another equivalent
characterisation in terms of the premonoidal structure of the monad.

\begin{thm}[Centre]\label{th:centralisability}
	Let $\CC$ be a symmetric monoidal category and $\TT$ a strong monad on it.
	The following are equivalent:
	\begin{enumerate}
		\item \label{cond:1} For any object $X$ of $\CC$, $\TT$ admits a terminal
			central cone at $X$;
		\item \label{cond:2} There exists a commutative submonad $\ZZ$ of $\TT$
			(which we call \emph{the centre} of $\TT$) such that the canonical
			embedding functor $\II:\CC_\ZZ\to\CC_\TT$ corestricts to an
			isomorphism of categories $\CC_\ZZ\cong Z(\CC_\TT)$;
		\item \label{cond:3} The corestriction of the Kleisli left adjoint
			$\JJ:\CC\to\CC_\TT$ to the premonoidal centre $\hat \JJ:\CC\to
			Z(\CC_\TT)$ also is a left adjoint.
	\end{enumerate}
\end{thm}
\begin{proof}
	We prove $1 \Rightarrow 2 \Rightarrow 3 \Rightarrow 1$.

	$(1\Rightarrow 2):$
	By Theorem~\ref{th:submonad-from-cone} and Theorem~\ref{th:iso-of-categories}.

	$(2\Rightarrow 3):$
	Let us consider the Kleisli left adjoint $\JJ^\ZZ$ associated to the
	monad $\ZZ$. All our hypotheses can be summarised by the diagram
	\[\input{main-theorem-2-to-3.tikz}\] where $\hat \II \colon \CC_\ZZ\cong
	Z(\CC_\TT)$ is the corestriction of $\II$. This diagram commutes,
	because $\ZZ$ is a submonad of $\TT$ (recall also that $\hat \JJ$ is
	the indicated corestriction of $\JJ$, see \secref{sub:premonoidal}).
	Since $\hat \II$ is an isomorphism, then $\hat \JJ = \hat \II \circ
	\JJ^\ZZ$ is the composition of two left adjoints and it is therefore
	also a left adjoint.

	$(3\Rightarrow 1):$
	Let $\mathcal R \colon Z(\CC_\TT)\to\CC$ be the right adjoint of $\hat \JJ$
	and let $\varepsilon$ be the counit of the adjunction. We show that the
	pair $(\mathcal R X, \varepsilon_X)$ is the terminal central cone of $\TT$ at
	$X$. First, since $\varepsilon_X$ is a morphism in $Z(\CC_\TT)$, it follows
	that it is central. Thus the pair $(\mathcal R X, \varepsilon_X)$ is a
	central cone of $\TT$ at $X$. Next, let $\Phi \colon Z(\CC_\TT)[\hat \JJ Y,
	X] \cong \CC[Y, \mathcal R X]$ be the natural bijection induced by the
	adjunction. If $f \colon Y \to \TT X$ is central, meaning a morphism of
	$Z(\CC_\TT)$, the diagram below left commutes in $Z(\CC_\TT)$, or
	equivalently, the diagram below right commutes in $\CC$:
	\[\input{adjoint-diagram-unique-centre.tikz}\] Note that the pair $(Y, f)$ is
	equivalently a central cone for $\TT$ at $X$ (by Proposition
	\ref{prop:central}). Thus $f$ uniquely factors through the counit
	$\varepsilon_X: \mathcal RX\to\TT X$ and therefore $(\mathcal
	RX,\varepsilon_X)$ is the terminal central cone of $\TT$ at $X$.
\end{proof}

This theorem shows that Definition \ref{def:centralisable} may be stated by
choosing any one of the above equivalent criteria. We note that the first
condition is the easiest to verify in practice. The second one is the
most useful for providing a computational interpretation, as we do in the
sequel. The third condition provides an important link to premonoidal
categories.

\begin{exa}
	\label{ex:free-monad}
	Given a monoid $(M,e,m)$, consider the free monad induced by $M$, also known
	as the \emph{writer monad}, which we write as $\TT = (- \times M) \colon \Set
	\to \Set$. The centre $\ZZ$ of $\TT$ is given by the commutative monad
	$(-\times Z(M)) \colon \Set \to \Set$, where $Z(M)$ is the centre of the
	monoid $M$ and where the monad data is given by the (co)restrictions of the
	monad data of $\TT$. Note that $\TT$ is a commutative monad iff $M$ is a
	commutative monoid. See also Example \ref{ex:counter}.
\end{exa}

\subsection{A Non-centralisable Monad}
\label{sub:bad-monad}

In $\Set$, the terminal central cones used to define the centre are defined by
taking appropriate subsets. One may wonder what happens if not all subsets of
a given set are objects of the category. The following example describes such
a situation, which gives rise to a non-centralisable strong monad.

\begin{exa}
	\label{ex:counter}
	Consider the Dihedral group $\mathbb{D}_4,$ which has $8$ elements. Its
	centre $Z(\mathbb{D}_4)$ is non-trivial and has 2 elements. Let $\CC$ be the
	full subcategory of $\Set$ with objects that are finite products of the set
	$\mathbb{D}_4$ with itself. This category has a cartesian structure, and the
	terminal object is the singleton set (which is the empty product). Notice
	that every object in this category has a cardinality that is a power of $8$.
	Therefore the cardinality of every homset of $\CC$ is a power of $8$. Since
	$\CC$ has a cartesian structure and since $\mathbb{D}_4$ is a monoid, we can
	consider the writer monad $\MM \eqdef (- \times \mathbb D_4) \colon \CC \to \CC$
	induced by $\mathbb{D}_4$, which can be defined in the same way as in Example
	\ref{ex:free-monad}. It follows that $\MM$ is a strong monad on $\CC$.
	However, it is easy to show that this monad is not centralisable. Assume (for
	contradiction) that there is a monad $\ZZ \colon \CC \to \CC$ such that
	$\CC_\ZZ\cong Z(\CC_\MM)$ (see Theorem \ref{th:centralisability}). Next,
	observe that the homset $Z(\CC_\MM)[1,1]$ has the same cardinality as the
	centre of the monoid $\mathbb D_4$, \emph{i.e.} its cardinality is $2$. However,
	$\CC_\ZZ$ cannot have such a homset since $\CC_\ZZ[X,Y] = \CC[X,\ZZ Y]$ which
	must have cardinality a power of $8$. Therefore there exists no such monad
	$\ZZ$ and $\MM$ is not centralisable.
\end{exa}

Besides this example and any further attempts at constructing non-centralisable
monads for this sole purpose, we do not know of any other strong monad in the
literature that is not centralisable.
In the next section, we present many
examples of centralisable monads and classes of centralisable monads which show
that our results are widely applicable.

\section{Examples of Centres of Strong Monads}
\label{sec:examples}

In this section, we show how we can make use of the mathematical results we
already established in order to reason about the centres of monads of interest.

\subsection{Categories whose Strong Monads are Centralisable}
\label{sub:examples}

We saw earlier that every (strong) monad on $\Set$ is centralisable. In fact,
this is also true for many other naturally occurring categories. For example,
in many categories of interest, the objects of the category have a suitable
notion of subobject (e.g. subsets in $\Set$, subspaces in $\mathbf{Vect}$) and
the centre can be constructed in a similar way to the one in $\Set$.

\begin{exa}
	Let $\DCPO$ be the category whose objects are directed-complete partial
	orders and whose morphisms are Scott-continuous maps between them. Every
	strong monad on $\DCPO$ with respect to its cartesian structure is
	centralisable. The easiest way to see this is to use Theorem
	\ref{th:centralisability} \eqref{cond:1}. Writing $\TT \colon \DCPO \to \DCPO$ for
	an arbitrary strong monad on $\DCPO$, the terminal central cone of $\TT$ at
	$X$ is given by the subdcpo $\ZZ X \subseteq \TT X$ which has the underlying
	set
	\[ \ZZ X \eqdef \left \{ t \in \TT X \ |\ \forall Y \in \Ob(\DCPO). \forall s \in \TT Y.\ \mu(\TT\tau'(\tau(t,s))) = \mu(\TT\tau(\tau'(t,s))) \right \} . \]
	That $\ZZ X$ (with the inherited order) is a subdcpo of $\TT X$ follows
	easily by using the fact that $\mu, \tau, \tau'$ and $\TT$ are
	Scott-continuous. Therefore, the construction is fully analogous to the one
	in $\Set$.
\end{exa}

\begin{exa}
	Let $\TOP$ be the category whose objects are topological
	spaces, and whose morphisms are continuous maps between them.
	Every strong monad on $\TOP$ with respect to its cartesian structure
	is centralisable.
	Using Theorem \ref{th:centralisability} \eqref{cond:1} and writing $\TT :
	\TOP \to \TOP$ for an arbitrary strong monad on $\TOP$, the terminal central
	cone of $\TT$ at $X$ is given by the space $\ZZ X \subseteq \TT X$ which has
	the underlying set 
	\[  \ZZ X \eqdef \left \{ t \in \TT X \ |\ \forall Y \in 	\Ob(\TOP). \forall s \in \TT Y. \mu(\TT\tau'(\tau(t,s))) = \mu(\TT\tau(\tau'(t,s))) \right \} \]
	and whose topology is the subspace topology
	inherited from $\TT X$.
\end{exa}

\begin{exa}
	\label{ex:meas}
	Every strong monad on the category $\Meas$ (whose objects are measurable
	spaces and the morphisms are measurable maps between them) is centralisable.
	The construction is fully analogous to the previous example, but instead of
	the subspace topology, we equip the underlying set with the subspace
	$\sigma$-algebra inherited from $\TT X$ (which is the smallest
	$\sigma$-algebra that makes the subset inclusion map measurable).
\end{exa}

\begin{exa}
	Let $\mathbf{Vect}$ be the category whose objects are vector spaces, and
	whose morphisms are linear maps between them. Every strong monad on
	$\mathbf{Vect}$ with respect to the usual symmetric monoidal structure is
	centralisable. One simply defines the subset $\ZZ X$ as in the other examples
	and shows that this is a linear subspace of $\TT X$. That this is the
	terminal central cone is then obvious.
\end{exa}

The above categories, together with the category $\Set$, are not meant to
provide an exhaustive list of categories for which all strong monads are
centralisable. Indeed, there are many more categories for which this is true.
The purpose of these examples is to illustrate how we may use Theorem
\ref{th:centralisability} \eqref{cond:1} to construct the centre of a strong
monad. Changing perspective, the proof of the next proposition
uses Theorem \ref{th:centralisability} \eqref{cond:3}.

\begin{prop}\label{prop:suffcond}
	Let $\mathbf{C}$ be a symmetric monoidal closed category that is total,
	\emph{i.e.} a locally small category whose Yoneda embedding has a left
	adjoint. Then all strong monads over $\mathbf{C}$ are centralisable.
\end{prop}
\begin{proof}
	Given any strong monad on $\mathbf{C}$, we first show that the
	co-restriction of the Kleisli inclusion is cocontinuous. We consider
	an initial cocone $\epsilon :\Delta_c \Rightarrow J $ over a diagram
	$J: D \to \mathbf{C}$ in $\mathbf{C}$. Its image
	$\hat{\mathcal{J}}\epsilon :\Delta_{c} \Rightarrow \hat{\mathcal{J}}
	\circ J $ is a cocone in $Z(\mathbf{C}_{\mathcal{T}})$. We show
	that it is initial. We consider another cocone $\epsilon' :\Delta_{c'}
	\Rightarrow \hat{\mathcal{J}} \circ J $ in
	$Z(\mathbf{C}_{\mathcal{T}})$. Since $\mathcal{J}$ is a left adjoint,
	it is cocontinuous and then $\mathcal{J}\epsilon :\Delta_{c}
	\Rightarrow \mathcal{J} \circ J $ is an initial cocone in
	$\mathbf{C}_{\mathcal{T}}$. So there is a unique arrow $h: c \to c' $
	in $\mathbf{C}_{\mathcal{T}}$ such that $h \circ \mathcal{J}\epsilon =
	\epsilon' $. We have to show that $h$ also is in
	$Z(\mathbf{C}_{\mathcal{T}})$, in other words, that the following
	diagram commutes for all $f:X\to Y$
	\begin{center}
		\input{suff-cond-com.tikz}
	\end{center}
	The cocone $\epsilon$ is initial in $\mathbf{C}$ and since the functors $-
	\otimes X$ are assumed to be cocontinuous, $\mathcal{J}(\epsilon \otimes X)$
	also is an initial cocone in $\mathbf{C}_{\mathcal{T}}$. Its components are
	then jointly epic and checking the commutativity of the diagram below amounts
	to check the commutativity of the following diagram for each component:
	\begin{center}
		\input{suff-cond-ct.tikz}
	\end{center}
	Since the composition $f\circ \mathcal{J}(g)$ in $\mathbf{C}_\TT $
	corresponds to $f\circ g $ in $\mathbf{C}$, this is equivalent to the
	following diagram in $\mathbf{C}$:
	\begin{center}
		\scalebox{0.8}{\input{suff-cond-c.tikz}}
	\end{center}
	It commutes, because: \textcolor{blue}{(1)} is the definition of $h$; \textcolor{blue}{(2)}
	is the exchange law; \textcolor{blue}{(3)} is the naturality of the strength;
	\textcolor{blue}{(4)} is again the definition of $h$ together with
	functoriality of $\TT(- \otimes Y)$; and \textcolor{blue}{(5)} is the fact
	that $\epsilon'_A $ is by definition central.
	We can then conclude that $h$ is central and so the corestriction
	$\hat{\mathcal{J}}$ is cocontinuous.
	Then by the adjoint functor theorem for total categories
	\cite{street1978yoneda}, $\hat{\mathcal{J}}$ is a left adjoint, and by
	Theorem \ref{th:centralisability} it follows that the corresponding strong
	monad is centralisable.
\end{proof}

\begin{exa}
	Any category which is the Eilenberg-Moore category of a commutative monad
	over $\Set$ is total \cite{kelly1986survey}. Furthermore it is symmetric
	monoidal closed \cite{keigher1978symmetric}, thus all strong monads on it are
	centralisable. This includes: the category $\Set_* $ of pointed sets and
	point preserving functions (algebras of the lift monad); the category
	$\mathbf{CMon}$ of commutative monoids and monoid homomorphisms (algebras of
	the commutative monoid monad); the category $\mathbf{Conv}$ of convex sets
	and linear functions (algebras of the distribution monad); and the category
	$\mathbf{Sup}$ of complete semilattices and sup-preserving functions
	(algebras of the powerset monad).
\end{exa}

\begin{exa}
	Any presheaf category $\Set^{\mathbf{C}^{\mathrm{op}}}$ over a small category
	$\mathbf{C}$ is total \cite{kelly1986survey} and cartesian closed, thus all strong monads on it (with respect to the cartesian structure) are
	centralisable. This includes: the category $\Set^{A^{\mathrm{op}}}$, where $A$ is the
	category with two objects and two parallel arrows, which can be seen as the
	category of directed multi-graphs and graph homomorphisms; the category
	$\Set^{G^{\mathrm{op}}}$, where $G$ is a group seen as a category, which can be seen
	as the category of $G$-sets (sets with an action of $G$) and equivariant
	maps; and the topos of trees $\Set^{\mathbb{N}^{\mathrm{op}}}$. If
	$\mathbf{C}$ is symmetric monoidal, then the Day convolution product makes
	$\Set^{\mathbf{C}^{\mathrm{op}}}$ symmetric monoidal closed \cite{day1970construction},
	hence all strong monads on it with respect to the Day convolution monoidal
	structure also are centralisable.
\end{exa}

\begin{exa}
	Any Grothendieck topos is cartesian closed and total, therefore it satisfies
	the conditions of Proposition \ref{prop:suffcond}.
\end{exa}

\subsection{Specific Examples of Centralisable Monads}
\label{sub:specific-examples}

In this subsection, we consider specific monads and construct their centres.

\begin{exa}
	\label{ex:commutative}
	Every commutative monad is naturally isomorphic to its centre.
\end{exa}

\begin{exa}
	\label{ex:continuation}
	Let $S$ be a set. The continuation monad is the endofunctor $\TT =
	[[-,S],S]:\Set\to\Set$, together with unit $\eta_X = x \mapsto \lambda f. f(x)$,
	multiplication 
	$\mu_X : F \mapsto\lambda g.F(\lambda h. h(g))$ and strength $\tau_{X,Y} = (x,f)
	\mapsto\lambda g. f(\lambda y. g(x,y))$. Note that, if $S$ is the empty set
	or a singleton set, then $\TT$ is commutative, so we are in the situation
	of Example \ref{ex:commutative}.  Otherwise, the centre of $\TT$
	at $X$ consists of the elements $\varphi\in \TT X$, such that $\forall Y, \forall
	\psi \in \TT Y, \forall g\in [X\times Y,S]$:
	\[ \psi(\lambda y.\varphi(\lambda x.  g(x,y))) = \varphi(\lambda x.\psi(\lambda y. 	g(x,y))) . \]
	Suppose that $Y=\{*\}$. Then $\psi$ is a function $S\to S$ and
	the condition becomes: \[\psi(\varphi(\lambda x. g(x))) = \varphi(\lambda
	x.\psi(g(x)))\] If $\varphi$ is constant, this does not hold for any
	$\psi$. So $\varphi$ is necessarily of the form $\lambda f.\sigma(f(z))$ with
	$\sigma:S\to S$ and $z\in X$, which has to satisfy $\forall \psi:S\to S,
	\forall g:X\to S$: \[\psi(\sigma(g(z))) = \sigma(\psi(g(z))) . \] 
	It follows that $\sigma$ can
	only be the identity, so $\varphi$ can only be $\eta_X(z) = \lambda
	f.f(z)$ for some $z\in X$. Indeed, these functions are solutions of the
	problem.  Thus, when $S$ is not trivial, $\ZZ X = \eta_X(X) \simeq X$ and
	the central submonad of $\TT$ is naturally isomorphic to the identity
	monad.
\end{exa}

Example \ref{ex:continuation} shows that the centre of a monad may be trivial
in the sense that it is precisely the image of the monadic unit and this is the
least it can be. At the other extreme, Example \ref{ex:commutative} shows that
the centre of a commutative monad coincides with itself, as one would expect.
Thus, the monads that have interesting centres are those monads which are
strong but not commutative, and which have non-trivial centres, such as the one
in Example \ref{ex:free-monad}. The following example highlights a monad that
has a non-trivial centre.

\begin{exa}
	\label{ex:list}
	Consider the well-known list monad $\TT \colon \Set\to\Set$ that is given
	by $\TT X = \bigsqcup_{n\geq 0} X^n.$ Then, the centre of $\TT$ is
	naturally isomorphic to the maybe monad, that is to say the endofunctor
	$\MM \colon \Set \to \Set$ such that $\MM X = X \uplus \{ * \}$.
\end{exa}

Another interesting example of a strong monad with a non-trivial centre is
provided next.

\begin{exa}
	\label{ex:semiring}
	Every semiring $(S,+,0,\cdot,1)$ induces a monad $\TT_S:\Set\to\Set$
	\cite{jakl2022bilinear}. This monad maps a set $X$ to the set of finite
	formal sums of the form $\sum s_i x_i$, where $s_i$ are elements of
	$S$ and $x_i$ are elements of $X$. The monad $\TT_S$ is commutative iff
	$S$ is commutative as a semiring. The centre
	$\ZZ$ of $\TT_S$ is induced by the commutative semiring $Z(S)$, \emph{i.e.}
	by the centre of $S$ in the usual sense. Therefore, $\ZZ = \TT_{Z(S)}.$
\end{exa}

\begin{exa}
	Any Lawvere theory $\lawT$ \cite{hyland2007lawvere}
	induces a finitary monad on $\Set$. The centre
	of this monad is the monad induced by the centre of $\lawT$
	in the sense of Lawvere theories \cite{wraith-lecture}.
	This is detailed in \secref{sub:lawvere}.
\end{exa}

\begin{exa}
	\label{ex:valuations-monad}
	The valuations monad $\mathcal V \colon \DCPO \to \DCPO$
	\cite{jones-plotkin,jones90} is similar in spirit to the Giry monad on
	measurable spaces \cite{giry}. It is an important monad in domain theory
	\cite{domain-theory} that is used to combine probability and recursion for
	dcpo's.
	Given a dcpo $X$, the valuations monad $\VVV$ assigns
	the dcpo $\VVV X$ of all Scott-continuous \emph{valuations} on $X$, which are
	Scott-continuous functions $\nu \colon \mathcal \sigma(X) \to [0,1]$ from the
	Scott-open sets of $X$ into the unit interval that satisfy some additional
	properties that make them suitable to model probability (details omitted
	here, see \cite{jones90} for more information).
	The category $\DCPO$ is cartesian closed and the valuations monad $\mathcal V
	\colon \DCPO \to \DCPO$ is strong, but its commutativity on $\DCPO$ has been
	an open problem since 1989 \cite{jones90,jones-plotkin,central-valuations,valuation-monads,valuation-statistical}. The
	difficulty in (dis)proving the commutativity of $\VVV$ boils down to
	(dis)proving the following Fubini-style equation
	\[ \int_X \int_Y \chi_{U}(x,y)d \nu d \xi = \int_Y \int_X \chi_{U}(x,y)d \xi d \nu \]
	holds for any dcpo's $X$ and $Y$, any Scott-open subset $U \in \sigma(X \times Y)$
	and any two valuations $\xi \in \VVV X$ and $\nu \in \VVV Y.$
	In the above equation, the notion of integration is given by the
	\emph{valuation integral} (see \cite{jones90} for more information).

	The \emph{central valuations monad} \cite{central-valuations}, is the submonad $\ZZ \colon \DCPO
	\to \DCPO$ that maps a dcpo $X$ to the dcpo $\ZZ X$ which has all
	\emph{central valuations} as elements. Equivalently:
	\begin{align*}
		\ZZ X \eqdef \Biggl\{ \xi \in \VVV(X)\ |\ \forall Y \in \Ob(\DCPO). \forall U \in \sigma(X \times Y). 
		\forall \nu \in \VVV(Y) . \\
		\int_X \int_Y \chi_{U}(x,y)d \nu d \xi = \int_Y \int_X \chi_{U}(x,y)d \xi d \nu \Biggr\} .
	\end{align*}
	But this is precisely the centre of $\VVV$, which can be seen using Theorem
	\ref{th:centralisability} \eqref{cond:1} after unpacking the definition of
	the monad data of $\VVV$. Therefore, we see that the main result of
	\cite{central-valuations} is a special case of our more general categorical
	treatment. We wish to note that the centre of $\VVV$ is quite large. It
	contains all three commutative submonads identified in
	\cite{valuation-monads} and all of them may be used to model lambda calculi
	with recursion and discrete probabilistic choice (see
	\cite{valuation-monads,central-valuations}).
\end{exa}

\begin{exa}
	\label{ex:expected-cost}
	The category of $\omega$-quasi Borel spaces \cite{sfpc} can be used as a
	model for the denotational semantics of a probabilistic programming
	language. This category admits several strong monads, including $P$ and
	$P_{\leq 1}$ for probability and sub-probability distributions,
	respectively, defined as continuation monads. It turns out that the monad
	$P$ is the centre of the monad $[ 0, \infty ] \times P_{\leq 1}$ which is
	used for expected cost semantics \cite{expected-cost}. This shows, in
	particular, that the central operations are the ones that have trivial
	expected cost and have probability mass equal to 1.
\end{exa}

\subsection{Relationship with Lawvere theories}
\label{sub:lawvere}

Commutants for Lawvere theories \cite{hyland2007lawvere} were defined
in Wraith's lecture notes \cite{wraith-lecture} but were only
studied in details by Lucyshyn-Wright \cite{rory2018convex} later.
The centre of a Lawvere theory is a special case of a commutant.

In a Lawvere theory $\lawT$, we say that $f:A^n\to A^{n'}$ and $g:A^m\to
A^{m'}$ commute if and only if $f^{m'}\circ g^n$ (also written $f\star g$) and
$g^{n'}\circ f^m$ (also written $g\star f$) are equal, up to isomorphism. If
$\lawS$ is a full subcategory of $\lawT$, one can define the commutant of
$\lawS$ in $\lawT$ as a full subcategory of $\lawT$ whose morphisms
commute with the morphisms of $\lawS$. This commutant is written $\lawS^\bot$,
and is also a Lawvere subtheory of $\lawT$. Considering this, $\lawT^\bot$ is
seen as the \emph{centre} of the Lawvere theory $\lawT$ \cite{wraith-lecture};
and any subtheory of $\lawT^\bot$ is a \emph{central subtheory} of $\lawT$.

What about monads? Models of a Lawvere theory $\lawT$ are
finite-product-preserving functors $\lawT\to\Set$ and they form a category
$\Mod(\lawT,\Set)$. This category is adjoint to $\Set$ through a forgetful and
free functors. These adjunctions give rise to a monad. This monad is on $\Set$,
and is therefore strong, centralisable and finitary since it originates from a
Lawvere theory. Thus given a Lawvere theory $\lawT$, we obtain a monad $\TT$
whose centre $\ZZ$ is a commutative submonad of $\TT$ and is finitary, which
means that there exists a corresponding Lawvere theory. This Lawvere theory is
a commutative subtheory of $\lawT$, as we explain next.

The connection between Lawvere theories and finitary monads
is extensively detailed in \cite{street1972formal,garner2014lawvere,garner2018enriched}.
To get a Lawvere theory out of a finitary monad $\ZZ$ on $\Set$,
one needs to look at the opposite category of a skeleton
of $\Set_\ZZ$ \cite{hyland2007lawvere}, for which we write $\mathfrak s\Set_\ZZ^{op}$.
This Lawvere theory is commutative because
$\Set_\ZZ$ is monoidal. Moreover, $\Set_\ZZ$ is embedded in
$\Set_\TT$, then $\mathfrak s\Set_\ZZ^{op}$ is embedded in
$\mathfrak s\Set_\TT^{op}$ ; the latter being equivalent
to $\lawT$.

\begin{thm}
	\label{th:linklawvere}
	Given a Lawvere theory $\lawT$, its $\Set$-monad $\TT$ is
	centralisable and its centre $\ZZ$
	has a corresponding Lawvere theory $\mathfrak s\Set_\ZZ^{op}$
	that is equivalent to $\lawT^\bot$.
\end{thm}
\begin{proof}
	This is a direct application of the point (2) of
	Theorem~\ref{th:centralisability}.
\end{proof}

\section{Central Submonads}
\label{sec:real-central}

So far, we focused primarily on \emph{the} centre of a strong monad. Now we
focus our attention on \emph{central submonads} of a strong monad which we
define by taking inspiration from the notion of central subgroup in group
theory. Just like central subgroups, central submonads are more general
compared to the centre. The centre of a strong monad, whenever it
exists, can be intuitively understood as the largest central submonad, so the
two notions are strongly related. We will later see that central submonads are
more interesting computationally.

\begin{thm}[Centrality]\label{th:centrality}
	Let $\CC$ be a symmetric monoidal category and $\TT$ a strong monad on it.
	Let $\SC$ be a strong submonad of $\TT$ with $\iota:\SC\Rightarrow\TT$ the
	strong submonad monomorphism. The following are equivalent:
	\begin{enumerate}
		\item[1)] \label{ccond:1} For any object $X$ of $\CC$,
			$(\SC X,\iota_X)$ is a central cone for $\TT$ at $X$;
		\item[2)] \label{ccond:2} the canonical embedding functor
			$\II:\CC_\SC\to\CC_\TT$ corestricts to an embedding
			of categories $\hat\II:\CC_\SC\to Z(\CC_\TT)$.
	\end{enumerate}
	Furthermore, these conditions imply that $\SC$ is a commutative submonad
	of $\TT$. Under the additional assumption that $\TT$ is centralisable,
	these conditions are also equivalent to:
	\begin{enumerate}
		\item [3)] $\SC$ is a submonad of the centre of $\TT$, and thus is
			commutative.
	\end{enumerate}
\end{thm}
\begin{proof}
	\

	$(1\Rightarrow 2):$
	The proof of Th.~\ref{th:iso-of-categories} contains the necessary
	steps for this proof. In detail, we know that all the components
	of $\iota$ are central, and we also know that postcomposing with a central
	morphism ensures centrality (see Lemma~\ref{lem:precompose-central}).

	$(2\Rightarrow 1):$
	The hypothesis ensures that $\hat\II(id_X)=\iota_X$ is central.
	The diagram in (\ref{eq:proof-commutative}) proves that the centre of a
	centralisable monad is commutative. Assuming (1) or (2) is true,
	then the same diagram where we replace $\ZZ$ by $\SC$ proves that $\SC$ is a
	commutative monad.

	$(1\Rightarrow 3):$
	Since each $\iota^\SC_X:\SC X \to \TT X$ factorises through
	the morphism $\iota^\ZZ_X \colon \ZZ X \to \TT X$ of the terminal central cone, it is straightforward to show that a strong monad morphism
	$\SC\Rightarrow\ZZ$ arises from these factorisations.

	$(3\Rightarrow 1):$
	Let us write $\ZZ$ for the centre of $\TT$. We write $\iota^\SC \colon \SC\Rightarrow\ZZ$
	and $\iota^\ZZ \colon \ZZ\Rightarrow\TT$ for the submonad morphisms. The
	components of $\iota^\ZZ$ are central, so $\iota^\ZZ\circ\iota^\SC$ is also central by
	Lemma~\ref{lem:precompose-central}. Therefore the components of the
	submonad morphism from $\SC$ to $\TT$ are central.
\end{proof}

\begin{defi}[Central Submonad]\label{def:central-sub}
	Given a strong submonad $\SC$ of $\TT$, we say that $\SC$ is a \emph{central
	submonad} of $\TT$ if it satisfies any one of the above equivalent criteria
	from Theorem \ref{th:centrality}.
\end{defi}

Just like the centre of a strong monad, any central submonad is also
commutative and the above theorem (Theorem~\ref{def:central-sub}) shows that
central submonads have a similar structure to the centre of a strong monad. The
final statement shows that we may see the centre (whenever it exists) as the
largest central submonad of $\TT$. The centre of a strong monad often does
exist (as we already argued), so the last criterion also provides a simple way
to determine whether a submonad is central or not.

\begin{exa}
	By the above theorem, every centre described in \secref{sec:examples} is a
	central submonad.
\end{exa}

\begin{exa}
	Let $\TT$ be a strong monad on a symmetric monoidal category $\CC,$ such that
	all unit maps $\eta_X \colon X \to \TT X$ are monomorphisms (this is often
	the case in practice). Then the identity monad on $\CC$ is a central
	submonad of $\TT.$
\end{exa}

\begin{exa}
	Given a monoid $M,$ let $\TT = (M \times -)$ be the monad on $\Set$ from
	Example~\ref{ex:free-monad}. Any submonoid $S$ of $Z(M)$ induces a central
	submonad $(S \times -)$ of $\TT$.
\end{exa}

\begin{exa}
	Given a semiring $R$, consider the monad $\TT_R$ from Example
	\ref{ex:semiring}. Any subsemiring $S$ of $Z(R)$ induces a central submonad
	$\TT_{S}$ of $\TT_R.$
\end{exa}

\begin{exa}
	A notion of \emph{central Lawvere subtheory} can be introduced in an obvious
	way. It induces a central submonad of the monad induced by the
	original Lawvere theory.
\end{exa}

\begin{exa}
	The three commutative submonads identified in \cite{valuation-monads} are
	central submonads of the valuations monad $\VV$ from Example
	\ref{ex:valuations-monad}, because each one of them is a commutative submonad
	of the centre of $\VV$ \cite{central-valuations}.
\end{exa}

\begin{rem}
	Given an arbitrary monoid $M$ (on $\Set$), there could be a commutative
	submonoid $S$ of $M$ that is not central (\emph{i.e.} its elements do not commute
	with all elements of $M$). The same holds for strong monads. For instance,
	let $M = \mathbb{D}_4$ (see Example \ref{ex:counter}) and let $S$ be the
	submonoid of $M$ that contains only the rotations (of which there are
	four). Then, $S$ is a commutative submonoid that is not central. By taking
	the free monads induced by these monoids (see Example \ref{ex:free-monad})
	on $\Set$, we get an example of a commutative submonad that is not central.
	Moreover, if we take $\DD$ to be the full subcategory of $\Set$ whose
	objects have cardinality that is different from two, then $\DD$ has a
	cartesian structure and the writer monads induced by $S$ and $M$ on $\DD$
	give an example of a non-centralisable strong monad that admits a
	commutative non-central submonad. In this situation, the identity
	monad on $\DD$ gives an example of a central (commutative) submonad
	even though the ambient monad (induced by $M$) is not centralisable.
\end{rem}

\section{Computational Interpretation}
\label{sec:computational}

In this section, we provide a computational interpretation of our ideas. We
consider a simply-typed lambda calculus together with a strong monad $\TT$ and
a \emph{central submonad} $\SC$ of $\TT$. We call this system the
\emph{Central Submonad Calculus (CSC)}. We describe its equational theories,
formulate appropriate categorical models for it, and we prove soundness,
completeness, and internal language results for our semantics.

\subsection{Syntactic Structure of the Central Submonad Calculus}
\label{sub:computational-syntax}

We begin by describing the types we use. The grammar of types (see Figure
\ref{fig:grammars}) are just the usual ones with one addition -- we extend the
grammar by adding the family of types $\SC A$. The type $\TT A$ represents the
type of monadic computations for our monad $\TT$ that produce values of type
$A$ (together with a potential side effect described by $\TT$). The type $\SC
A$ represents the type of \emph{central} monadic computations for our monad
$\TT$ that produce values of type $A$ (together with a potential \emph{central}
side effect that is in the submonad $\SC$). Some terms and typing rules can
be expressed in the same way for types of the form $\SC A$ or $\TT A$ and in
this case we simply write $\XX A$ to indicate that $\XX$ may range over $\{\SC,
\TT \}.$

The grammar of terms and their typing rules are described in Figure
\ref{fig:grammars}. The first six rules in Figure \ref{fig:grammars} are just
the usual typing rules for a simply-typed lambda calculus with pair types.
Contexts are considered up to permutation and without repetition and all
judgements we consider are implicitly closed under weakening (which is
important when adding constants). The $\mathtt{ret}_\XX\ M$ term is used as an
introduction rule for the monadic types and it allows us to see the pure (\emph{i.e.}
non-effectful) computation described by the term $M$ as a monadic one. The
term $\iota M$ allows us to view a \emph{central} monadic computation as a
monadic (not necessarily central) one. Semantically, it corresponds to applying
the $\iota$ submonad inclusion we saw in previous sections. Finally, we have
two terms for monadic sequencing that use the familiar $\mathtt{do}$-notation.
The monadic sequencing of two central computations remains central, which is
represented via the $\mathtt{do}_\SC$ terms; the $\mathtt{do}_\TT$ terms are
used for monadic sequencing of (not necessarily central) computations.

\begin{figure}
	\noindent $\text{(Types)}\quad A, B ~~::= 1 \alt A \to B\alt A\times B
	\alt \SC A \alt \TT A$ \\
	$\quad$\\
	$\text{(Terms)} \quad M,N ~~ ::= x \alt * \alt \lambda x^A.M \alt MN \alt \pv M N$ \\
	$\text{ }\qquad\alt \pi_i M \alt \sret M \alt \iota M \alt \tret M$\\
	$\text{ }\qquad\alt \zdo x M N \alt \tdo x M N $
	$\quad$\\
	\[\begin{array}{c}
		\infer{
			\Gamma,x:A\vdash x:A}{}
		\qquad
		\infer{
			\Gamma\vdash MN:B
		}{
			\Gamma\vdash M:A\to B
			&
			\Gamma\vdash N:A}
		\\[1.5ex]
		\infer{
			\Gamma\vdash *:1}{}
		\qquad
		\infer{\Gamma\vdash\lambda x^A.M:A\to B}{\Gamma,x:A\vdash M:B}
		\qquad
		\infer{
			\Gamma\vdash\pi_i M \colon A_i}{\Gamma\vdash M:A_1\times A_2}
		\\[1.5ex]
		\infer{
			\Gamma\vdash \pv{M}{N}: A\times B
		}{
			\Gamma\vdash M:A
			&
			\Gamma\vdash N:B
		}
		\qquad
		\infer{
			\Gamma\vdash \xret M :\XX A}{\Gamma\vdash M:A}
		\\[1.5ex]
		\infer{
			\Gamma\vdash \iota M :\TT A}{\Gamma\vdash M:\SC A}
		\qquad
		\infer{
			\Gamma\vdash\xdo x M N \colon \XX B
		}{
			\Gamma\vdash M:\XX A
			&
			\Gamma, x:A\vdash N:\XX B
		}
	\end{array}\]
	\caption{Grammars and typing rules.}
	\label{fig:grammars}
\end{figure}

\subsection{Equational Theories of the Central Submonad Calculus}
\label{sub:theories}

Next, we describe equational theories for our calculus. We follow the
vocabulary and the terminology in \cite{maietti2005relating} in order to
formulate an appropriate notion of $\CSC$-theory.

\begin{defi}[$\CSC$-theory]
	A $\CSC$-theory is an extension of the Central Submonad Calculus (see
	\secref{sub:computational-syntax}) with new ground types, new term constants
	(which we assume are well-formed in any context, including the empty one) and
	new equalities between types and between terms.
\end{defi}

In a $\CSC$-theory, we have four types of judgements: the judgement $ \vdash A :
\typ$ indicates that $A$ is a (simple) type; the judgement $\vdash A = B :
\typ$ indicates that types $A$ and $B$ are equal; the judgement $\Gamma \vdash
M \colon A$ indicates that $M$ is a well-formed term of type $A$ in context
$\Gamma$, as usual; finally, the judgement $\Gamma \vdash M = N \colon A$ indicates
that the two well-formed terms $M$ and $N$ are equal.

Type judgements and term judgements are described in Figure \ref{fig:grammars}
and type equality judgements in Figure \ref{fig:type-rules}. Following the
principle of judgemental equality, we add type conversion rules in Figure
\ref{fig:variable-conversion-rule}. The rules in Figure \ref{fig:moggi-rules}
are the usual rules that describe the equational theory of the simply-typed
lambda calculus. 
As often done by many authors, we implicitly identify terms that are
$\alpha$-equivalent. The rules for $\beta$-equivalence and $\eta$-equivalence
are explicitly specified.

In Figure \ref{fig:rules}, we present the equational rules for monadic
computation. The rules on the first three lines -- \emph{(ret.eq), (do.eq),
$(\XX.\beta)$, $(\XX.\eta)$, ($\XX$.assoc)} -- axiomatise the structure of a
strong monad. Because of this, these rules are stated for both monads $\TT$ and
$\SC.$ The rules \emph{($\iota$.mono), ($\iota \SC$.ret)} and \emph{($\iota
\SC$.comp)} are used to axiomatise the structure of $\SC$ as a submonad of
$\TT.$ Intuitively, these rules can be understood as specifying that central
monadic computations can be seen as (general) monadic computations of the
ambient monad $\TT$. The remainder of the rules are used to axiomatise the
behaviour of $\SC$ as a \emph{central} submonad of $\TT$. The rule
$(\SC.central)$ is undoubtedly the most important one, because it ensures that
central computations commute with any other (not necessarily central)
computation when performing monadic sequencing with the $\TT$ monad.

\begin{figure*}
	\[\begin{array}{c}
		\infer{
			\vdash A = A \colon \typ
		}{
			\vdash A \colon \typ
		}
		\qquad
		\infer{
			\vdash B=A:\typ
		}{
			\vdash A=B:\typ
		}
		\qquad
		\infer{
			\vdash A=C:\typ
		}{
			\vdash A=B:\typ
			&
			\vdash B=C:\typ
		}
		\\[1.5ex]
		\infer{
			\vdash A \times B = A'\times B' \colon \typ
		}{
			\vdash A = A' :\typ
			&
			\vdash B = B' :\typ
		}
		\qquad
		\infer{
			\vdash A \to B = A'\to B' \colon \typ
		}{
			\vdash A = A' :\typ
			&
			\vdash B = B' :\typ
		}
		\\[1.5ex]
		\infer{
			\vdash \XX A = \XX B :\typ
		}{
			\vdash A = B:\typ
		}
	\end{array}\]
	\caption{Equational rules for types.}
	\label{fig:type-rules}
\end{figure*}

\begin{figure*}
	\[\begin{array}{c}
		\infer{
			\Gamma, x \colon B \vdash M \colon D
		}{
			\vdash A=B:\typ
			&
			\vdash C=D:\typ
			&
			\Gamma, x \colon A \vdash M \colon C
		}
		\\[1.5ex]
		\infer{
			\Gamma, x \colon B \vdash M = N \colon D
		}{
			\vdash A=B:\typ
			&
			\vdash C=D:\typ
			&
			\Gamma, x \colon A \vdash M = N \colon C
		}
	\end{array}\]
	\caption{Type conversion rules.}
	\label{fig:variable-conversion-rule}
\end{figure*}

\begin{figure*}
	\[\begin{array}{c}
		\infer[(refl)]{
			\Gamma\vdash M=M:A
		}{
			\Gamma\vdash M:A
		}
		\qquad
		\infer[(symm)]{
			\Gamma\vdash M=N:A
		}{
			\Gamma\vdash N=M:A
		}
		\\[1.5ex]
		\infer[(trans)]{
			\Gamma\vdash M=P:A
		}{
			\Gamma\vdash M=N:A
			&
			\Gamma\vdash N=P:A
		}
		\qquad
		\infer[(1.\eta)]{
			\Gamma,x:1\vdash *=x:1
		}{}
		\\[1.5ex]
		\infer[(subst)]{
			\Gamma\vdash N[M/x] = P[M/x] \colon B
		}{
			\Gamma\vdash M \colon A
			&
			\Gamma, x:A\vdash N = P \colon B
		}
		\qquad
		\infer[(\pv{}{} .eq)]{
			\Gamma\vdash \pv M N = \pv{M'}{N'} \colon A\times B
		}{
			\Gamma\vdash M=M' \colon A
			&
			\Gamma\vdash N=N' \colon B
		}
		\\[1.5ex]
		\infer[(\times.\beta)]{
			\Gamma\vdash\pi_i\pv{M_1}{M_2} = M_i \colon A_i
		}{
			\Gamma\vdash M_1 \colon A_1
			&
			\Gamma\vdash M_2 \colon A_2
		}
		\qquad
		\infer[(\times.\eta)]{
			\Gamma\vdash \pv{\pi_1 M}{\pi_2 M} = M \colon A\times B
		}{
			\Gamma\vdash M \colon A\times B
		}
		\\[1.5ex]
		\infer[(app.eq)]{
			\Gamma\vdash MN=M'N' \colon B
		}{
			\Gamma\vdash M=M' \colon A\to B
			&
			\Gamma\vdash N=N' \colon A
		}
		\qquad
		\infer[(\lambda.eq)]{
			\Gamma\vdash \lambda x^A.M = \lambda x^A.N \colon A\to B
		}{
			\Gamma, x:A \vdash M = N \colon B
		}
		\\[1.5ex]
		\infer[(\lambda.\beta)]{
			\Gamma\vdash (\lambda x^A. M) N = M[N/x] \colon B
		}{
			\Gamma,x:A\vdash M \colon B
			&
			\Gamma\vdash N \colon A
		}
		\qquad
		\infer[(\lambda.\eta)]{
			\Gamma\vdash \lambda x^A. Mx = M \colon A\to B
		}{
			\Gamma\vdash M \colon A\to B
		}
		\\[1.5ex]
		\infer[(weak)]{
			\Gamma,x:A\vdash M=N \colon B
		}{
			\Gamma\vdash M=N \colon B
		}
	\end{array}\]
	\caption{Equational rules of the simply-typed $\lambda$-calculus.}
	\label{fig:moggi-rules}
\end{figure*}

\begin{figure*}
	\resizebox{\hsize}{!}{
		$
		\begin{array}{c}
			\infer[(ret.eq)]{
				\Gamma\vdash \xret M = \xret N \colon \XX A
			}{
				\Gamma\vdash M = N \colon A
			}
			\qquad
			\infer[(do.eq)]{
				\Gamma\vdash \xdo x M N = \xdo{x}{M'}{N'} \colon \XX B
			}{
				\Gamma\vdash M=M' \colon \XX A
				&
				\Gamma, x:A\vdash N=N' \colon \XX B
			}
			\\[1.5ex]
			\infer[(\XX.\beta)]{
				\Gamma\vdash \xdo{x}{\xret M}{N} = N[M/x] \colon \XX B
			}{
				\Gamma\vdash M \colon A
				&
				\Gamma,x:A\vdash N \colon \XX B
			}
			\qquad
			\infer[(\XX.\eta)]{
				\Gamma\vdash \xdo{x}{M}{\xret x} = M \colon \XX A
			}{
				\Gamma\vdash M \colon \XX A
			}
			\\[1.5ex]
			\infer[(\XX.assoc)]{
				\Gamma\vdash \xdo{y}{(\xdo x M N)}{P} = \xdo{x}{M}{\xdo y N P} \colon \XX C
			}{
				\Gamma\vdash M \colon \XX A
				&
				\Gamma\vdash N \colon \XX B
				&
				\Gamma,x:A,y:B\vdash P \colon \XX C
			}
			\\[1.5ex]
			\infer[(\SC.central)]{
				\Gamma \vdash \tdo{x}{\iota M}{\tdo{y}{N}{P}} = \tdo{y}{N}{\tdo{x}{\iota M}{P}} \colon \TT C
			}{
				\Gamma\vdash M:\SC A
				&
				\Gamma\vdash N \colon \TT B
				&
				\Gamma,x:A,y:B\vdash P:\TT C
			}
			\\[1.5ex]
			\infer=[(\iota.mono)]{
				\Gamma\vdash\iota M=\iota N:\TT A
			}{
				\Gamma\vdash M=N:\SC A
			}
			\qquad
			\infer[(\iota\SC.ret)]{
				\Gamma\vdash \iota~\sret M = \tret M \colon \TT A
			}{
				\Gamma\vdash M \colon A
			}
			\\[1.5ex]
			\infer[(\iota\SC.comp)]{
				\Gamma\vdash \tdo{x}{\iota M}{\iota N} = \iota~\zdo x M N \colon \TT B
			}{
				\Gamma \vdash M \colon \SC A
				&
				\Gamma, x:A \vdash N \colon \SC B
			}
		\end{array}
		$
		}
	\caption{Equational rules for terms of monadic types of $\mathrm{CSC}$.}
	\label{fig:rules}
\end{figure*}

\begin{exa}
	\label{ex:theory-writer}
	Let us consider an example of a $\CSC$-theory. Given a monoid $(M,e,m)$ we now
	axiomatise the writer monad induced by $M$. A theory for this monad does not
	add any new types, but it adds constants for each element $c$ of $M$ with typing
	judgement
	$\Gamma\vdash\tact:\TT 1.$
	In this specific theory, we may think of the side-effect computed by monadic
	sequencing as being simply an element of $M$. The term $\tact$ can be
	understood as performing the monoid multiplication on the right with argument
	$c$, \emph{i.e.} it applies the function $m(-,c)$ to whatever is the current state
	of the program.

	Let $S$ be a submonoid of the centre $Z(M)$ of $M$. This makes $S$ a
	\emph{central} submonoid of $M$ (this can be defined in a similar way to
	central subgroups). We enrich the theory with the following constant and
	rule for each $s$ in $S$:
	\[
		\infer{\Gamma\vdash\zact[s]:\SC 1}{}
		\qquad
		\infer{\Gamma\vdash\iota~\zact[s] = \tact[s]:\TT 1}{}
	\]
	The application of $\xret$ is equivalent to acting on the monoid
	data with the neutral element:
	\[
		\infer{\Gamma\vdash\xret * = \xact[e] \colon \SC 1}{}
	\]
	Of course, the actions compose:
	\[
		\infer{
			\begin{array}{c}
				\Gamma\vdash\xdo{*}{\xact}{\xdo{*}{\xact[c']}{M}} \\
				= \xdo{*}{\xact[m(c,c')]}{M}:\XX A
			\end{array}
		}{
			\Gamma\vdash M:\XX A
		}
	\]
	where we have used some (hopefully obvious) syntactic sugar.
	We write $\Th_M$ to refer to this theory.
\end{exa}

\begin{rem}
	\label{rem:why-not-centre}
	As we have now seen, the equational theories of central submonads admit a
	presentation that is similar in spirit to that of the simply-typed
	$\lambda$-calculus. However, that is not the case with \emph{the} centre of
	a strong monad. The reason is that the theory $\Th$ can introduce a central
	effect -- one that commutes with all others -- as a constant $c$ that is not
	assigned the type $\SC A$, but the type $\TT A$, for some $A$. However, the
	centre, being the largest central submonad, must contain all such effects, so
	the constant $c$ has to be equal to a term of the form $\iota c'$. One
	solution to this problem would be to use a more expressive logic and
	introduce a rule as follows:
	given $c \colon \TT A$ and $x:A,y:B\vdash P \colon \TT C$, such that
	\[ \forall N:\TT B.~\vdash \tdo{x}{c}{\tdo{y}{N}{P}} = \tdo{y}{N}{\tdo{x}{c}{P}} \colon \TT C \]
	then $\exists c':\SC A.~\vdash c = \iota c' \colon \TT A.$
	However, the addition of such
	a rule seems unnecessary to prove our main point and it increases the
	complexity of the logic due to the presence of existential and universal quantifiers.
	Because of this, our choice is to focus on central
	submonads. Another reason to prefer central submonads over the centre is
	that they are more general and it is not required to identify \emph{all}
	central effects (which would be the case for the centre). Overall, our choice
	for central submonads is motivated by the advantages they provide in terms of
	generality, simplicity and practicality of their equational theories compared
	to the centre.
\end{rem}

Now that we have introduced theories, we explain how they can be translated
into one another in an appropriate way.

\begin{defi}[$\mathrm{CSC}$-translation]
	A \emph{translation} $V$ between two $\CSC$-theories $\Th$ and $\Th'$ is
	a function that maps types of $\Th$ to types of $\Th'$ and terms of $\Th$ to
	terms of $\Th'$ that preserves the provability of all type judgements, term
	judgements, type equality judgements and term equality judgements.
	Moreover, such a translation is required to satisfy the following
	structural requirements on types:
	\[\begin{array}{c}
		V(1)=1
		\qquad
		V(\TT A) = \TT V(A)
		\qquad
		V(\SC A) = \SC V(B)
		\\
		V(A\to B) = V(A)\to V(B)
		\qquad
		V(A\times B) = V(A)\times V(B)
	\end{array}\]
	and on terms:
	\[\begin{array}{c}
		V(*) = *
		\\
		V(\lambda x^A.M) = \lambda x^A.V(M)
		\qquad
		V(MN) = V(M)V(N)
		\\
		V(\pv M N) = \pv{V(M)}{V(N)}
		\qquad
		V(\pi_i M) = \pi_i V(M)
		\\
		V(\iota M) = \iota V(M)
		\qquad
		V(\xret M) = \xret V(M)
		\\
		V(\xdo x M N) = \xdo{x}{V(M)}{V(N)}
	\end{array}\]
\end{defi}

\begin{rem}
	The above equations do not imply preservation of the relevant judgements for
	\emph{constants}. Because of this, the first part of the definition also is
	necessary.
\end{rem}

Of course, it is easy to see that $\CSC$-theories and $\CSC$-translations form a
category. However, in order to precisely state our main result, we have to
consider the 2-categorical structure of $\CSC$-theories. Intuitively, we may view
every $\CSC$-theory as a category itself (with types as objects
and terms as morphisms) and every $\CSC$-translation as a functor that strictly
preserves the relevant structure. Then, intuitively, an appropriate
notion of a 2-morphism would be a natural transformation between such functors.
This is made precise (in non-categorical terms) by our next definition.

\begin{defi}[$\CSC$-translation Transformation]\label{def:trans-trans}
	Given two $\CSC$-theories $\Th$ and $\Th'$,
	and two $\CSC$-translations $V$ and $V'$ between
	them, a \emph{$\CSC$-translation transformation} $\alpha:V\Rightarrow V'$
	is a type-indexed family of term judgements $x:V(A)\vdash \alpha_A \colon V'(A)$
	such that, for any valid judgement $x:A\vdash f:B$ in $\Th$
	\[ x:V(A)\vdash \alpha_B[V(f)/x] = V'(f)[\alpha_A/x] \colon V'(B) \]
	also is derivable in $\Th'.$
\end{defi}

\begin{prop}
	$\mathrm{CSC}$-theories, $\mathrm{CSC}$-translations
	and $\mathrm{CSC}$-translation transformations
	form a $2$-category $\TR(\mathrm{CSC})$.
\end{prop}
\begin{proof}
	Direct with Definition~\ref{def:trans-trans}.
\end{proof}

\subsection{Categorical Models of CSC}
\label{sub:models}

Now we describe what are the appropriate categorical models for providing a
semantic interpretation of our calculus.

\begin{defi}[$\CSC$-model]
	A \emph{$\CSC$-model} is a cartesian closed category $\CC$ equipped with both a
	strong monad $\TT$ and a central submonad $\SC^\TT$ of $\TT$ with submonad
	monomorphism written as $\iota^\TT \colon \SC^\TT \naturalto \TT.$ We often
	use a quadruple $(\CC,\TT,\SC^\TT, \iota^\TT)$ to refer to a $\CSC$-model.
\end{defi}

We soon show that $\CSC$-models correspond to $\CSC$-theories in a precise way.
This correspondence covers $\CSC$-translations too and for
this we introduce our next definition.

\begin{defi}[$\CSC$-model Morphism]
	Given two $\CSC$-models $(\CC,\TT,\SC^\TT, \iota^\TT)$ and $(\DD,\MM,\SC^\MM,
	\iota^\MM)$, a \emph{$\CSC$-model morphism} is a strict cartesian closed functor
	$F:\CC\to\DD$ that satisfies the following additional coherence properties:
	\[\begin{array}{c}
		F(\TT X) = \MM(FX)
		\qquad
		F(\SC^\TT X) = \SC^\MM (FX)
		\\[1.5ex]
		F\iota^\TT_X = \iota^\MM_{FX}
		\qquad
		F\eta^\TT_X = \eta^\MM_{FX}
		\\[1.5ex]
		F\mu^\TT_X = \mu^\MM_{FX}
		\qquad
		F\tau^\TT_{X,Y} = \tau^\MM_{FX,FY}.
	\end{array}\]
\end{defi}

Notice that a $\CSC$-model morphism \emph{strictly} preserves all of the relevant
categorical structure. This is done on purpose so that we can establish an
exact correspondence with $\CSC$-translations, which also strictly preserve the
relevant structure. To match the notion of a $\CSC$-translation transformation, we
just have to consider natural transformations between $\CSC$-model morphisms.

\begin{prop}
	$\CSC$-models, $\CSC$-model morphisms and natural transformations between
	them form a $2$-category $\Mod(\CSC)$.
\end{prop}
\begin{proof}
	Direct.
\end{proof}

\subsection{Semantic Interpretation}
\label{sub:denotational}

Now we explain how to introduce a denotational semantics for our theories using
our models. An interpretation of a $\CSC$-theory $\Th$ in a $\CSC$-model $\CC$
is a function $\sem{-}$ that maps types of $\Th$ to objects of $\CC$ and
well-formed terms of $\Th$ to morphisms of $\CC$. We provide the details below.

For each ground type $G$, we assume there is an appropriate corresponding
object $\sem{G}$ of $\CC$. The remaining types are interpreted as objects in
$\CC$ as follows:
$
\sem{1} \eqdef 1 ;
\sem{A\to B} \eqdef \sem B^{\sem A} ;
\sem{A\times B} \eqdef \sem A\times\sem B ;
\sem{\SC A} \eqdef \SC\sem A ;
\sem{\TT A} \eqdef \TT\sem A .
$
Variable contexts $\Gamma = x_1 \colon A_1 \dots x_n \colon A_n$ are interpreted as usual as
$\sem\Gamma\defeq\sem{A_1}\times\dots\times \sem{A_n}$. Terms are interpreted
as morphisms $ \sem{\Gamma\vdash M:A} \colon \sem\Gamma \to \sem A $ of $\CC$. When
the context and the type of a term $M$ are understood, then we simply write
$\sem M$ as a shorthand for $\sem{\Gamma\vdash M:A}$. The interpretation of
term constants and the terms of the simply-typed $\lambda$-calculus is defined
in the usual way (details omitted). The interpretation of the monadic terms is
given by:
\begin{align*}
	\sem{\Gamma\vdash \xret M :\XX A} &= \eta^\XX_{\sem A}\circ\sem M \\
	\sem{\Gamma\vdash \iota M :\TT A} &= \iota_{\sem A}\circ\sem M \\
	\sem{\Gamma\vdash\xdo x M N \colon \XX B} &= \mu^\XX_{\sem B}\circ\XX\sem
	N\circ\tau^\XX_{\sem\Gamma,\sem A}\circ\pv{\iid}{\sem M}
\end{align*}
where we use $\XX$ to range over $\TT$ or its central submonad $\SC.$

\begin{defi}[Soundness and Completeness]
	An interpretation $\sem{-}$ of a $\CSC$-theory $\Th$ in a $\CSC$-model $\CC$
	is said to be \emph{sound} if for any type equality judgement $\vdash
	A=B:\typ$ in $\Th$, we have that $\sem A=\sem B$ in $\CC$, and for any
	equality judgement $\Gamma\vdash M=N :A$ in $\Th$, we have that
	$\sem{\Gamma\vdash M \colon A}=\sem{\Gamma\vdash N:A}$ in $\CC$. An
	interpretation $\sem{-}$ is said to be \emph{complete} when $\vdash A=B:\typ$
	iff $\sem A = \sem B$ and $\Gamma\vdash M=N \colon A$ iff $\sem{\Gamma\vdash
	M \colon A}=\sem{\Gamma\vdash N \colon A}.$ If, moreover, the interpretation
	is clear from context, then we may simply say that the model $\CC$ itself is
	sound and complete for the $\CSC$-theory $\Th$.
\end{defi}

\begin{rem}
	There are different definitions of what constitutes a ``model'' in the
	literature. For example, a ``model'' in \cite{crole1994} corresponds to a
	sound interpretation in our sense.
\end{rem}

\begin{exa}
	A categorical model for the $\CSC$-theory $\Th_M$ of
	Example~\ref{ex:theory-writer} is given by the category $\Set$ together with
	the writer monad $\TT \eqdef (-\times M) \colon \Set \to \Set$ and the central
	submonad $\SC \eqdef (-\times S) \colon \Set \to \Set$. More specifically, the
	monad data for $\TT$ is given by:
	\begin{align*}
		& \eta_A   \colon A \to A\times M :: a \mapsto (a,e) \\
		& \mu_A   \colon (A\times M)\times M \to A\times M :: ((a,c),c') \mapsto (a,m(c,c')) \\
		& \tau_{A,B} \colon A\times(B\times M) \to (A\times B)\times M ::\\
		& \qquad\quad (a,(b,c)) \mapsto ((a,b),c)
	\end{align*}
	and the monad data for $\SC$ is defined in the same way by (co)restricting to
	the submonoid $S$. The interpretation of the term constants is given by:
	\begin{align*}
		\sem{\Gamma\vdash \tact:\TT 1} &\colon \sem\Gamma \to 1\times M :: \gamma \mapsto (*,c) \\
		\sem{\Gamma\vdash \zact:\SC 1} &\colon \sem\Gamma \to 1\times S :: \gamma \mapsto (*,c)
	\end{align*}
	This interpretation of the theory $\Th_M$ is sound and complete.
\end{exa}

\subsection{Equivalence between Theories and Models}
\label{sub:equivalence}

Our final result in this article is to show that $\CSC$-theories and
$\CSC$-models are strongly related. To do this, we define the \emph{syntactic
$\CSC$-model} $S(\Th)$ of $\CSC$-theory $\Th$, and the \emph{internal language}
$L(\CC)$ that maps a $\CSC$-model $\CC$ to its internal language viewed as a
$\CSC$-theory. These two assignments give rise to our desired equivalence
(Theorem \ref{th:internal-language}).

\paragraph{The Syntactic $\CSC$-model.}
\label{sub:syntactic-fun}
Assume throughout the subsection that we are given a $\mathrm{CSC}$-theory
$\Th$. We show how to construct a sound and complete model $S(\Th)$ of $\Th$
by building its categorical data using the syntax provided by $\Th.$

\begin{defi}[Syntactic Category]
	Let $S(\Th)$ be the category whose objects are the types of $\Th$ modulo type
	equality, \emph{i.e.} the objects are equivalence classes $[A]$ of types with $A'
	\in [A]$ iff $\vdash A' = A \colon \typ$ in $\Th.$ The morphisms $S(\Th)([A],
	[B])$ are equivalence classes of judgements $[x:A\vdash f:B],$ where $(x:A'
	\vdash f' \colon B') \in [x:A\vdash f \colon B]$ iff $\vdash A' = A \colon
	\typ$ and $\vdash B' = B \colon \typ$ and $x:A \vdash f = f' \colon B.$
	Identities are given by $[x \colon A \vdash x \colon A]$ and composition is
	defined by
	\[ [y \colon B \vdash g \colon C] \circ [x \colon A \vdash f \colon B'] = [x
	\colon A \vdash g[f/y] \colon C] , \]
	with $B' \in [B].$
\end{defi}

\begin{lem}
	\label{lem:syntactic-category}
	The above definition is independent of the choice of representatives and the
	syntactic category $S(\Th)$ is a well-defined cartesian closed category.
\end{lem}
\begin{proof}
	Suppose we are given two morphisms $f \colon A\to B$ and $g \colon B\to C$ and a choice
	$[x \colon A'\vdash f' \colon B'_f] = f$ and $[y \colon B'_g\vdash g' \colon
	C']= g$. Note that $B=[B'_f]=[B'_g]$, and in particular $y \colon B'_f\vdash
	g' \colon C'$ is derivable with $[y \colon B'_g\vdash g' \colon C']=[y \colon
	B'_f\vdash g' \colon C']$. Thus, $x \colon A'\vdash g'[f'/y] \colon C'$ is
	derivable. We now prove that the choice $[x \colon A'\vdash f' \colon B'_f]
	= f$ and $[y \colon B'_f\vdash g' \colon C']= g$ does not matter. We
	consider new term judgments for some terms $f''$ and $g''$ such that $[x
	\colon A'\vdash f' \colon B'_f] = [x \colon A''\vdash f'' \colon B''_f]$ and
	$[y \colon B'_f\vdash g' \colon C'] = [y \colon B''_f\vdash g'' \colon C'']$.
	By definition, $[A']=[A'']$, $[B'_f]=[B''_f]$ and $[C']=[C'']$, and we wish
	to prove that $[x \colon A'\vdash g'[f'/y] \colon C'] = [x \colon A''\vdash
	g''[f''/y] \colon C'']$.

	\[
		\begin{array}{l}
			\Pi_2 = \left\{
				\begin{array}{l}
					\infer[(\lambda.eq)]{
						x \colon A'\vdash (\lambda y^{B'_f}.g')f' =
						(\lambda y^{B''_f}.g'')f'' \colon C'
					}{
						\infer{
							x \colon A'\vdash \lambda y^{B'_f}.g'
							= \lambda y^{B''_f}.g'' \colon
							B'_f\to C'
						}{
							x \colon A,y \colon B'\vdash g'=g'' \colon C'
						}
						&
						x \colon A'\vdash f'=f'' \colon C'
					}
				\end{array} \right.
		\end{array}
	\]
	\[
		\begin{array}{l}
			\Pi_1 = \left\{
				\begin{array}{l}
					\infer[(trans)]{
						x \colon A'\vdash (\lambda y^{B'_f}.g')f' = g''[f''/y] \colon C'
					}{
						\Pi_2
						&
						\infer[(\lambda.\beta)]{
							x \colon A'\vdash (\lambda y^{B''_f}.g'')f'' =
							g''[f''/y] \colon C'
						}{
							x \colon A',y \colon B''_f \vdash g'' \colon C'
							&
							x \colon A' \vdash f'' \colon B''_f
						}
				} \end{array} \right.
		\end{array}
	\]
	\[
		\infer[(trans)]{
			x \colon A'\vdash g'[f'/y] = g''[f''/y] \colon C'
		}{
			\infer[(\lambda.\beta)]{
				x \colon A'\vdash g'[f'/y] = (\lambda y^{B'_f}.g')f' \colon C'
			}{
				x \colon A',y \colon B'_f\vdash g' \colon C'
				&
				x \colon A'\vdash f' \colon C'
			}
			&
			\Pi_1
		}
	\]

	Thus, it is safe to define $g\circ f$ as $[x \colon A'\vdash g'[f'/y] \colon C']$.

	Given a choice of $A'$ in $[A]$, the morphism $[x \colon A'\vdash x \colon A']$ is the
	identity morphism for the type $[A]$. Considering $[x \colon A'\vdash
	f \colon B']$ and $[y \colon C'\vdash g \colon A']$, we have:
	\[
		[x \colon A'\vdash f \colon B']\circ [x \colon A'\vdash x \colon A']
		= [x \colon A' \vdash f[x/x] \colon B']
		= [x \colon A' \vdash f \colon B'],\]
	and
	\[ [x \colon A'\vdash x \colon A'] \circ [y \colon C'\vdash g \colon A']
	= [y \colon C'\vdash x[g/x] \colon A']
	= [y \colon C'\vdash g \colon A']. \]
	One can notice that, for example, $x \colon A'\vdash f \colon B'$ has
	conveniently be chosen with the right type $A'$. It is
	authorised, because we have proven above that the choice of
	representative does not matter in composition.

	The cartesian closure is a standard result for a syntactic
	category from a simply-typed $\lambda$-calculus, and it is
	preserved in our context.
\end{proof}

\begin{rem}
	Note that by using \emph{Scott's trick} \cite{scott-trick} we can take
	quotients without having to go up higher in the class hierarchy, so foundational
	issues can be avoided.
\end{rem}

\begin{lem}[\cite{awodey-bauer-lecture}]
	\label{lem:strong-syntactic}
	The following assignments:
	\begin{align*}
		\TT([A])  &= [\TT A] \\
		\TT( [x \colon A \vdash f \colon B] ) &= [y \colon \TT A \vdash \tdo x y {\tret f} \colon \TT B] \\
		\eta_{[A]}   &= [ x \colon A\vdash \tret x \colon \TT A ] \\
		\mu_{[A]}   &= [ x \colon \TT\TT A \vdash \tdo y x y \colon \TT A ] \\
		\tau_{[A],[B]} &= [ x \colon A\times \TT B \vdash
		\tdo{y}{\pi_2 x}{\tret\pv{\pi_1 x}{y}} \colon \TT(A\times B) ]
	\end{align*}
	are independent of the choice of representatives and define a strong monad $(\TT, \eta, \mu, \tau)$ on $S(\Th)$.
\end{lem}

\begin{lem}\label{lem:syntactic-central-submonad}
	In a similar way to Lemma \ref{lem:syntactic-category}, we can define a
	strong monad $(\SC, \eta^\SC, \mu^\SC, \tau^\SC)$ on $S(\Th)$ by using the
	corresponding monadic primitives. Then the assignment:
	\[
		\iota_{[A]} = [x \colon \SC A\vdash \iota x \colon \TT A]
	\]
	is independent of the choice of representative and gives a strong submonad
	monomorphism $\iota \colon \SC \naturalto \TT$ that makes $\SC$ a central submonad
	of $\TT.$
\end{lem}
\begin{proof}
	In all the following proofs, we consider convenient members of equivalence
	classes, because the choice of representative does not change the result,
	thanks to Lemma~\ref{lem:syntactic-category}.

	We prove that $\iota$ is a submonad morphism:

	$\begin{array}{cl}
		&\iota_A\circ\eta^\SC_A \\
		\stackrel{def.}{=}
		& [y \colon \SC A\vdash \iota y \colon \TT A] \circ [x \colon A\vdash \zret x \colon \SC A] \\
		\stackrel{comp.}{=}
		& [x \colon A\vdash \iota~\zret x \colon \TT A] \\
		\stackrel{(\iota\SC.ret)}{=} & [x \colon A\vdash \tret x \colon \TT A] \\
		\stackrel{def.}{=} & \eta^\TT_A
	\end{array}$
		\\[3.5ex]
	$\begin{array}{cl}
		&\mu^\TT_A\circ\TT\iota_A\circ\iota_{\SC A} \\
		\stackrel{def.}{=}
		& [z \colon \TT\TT A\vdash \tdo{y}{z}{y} \colon \TT A] \\
		&\circ~[y' \colon \TT\SC
		A\vdash \tdo{x}{y'}{\tret\iota x} \colon \TT\TT A] \circ [x' \colon \SC\SC
		A \vdash \iota x' \colon \TT\SC A] \\
		\stackrel{comp.}{=}
		& [x' \colon \SC\SC A\vdash\tdo{y}{\left(\tdo{x}{\iota x'}{\tret \iota
		x}\right)}{y} \colon \TT A] \\
		\stackrel{(\TT.assoc)}{=} & [x' \colon \SC\SC A\vdash \tdo{x}{\iota
		x'}{\tdo{y}{\tret \iota x}{y}} \colon \TT A] \\
		\stackrel{(\TT.\beta)}{=} & [x' \colon \SC\SC A\vdash \tdo{x}{\iota
		x'}{\iota x} \colon \TT A] \\
		\stackrel{(\iota\SC.comp)}{=} & [x' \colon \SC\SC A\vdash
		\iota~\zdo{x}{x'}{x} \colon \TT A] \\
		\stackrel{comp.}{=}
		& [y \colon \SC A\vdash \iota y \colon \TT A] \circ [x' \colon \SC\SC A\vdash
		\zdo{x}{x'}{x} \colon \SC A] \\
		\stackrel{def.}{=} & \iota_A\circ\mu^\SC_A
	\end{array}$
		\\[3.5ex]
	$\begin{array}{cl}
		&\iota_{A\times B}\circ\tau^\SC_{A,B} \\
		\stackrel{def.}{=}
		& [x \colon \SC(A\times B)\vdash \iota x \colon \TT(A\times B)] \\
		&\circ~
		[z \colon A\times\SC B \vdash \zdo{y}{\pi_2 z}{\zret\pv{\pi_1 z}{y}} \colon
		\SC(A\times B) ] \\
		\stackrel{comp.}{=} & [z \colon A\times\SC B \vdash
		\iota\left(\zdo{y}{\pi_2 z}{\zret\pv{\pi_1
		z}{y}}\right) \colon \TT(A\times B)] \\
		\stackrel{(\iota\SC.comp)}{=} & [z \colon A\times\SC
		B\vdash\tdo{y}{\iota~\pi_2 z}{\iota~\zret\pv{\pi_1
		z}{y}} \colon \TT(A\times B)] \\
		\stackrel{(\iota\SC.ret)}{=} & [z \colon A\times\SC
		B\vdash\tdo{y}{\iota~\pi_2 z}{\tret\pv{\pi_1 z}{y}} \colon \TT(A\times
		B)] \\
		\stackrel{(\times.\beta)}{=} & [z \colon A\times\SC
		B\vdash\tdo{y}{\pi_2\pv{\pi_1 z}{\iota \pi_2 z}}{\tret\pv{\pi_1
		\pv{\pi_1 z}{\iota \pi_2 z}}{y}} \colon \TT(A\times B)] \\
		\stackrel{comp.}{=} & [x \colon A\times\TT B \vdash \tdo{y}{\pi_2
		x}{\tret\pv{\pi_1 x}{y}} \colon \TT(A\times B)] \\
		&\circ~ [z \colon A\times\SC
		B\vdash \pv{\pi_1 z}{\iota \pi_2 z} \colon A\times \TT B] \\
		\stackrel{def.}{=} & \tau^\TT_{A,B}\circ(A\times\iota_B)
	\end{array}$

	Moreover, $\iota$ is a monomorphism because of the $(\iota.mono)$ rule.

	Finally,
	$\ZZ$ is a central submonad of $\TT$:
	\[\begin{array}{cl}
		& dst_{A,B}\circ(\iota\times\TT B) \\
		\stackrel{def.+comp.}{=} & [z \colon \SC A\times\TT B\vdash \\
		&\hspace{-.5cm} \tdo{x}{\left(\tdo{y}{\iota~\pi_1 z}{\tret\left(\tdo{y'}{\pi_2
		z}{\tret\pv{y}{y'}}\right)}\right)}{x} \colon \TT(A\times B)] \\
		\stackrel{(\TT.assoc)}{=} & [z \colon \SC A\times\TT
		B\vdash \\
		&\hspace{-.5cm} \tdo{y}{\iota~\pi_1 z}{\tdo{x}{\tret\left(\tdo{y'}{\pi_2
		z}{\tret\pv{y}{y'}}\right)}{x}} \colon \TT(A\times B)] \\
		\stackrel{(\TT.\beta)}{=} & [z \colon \SC A\times\TT
		B\vdash\tdo{y}{\iota~\pi_1 z}{\tdo{y'}{\pi_2
		z}{\tret\pv{y}{y'}}} \colon \TT(A\times B)] \\
		\stackrel{(\SC.central)}{=} & [z \colon \SC A\times\TT
		B\vdash\tdo{y'}{\pi_2 z}{\tdo{y}{\iota~\pi_1
		z}{\tret\pv{y}{y'}}} \colon \TT(A\times B)] \\
		\stackrel{(\TT.\beta)}{=} & [z \colon \SC A\times\TT
		B\vdash \\
		&\hspace{-.5cm} \tdo{y'}{\pi_2 z}{\tdo{x}{\tret\left(\tdo{y}{\iota~\pi_1
		z}{\tret\pv{y}{y'}}\right)}{x}} \colon \TT(A\times B)] \\
		\stackrel{(\TT.assoc)}{=} & [z \colon \SC A\times\TT
		B\vdash \\
		&\hspace{-.5cm} \tdo{x}{\left(\tdo{y'}{\pi_2
		z}{\tret\left(\tdo{y}{\iota~\pi_1
		z}{\tret\pv{y}{y'}}\right)}\right)}{x} \colon \TT(A\times B)] \\
		\stackrel{comp.+def.}{=} & dst'_{A,B}\circ(\iota\times\TT B)
	\end{array}\]
\end{proof}

Now we can prove our completeness result.

\begin{thm}[Completeness]
	\label{th:completeness}
	The quadruple $(S(\Th), \TT, \SC, \iota)$ is a sound and complete $\CSC$-model
	for the $\CSC$-theory $\Th$.
\end{thm}
\begin{proof}
	There exists an (obvious) interpretation $\sem{-}$ of $\Th$ into $S(\Th)$
	which follows the structure outlined in \secref{sub:denotational}. Standard
	arguments then show that $\Gamma\vdash M = N \colon A$ in $\Th$ iff
	$\sem{\Gamma\vdash M \colon A} = \sem{\Gamma\vdash N \colon A}$ in $S(\Th)$.
\end{proof}

\begin{rem}
	Note that the obvious canonical interpretation of $\Th$ in $S(\Th)$ is
	initial as one may expect: any sound interpretation of $\Th$ in a
	$\CSC$-model $\CC$ factorises uniquely through the canonical interpretation
	via a $\CSC$-model morphism.
\end{rem}

\paragraph{Internal Language.}
\label{sub:internal-language}
With completeness proven, we now wish to establish an internal language result.

\begin{defi}[Internal Language]
	\label{def:language}
	Given a $\CSC$-model $\CC$, we define a $\CSC$-theory $L(\CC)$
	as follows:
	\begin{itemize}
		\item For each object $A$ of $\CC$ we add a ground type which we name $A^*$.
		\item Every ground type $A^*$ is interpreted in $\CC$ by setting $\sem{A^*}
			\eqdef A.$ This uniquely determines an interpretation on all types.
		\item If $A$ and $B$ are two (not necessarily ground) types, we add a type
			equality $\vdash A = B \colon \typ$ iff $\sem A = \sem B$.
		\item For every morphism $f \colon A \to B$ in $\CC$, we add a term constant
			$\vdash c_f \colon A^* \to B^*.$ Its interpretation in $\CC$ is defined to be
			$\sem{c_f} \eqdef \mathrm{curry}(f \circ \cong) \colon 1 \to B^A$, \emph{i.e.} it is
			defined by currying the morphism $f$ in the obvious way. This uniquely
			determines an interpretation on all well-formed terms.
		\item New term equality axioms $\Gamma\vdash M=N \colon B$ iff $\sem{\Gamma \vdash
			M \colon B} = \sem{\Gamma \vdash N \colon B}.$
	\end{itemize}
\end{defi}

\begin{thm}
	For any $\CSC$-model $\CC$ the above definition gives a well-defined $\CSC$-theory
	$L(\CC).$ Moreover, the model $\CC$ is sound and complete for $L(\CC).$
\end{thm}
\begin{proof}
	Well-definedness is straightforward and follows by a simple induction
	argument using the fact that the semantic interpretation $\sem{-}$ defined
	in \secref{sub:denotational} is always sound. Completeness is then
	immediate by the last condition in Definition~\ref{def:language}.
\end{proof}

\paragraph{Equivalence Theorem.}
Finally, we show that both the construction of the syntactic category and the
assignment of the internal language give rise to appropriate equivalences.

\begin{thm}
	\label{th:internal-language}
	The relationship between the internal language and the syntactic model enjoys
	the following properties in the 2-categories $\Mod(\CSC)$ and $\TR(\CSC)$,
	respectively:
	\begin{enumerate}
		\item For any $\CSC$-model $\CC$, we have that $\CC \simeq SL(\CC)$,
		\emph{i.e.} there exist $\CSC$-model morphisms $F \colon \CC \to SL(\CC)$
	and $G \colon SL(\CC) \to \CC$ such that $F \circ G \cong \Id$ and $\Id
\cong G \circ F.$ \item For any $\CSC$-theory $\Th$, we have that $\Th \simeq
	LS(\Th)$, \emph{i.e.} there exist $\CSC$-translations $V \colon \Th \to
			LS(\Th)$ and $W \colon LS(\Th) \to \Th$ such that $V \circ W \cong \Id$
			and $\Id \cong W \circ V.$
	\end{enumerate}
\end{thm}
\begin{proof}
	Given an object $\CC$ of $\Mod(\CSC)$, we wish to prove that $\CC$ is
	equivalent to $SL(\CC)$. To do so, we introduce two strict cartesian
	closed functors $F \colon \CC\to SL(\CC)$ and $G \colon SL(\CC)\to\CC$, such that
	there are isomorphisms $\Id \Rightarrow GF$ and $FG\Rightarrow \Id$.
	\begin{itemize}
		\item $F$ maps an object $A$ of $\CC$ to $[A^*]$. It maps a
			morphism $f \colon A \to B$ to $[x:A^*\vdash c_f x \colon B^*]$.
		\item $G$ maps an object $[A]$ to $\sem{A}$, the interpretation of the type
			$A$ in $\CC$, because the choice of representative of $[A]$ does not
			change the interpretation. The functor $G$ maps a morphism $[x \colon A \vdash g
			\colon B]$ to $\sem{x \colon A\vdash g \colon B}$.
	\end{itemize}
	Then it is easy to check that $GF = \Id$ and $FG = \Id.$ Therefore $\CC$ is
	isomorphic to $SL(\CC)$. Furthermore, given a $\CSC$-theory $\Th$, we wish
	to prove that $\Th$ is equivalent to $LS(\Th)$. To do so, we introduce two
	$\CSC$-translations $V \colon \Th\to LS(\Th)$ and $W \colon LS(\Th)\to\Th$
	such that there are isomorphic $\CSC$-translation transformations
	$VW\Rightarrow \Id$ and $\Id \Rightarrow WV$.
	\begin{itemize}
		\item $V$ maps a type $A$ in $\Th$ to $[A]^*$, and term
			judgements $x \colon A\vdash f \colon B$ to $x \colon [A]^*\vdash
			c_{[x \colon A\vdash f \colon B]} x \colon [B]^*$.
		\item Observe that for each type $A$ in $LS(\Th)$, there is a
			type of the form $[B]^*$ such that $\vdash A=[B]^*
			 \colon \typ$ in $LS(\Th)$. We define $W(A) \eqdef B$ (the
			choice of $B$ does not matter). Then, for term
			constants we define $W( \vdash c_{[x \colon A \vdash f \colon B]}
			\colon B^*) \eqdef (\vdash \lambda x. f \colon A \to B)$ and
			this uniquely determines the action of $W$ on the
			remaining terms (the choice of $f$ does not matter).
	\end{itemize}
	Given a type $A$ in $\Th$, $x \colon W(V(A))\vdash x \colon A$ is derivable
	in $\Th$ because $\vdash W(V(A))=A \colon \typ$, and $\alpha_A \colon x
	\colon W(V(A))\vdash x \colon A$ defines an isomorphic $\CSC$-translation
	transformation: postcomposing (resp. composing) it with $x \colon A\vdash x
	\colon W(V(A))$ gives $x \colon W(V(A))\vdash x \colon W(V(A))$ (resp. $x
	\colon A\vdash x \colon A$). Given a type $A'$ in $LS(\Th)$, the same is true
	for $\beta_{A'} = x \colon A'\vdash x \colon V(W(A'))$. Thus, for every
	$\CSC$-theory, $\Th$ is equivalent to $LS(\Th)$.
\end{proof}

\begin{rem}
	We introduced type equalities so that we can prove Theorem
	\ref{th:internal-language}. This is also the approach taken in
	\cite{maietti2005relating} and without this, technical difficulties arise.
	Theory translations are defined strictly (up to equality, not up to
	isomorphism) and in order to match this with the corresponding notion of
	model morphism, we use type equalities. Without type equalities, the symmetry
	within Theorem~\ref{th:internal-language} can only be established if we make
	further changes. One potential solution would be to weaken the notion of
	theory translation by requiring that it preserves types up to type
	isomorphism (\emph{i.e.} make it strong instead of strict), but this is technically
	cumbersome.
\end{rem}

\section{Conclusion and Future Work}
\label{sec:conclude}

We showed that, under some mild assumptions, strong monads indeed admit a
centre, which is a commutative submonad, and we provided three equivalent
characterisations for the existence of this centre (Theorem
\ref{th:centralisability}) which also establish important links to the theory
of premonoidal categories. In particular, every (canonically strong) monad on
$\Set$ is centralisable (\secref{sub:sets}) and we showed that the same is true
for many other categories of interest (\secref{sub:examples}) and we identified
specific monads with interesting centres (\secref{sub:specific-examples}).
More generally, we considered central submonads and we provided a computational
interpretation of our ideas (\secref{sec:computational}) which has the added
benefit of allowing us to easily keep track of which monadic operations are
central, \emph{i.e.} which effectful operations commute under monadic sequencing with
any other (not necessarily central) effectful operation. We cemented our
semantics by proving soundness, completeness and internal language results.

One direction for future work is to consider a theory of \emph{commutants} or
\emph{centralisers} for monads (in the spirit of
\cite{commutants,garner2016commutativity}) and to develop a computational
interpretation with the expected properties (soundness, completeness and
internal language).
Another opportunity for future work includes studying the relationship between
the centres of strong monads and distributive laws. In particular, given two
strong monads and a strong/commutative distributive law between them, can we
show that the distributive law also holds for their centres (or for some
central submonads)? If so, this would allow us to use the distributive law to
combine not just the original monads, but their centres/central submonads as
well. Moreover, the interaction of the centre with operations on monadic
theories can be investigated.

Our definition of central submonads makes essential use of the notion of
monomorphism of strong monads. Another possibility for future work is to
investigate an alternative approach where we consider an appropriate class of
factorisation systems instead of monomorphisms to define central submonads. Yet
another possibility for future work is to investigate if central submonads of a
given strong monad have some interesting poset structure.

A natural generalisation of monads is the notion of \emph{arrows} -- or
\emph{strong promonads}. A promonad is a monoid in the category of profunctors,
and profunctors are to functors what relations are to functions. Arrows give
a more general framework to study computational effects, and are
particularly meaningful for effects in reversible computing
\cite{alimarine2005arrows, heunen2018arrows}. Our final proposal for future work
is to study equational theories and internal languages for arrows.

\bibliographystyle{alphaurl}
\bibliography{ref}

\end{document}

%% file: tikzit.tikzstyles
\tikzstyle{empty}=[shape=circle, tikzit fill={rgb,255: red,191; green,191; blue,191}]
\tikzstyle{dot}=[fill=black, draw=black, shape=circle]

\tikzstyle{to}=[draw=black, ->]
\tikzstyle{equal-arrow}=[{-}, double equal sign distance]
\tikzstyle{hook}=[right hook->, draw=black, tikzit draw=magenta]
\tikzstyle{dashed-arrow}=[dashed, ->]
\tikzstyle{mono}=[draw=black, to reversed->]

%% file: figures/monad-def.tikz
\begin{tikzpicture}
	\begin{pgfonlayer}{nodelayer}
		\node [style=empty] (0) at (-2.25, 2) {$\TT^3 X$};
		\node [style=empty] (1) at (2.25, 2) {$\TT^2 X$};
		\node [style=empty] (2) at (-2.25, -2) {$\TT^2 X$};
		\node [style=empty] (3) at (2.25, -2) {$\TT X$};
		\node [style=empty] (4) at (-3.25, 0) {$\TT\mu_X$};
		\node [style=empty] (5) at (0, 2.45) {$\mu_{\TT X}$};
		\node [style=empty] (6) at (3, 0) {$\mu_X$};
		\node [style=empty] (7) at (0, -2.5) {$\mu_X$};
		\node [style=none] (8) at (-2.25, 1.25) {};
		\node [style=none] (9) at (-2.25, -1.25) {};
		\node [style=none] (10) at (2.25, -1.25) {};
		\node [style=none] (11) at (2.25, 1.25) {};
	\end{pgfonlayer}
	\begin{pgfonlayer}{edgelayer}
		\draw [style=to] (2) to (3);
		\draw [style=to] (0) to (1);
		\draw [style=to] (8.center) to (9.center);
		\draw [style=to] (11.center) to (10.center);
	\end{pgfonlayer}
\end{tikzpicture}

%% file: figures/monad-def2.tikz
\begin{tikzpicture}
	\begin{pgfonlayer}{nodelayer}
		\node [style=empty] (0) at (-2.25, 2) {$\TT X$};
		\node [style=empty] (1) at (2.25, 2) {$\TT^2 X$};
		\node [style=empty] (2) at (-2.25, -2) {$\TT^2 X$};
		\node [style=empty] (3) at (2.25, -2) {$\TT X$};
		\node [style=empty] (4) at (-3.125, 0) {$\eta_{\TT X}$};
		\node [style=empty] (5) at (0, 2.5) {$\TT\eta_X$};
		\node [style=empty] (6) at (3, 0) {$\mu_X$};
		\node [style=empty] (7) at (0, -2.5) {$\mu_X$};
		\node [style=none] (8) at (-2.25, 1.25) {};
		\node [style=none] (9) at (-2.25, -1.25) {};
		\node [style=none] (10) at (2.25, 1.25) {};
		\node [style=none] (11) at (2.25, -1.25) {};
	\end{pgfonlayer}
	\begin{pgfonlayer}{edgelayer}
		\draw [style=to] (2) to (3);
		\draw [style=to] (0) to (1);
		\draw [style=equal-arrow] (0) to (3);
		\draw [style=to] (8.center) to (9.center);
		\draw [style=to] (10.center) to (11.center);
	\end{pgfonlayer}
\end{tikzpicture}

%% file: figures/def-central-morphism.tikz
\begin{tikzpicture}
	\begin{pgfonlayer}{nodelayer}
		\node [style=empty] (0) at (-2.5, 2) {$X\otimes X'$};
		\node [style=empty] (1) at (-2.5, -2.25) {$X\otimes Y'$};
		\node [style=empty] (2) at (4.5, 2) {$Y\otimes X'$};
		\node [style=empty] (3) at (4.5, -2.25) {$Y\otimes Y'$};
		\node [style=empty] (4) at (-2.5, 1.5) {};
		\node [style=empty] (5) at (-2.5, -1.75) {};
		\node [style=empty] (6) at (4.5, 1.5) {};
		\node [style=empty] (7) at (4.5, -1.75) {};
		\node [style=empty] (16) at (1, 2.5) {$f\otimes_l X'$};
		\node [style=empty] (17) at (-3.875, 0) {$X\otimes_r f'$};
		\node [style=empty] (18) at (5.875, 0) {$Y\otimes_r f'$};
		\node [style=empty] (19) at (1, -2.75) {$f\otimes_l Y'$};
	\end{pgfonlayer}
	\begin{pgfonlayer}{edgelayer}
		\draw [style=to] (0) to (2);
		\draw [style=to] (6) to (7);
		\draw [style=to] (4) to (5);
		\draw [style=to] (1) to (3);
	\end{pgfonlayer}
\end{tikzpicture}

%% file: figures/def-central-cone.tikz
\begin{tikzpicture}
	\begin{pgfonlayer}{nodelayer}
		\node [style=empty] (0) at (-6, 4) {$Z\otimes\TT Y$};
		\node [style=empty] (1) at (1, 4) {$\TT X\otimes\TT Y$};
		\node [style=empty] (2) at (8, 4) {$\TT (X\otimes\TT Y)$};
		\node [style=empty] (3) at (8, 0.5) {$\TT^2 (X\otimes Y)$};
		\node [style=empty] (4) at (8, -3) {$\TT (X\otimes Y)$};
		\node [style=empty] (5) at (-6, 0.5) {$\TT X\otimes\TT Y$};
		\node [style=empty] (6) at (-6, -3) {$\TT(\TT X\otimes Y)$};
		\node [style=empty] (7) at (1, -3) {$\TT^2 (X\otimes Y)$};
		\node [style=empty] (8) at (-6, 3.5) {};
		\node [style=empty] (9) at (-6, 1) {};
		\node [style=empty] (10) at (-6, 0) {};
		\node [style=empty] (11) at (-6, -2.5) {};
		\node [style=empty] (12) at (8, 3.5) {};
		\node [style=empty] (13) at (8, 1) {};
		\node [style=empty] (14) at (8, 0) {};
		\node [style=empty] (15) at (8, -2.5) {};
		\node [style=empty] (16) at (-2.5, 4.5) {$\iota\otimes\TT Y$};
		\node [style=empty] (17) at (4.5, 4.5) {$\tau'_{X,\TT Y}$};
		\node [style=empty] (18) at (9.25, 2.25) {$\TT\tau_{X,Y}$};
		\node [style=empty] (19) at (9.25, -1.25) {$\mu_{X\otimes Y}$};
		\node [style=empty] (20) at (-7.25, 2.25) {$\iota\otimes\TT Y$};
		\node [style=empty] (21) at (-7.25, -1.25) {$\tau_{\TT X,Y}$};
		\node [style=empty] (22) at (-2.5, -3.5) {$\TT\tau'_{X,Y}$};
		\node [style=empty] (23) at (4.5, -3.5) {$\mu_{X\otimes Y}$};
	\end{pgfonlayer}
	\begin{pgfonlayer}{edgelayer}
		\draw [style=to] (0) to (1);
		\draw [style=to] (1) to (2);
		\draw [style=to] (6) to (7);
		\draw [style=to] (7) to (4);
		\draw [style=to] (8) to (9);
		\draw [style=to] (10) to (11);
		\draw [style=to] (14) to (15);
		\draw [style=to] (12) to (13);
	\end{pgfonlayer}
\end{tikzpicture}

%% file: figures/central-cone-to-morph.tikz
\begin{tikzpicture}
	\begin{pgfonlayer}{nodelayer}
		\node [style=empty] (0) at (-12, 1.5) {$X\otimes X'$};
		\node [style=empty] (1) at (-12, -9.5) {$\TT(X\otimes Y')$};
		\node [style=empty] (2) at (-4, 1.5) {$\TT Y\otimes X'$};
		\node [style=empty] (3) at (-4, -9.5) {$\TT(\TT Y\otimes Y')$};
		\node [style=empty] (4) at (-12, 1) {};
		\node [style=empty] (5) at (-12, -3.5) {};
		\node [style=empty] (6) at (12, 1) {};
		\node [style=empty] (7) at (12, -3.5) {};
		\node [style=empty] (8) at (-8, 2) {$f\otimes X'$};
		\node [style=empty] (9) at (-13.25, -1.25) {$X\otimes f'$};
		\node [style=empty] (10) at (8, 2) {$\TT(Y\otimes f')$};
		\node [style=empty] (11) at (-8, -10.25) {$\TT(f\otimes Y')$};
		\node [style=empty] (12) at (4, 1.5) {$\TT (Y\otimes X')$};
		\node [style=empty] (13) at (0, 2) {$\tau'_{Y,X'}$};
		\node [style=empty] (14) at (-12, -4) {$X\otimes\TT Y'$};
		\node [style=empty] (15) at (-12, -4.5) {};
		\node [style=empty] (16) at (-12, -9) {};
		\node [style=empty] (17) at (12, -4) {$\TT^2(Y\otimes Y')$};
		\node [style=empty] (18) at (12, -4.5) {};
		\node [style=empty] (19) at (4, -9.5) {$\TT^2(Y\otimes Y')$};
		\node [style=empty] (20) at (12, -9) {};
		\node [style=empty] (21) at (13.5, -1.25) {$\TT\tau_{Y,Y'}$};
		\node [style=empty] (22) at (0, -10.25) {$\TT\tau'_{Y,Y'}$};
		\node [style=empty] (23) at (-13.25, -6.75) {$\tau_{X,Y'}$};
		\node [style=empty] (24) at (12, 1.5) {$\TT (Y\otimes\TT Y')$};
		\node [style=empty] (25) at (13.5, -6.75) {$\mu_{Y\otimes Y'}$};
		\node [style=empty] (26) at (12, -9.5) {$\TT(Y\otimes Y')$};
		\node [style=empty] (27) at (8, -10.25) {$\mu_{Y\otimes Y'}$};
		\node [style=empty] (28) at (-4, -4) {$\TT Y\otimes\TT Y'$};
		\node [style=empty] (29) at (-4, -4.5) {};
		\node [style=empty] (30) at (-4, -9) {};
		\node [style=empty] (31) at (-8, -4.5) {$f\otimes\TT Y'$};
		\node [style=empty] (32) at (-5.25, -6.75) {$\tau_{\TT Y,Y'}$};
		\node [style=empty] (33) at (-2.25, -2) {};
		\node [style=empty] (34) at (11.25, 1) {};
		\node [style=empty] (36) at (-4, 1) {};
		\node [style=empty] (37) at (-4, -1.5) {};
		\node [style=empty] (38) at (-5.5, -0.25) {$\TT Y\otimes f'$};
		\node [style=empty] (39) at (4.25, -1.5) {$\tau'_{Y,\TT Y'}$};
		\node [style=none] (40) at (-9.25, -0.5) {\textcolor{blue}{(1)}};
		\node [style=none] (41) at (0.75, 0.25) {\textcolor{blue}{(2)}};
		\node [style=none] (42) at (-9, -7) {\textcolor{blue}{(3)}};
		\node [style=none] (43) at (3.75, -6) {\textcolor{blue}{(4)}};
		\node [style=empty] (44) at (-4, -2.25) {$\TT Y\otimes\TT Y'$};
		\node [style=empty] (45) at (-8.25, -2.5) {$f\otimes\TT Y'$};
	\end{pgfonlayer}
	\begin{pgfonlayer}{edgelayer}
		\draw [style=to] (0) to (2);
		\draw [style=to] (6) to (7);
		\draw [style=to] (4) to (5);
		\draw [style=to] (1) to (3);
		\draw [style=to] (2) to (12);
		\draw [style=to] (15) to (16);
		\draw [style=to] (18) to (20);
		\draw [style=to] (3) to (19);
		\draw [style=to] (12) to (24);
		\draw [style=to] (19) to (26);
		\draw [style=to] (14) to (28);
		\draw [style=to] (29) to (30);
		\draw [style=to] (33) to (34);
		\draw [style=to] (36) to (37);
		\draw [style=to] (14) to (44);
	\end{pgfonlayer}
\end{tikzpicture}

%% file: figures/central-morph-to-cone.tikz
\begin{tikzpicture}
	\begin{pgfonlayer}{nodelayer}
		\node [style=empty] (0) at (-6, 3) {$Z\otimes\TT Y$};
		\node [style=empty] (1) at (4.75, 3) {$\TT X\otimes\TT Y$};
		\node [style=empty] (2) at (14, 3) {$\TT (X\otimes\TT Y)$};
		\node [style=empty] (3) at (14, -2) {$\TT^2 (X\otimes Y)$};
		\node [style=empty] (4) at (14, -7) {$\TT (X\otimes Y)$};
		\node [style=empty] (5) at (-6, -2) {$\TT X\otimes\TT Y$};
		\node [style=empty] (6) at (-6, -7) {$\TT(\TT X\otimes Y)$};
		\node [style=empty] (7) at (3, -7) {$\TT^2 (X\otimes Y)$};
		\node [style=empty] (8) at (-6, 2.5) {};
		\node [style=empty] (9) at (-6, -1.5) {};
		\node [style=empty] (10) at (-6, -2.5) {};
		\node [style=empty] (11) at (-6, -6.5) {};
		\node [style=empty] (12) at (14, 2.5) {};
		\node [style=empty] (13) at (14, -1.5) {};
		\node [style=empty] (14) at (14, -2.5) {};
		\node [style=empty] (15) at (14, -6.5) {};
		\node [style=empty] (16) at (0.5, 3.5) {$f\otimes\TT Y$};
		\node [style=empty] (17) at (10.25, 3.5) {$\tau'_{X,\TT Y}$};
		\node [style=empty] (18) at (15.25, 0.5) {$\TT\tau_{X,Y}$};
		\node [style=empty] (19) at (15.25, -4.5) {$\mu_{X\otimes Y}$};
		\node [style=empty] (20) at (-7.25, 0.5) {$f\otimes\TT Y$};
		\node [style=empty] (21) at (-7.25, -4.5) {$\tau_{\TT X,Y}$};
		\node [style=empty] (22) at (-2, -7.5) {$\TT\tau'_{X,Y}$};
		\node [style=empty] (23) at (9, -7.5) {$\mu_{X\otimes Y}$};
		\node [style=empty] (25) at (-5, 2.5) {};
		\node [style=empty] (26) at (0, 0.25) {};
		\node [style=empty] (27) at (-1, 1.75) {$f\otimes\TT Y$};
		\node [style=empty] (28) at (2, 0.5) {};
		\node [style=empty] (29) at (4.5, 2.5) {};
		\node [style=empty] (30) at (1.5, 0) {$\TT X\otimes\TT Y$};
		\node [style=empty] (31) at (11, 0) {$\TT (X\otimes\TT Y)$};
		\node [style=empty] (32) at (6.75, 0.5) {$\tau'_{X,\TT Y}$};
		\node [style=empty] (33) at (11.25, 0.5) {};
		\node [style=empty] (34) at (13.5, 2.5) {};
		\node [style=empty] (37) at (0.5, -5) {$\TT(Z\otimes Y)$};
		\node [style=empty] (38) at (0.5, -2.5) {};
		\node [style=empty] (39) at (0.5, -4.5) {};
		\node [style=empty] (40) at (-0.75, -3.75) {$\tau_{Z,Y}$};
		\node [style=empty] (41) at (-4.5, -6.5) {};
		\node [style=empty] (42) at (-0.75, -5.5) {};
		\node [style=none] (43) at (-3.5, -4) {\textcolor{blue}{(1)}};
		\node [style=empty] (44) at (0.5, -2) {$Z\otimes\TT Y$};
		\node [style=empty] (45) at (-5.5, 2.25) {};
		\node [style=empty] (46) at (-0.25, -1.5) {};
		\node [style=empty] (47) at (-2.75, -1.5) {$f\otimes\TT Y$};
		\node [style=empty] (48) at (-3.75, -5.5) {$\TT(f\otimes Y)$};
		\node [style=none] (49) at (7, -3.5) {\textcolor{blue}{(2)}};
	\end{pgfonlayer}
	\begin{pgfonlayer}{edgelayer}
		\draw [style=to] (0) to (1);
		\draw [style=to] (1) to (2);
		\draw [style=to] (6) to (7);
		\draw [style=to] (7) to (4);
		\draw [style=to] (8) to (9);
		\draw [style=to] (10) to (11);
		\draw [style=to] (14) to (15);
		\draw [style=to] (12) to (13);
		\draw [style=to] (25) to (26);
		\draw [style=equal-arrow] (28) to (29);
		\draw [style=to] (30) to (31);
		\draw [style=equal-arrow] (33) to (34);
		\draw [style=to] (38) to (39);
		\draw [style=equal-arrow] (45) to (46);
		\draw [style=to] (44) to (5);
		\draw [style=to] (42) to (41);
	\end{pgfonlayer}
\end{tikzpicture}

%% file: figures/central-precomposing.tikz
\begin{tikzpicture}
	\begin{pgfonlayer}{nodelayer}
		\node [style=empty] (0) at (-7, 3) {$X\otimes\TT X'$};
		\node [style=empty] (1) at (0, 3) {$\TT Y\otimes\TT X'$};
		\node [style=empty] (2) at (7, 3) {$\TT (Y\otimes\TT X')$};
		\node [style=empty] (3) at (7, 0) {$\TT^2 (Y\otimes X')$};
		\node [style=empty] (4) at (7, -3) {$\TT (Y\otimes X')$};
		\node [style=empty] (5) at (-7, 0) {$\TT Y\otimes\TT X'$};
		\node [style=empty] (6) at (-7, -3) {$\TT(\TT Y\otimes X')$};
		\node [style=empty] (7) at (0, -3) {$\TT^2 (Y\otimes X')$};
		\node [style=empty] (8) at (-7, 2.5) {};
		\node [style=empty] (9) at (-7, 0.5) {};
		\node [style=empty] (10) at (-7, -0.5) {};
		\node [style=empty] (11) at (-7, -2.5) {};
		\node [style=empty] (12) at (7, 2.5) {};
		\node [style=empty] (13) at (7, 0.5) {};
		\node [style=empty] (14) at (7, -0.5) {};
		\node [style=empty] (15) at (7, -2.5) {};
		\node [style=empty] (16) at (-3.5, 3.5) {$f\otimes\TT X'$};
		\node [style=empty] (17) at (3.5, 3.75) {$\tau'_{Y,\TT X'}$};
		\node [style=empty] (18) at (8.25, 1.5) {$\TT\tau_{Y,X'}$};
		\node [style=empty] (19) at (8.25, -1.5) {$\mu_{Y\otimes X'}$};
		\node [style=empty] (20) at (-8.75, 1.5) {$f\otimes\TT X'$};
		\node [style=empty] (21) at (-8.25, -1.5) {$\tau_{\TT Y,X'}$};
		\node [style=empty] (22) at (-3.5, -3.75) {$\TT\tau'_{Y,X'}$};
		\node [style=empty] (23) at (3.5, -3.5) {$\mu_{Y\otimes X'}$};
		\node [style=empty] (24) at (-14, 3) {$Z\otimes\TT X'$};
		\node [style=empty] (27) at (-10.5, 3.5) {$g\otimes\TT X'$};
	\end{pgfonlayer}
	\begin{pgfonlayer}{edgelayer}
		\draw [style=to] (0) to (1);
		\draw [style=to] (1) to (2);
		\draw [style=to] (6) to (7);
		\draw [style=to] (7) to (4);
		\draw [style=to] (8) to (9);
		\draw [style=to] (10) to (11);
		\draw [style=to] (14) to (15);
		\draw [style=to] (12) to (13);
		\draw [style=to] (24) to (0);
	\end{pgfonlayer}
\end{tikzpicture}

%% file: figures/central-postcomposing.tikz
\begin{tikzpicture}
	\begin{pgfonlayer}{nodelayer}
		\node [style=empty] (0) at (-13, 13) {$X\otimes\TT X'$};
		\node [style=empty] (1) at (6.75, 13) {$\TT Y\otimes\TT X'$};
		\node [style=empty] (2) at (13.75, 13) {$\TT Z\otimes\TT X'$};
		\node [style=empty] (3) at (13.75, 3) {$\TT^2 (Z\otimes X')$};
		\node [style=empty] (4) at (13.75, -3.25) {$\TT (Z\otimes X')$};
		\node [style=empty] (5) at (-13, -3.25) {$\TT Z\otimes\TT X'$};
		\node [style=empty] (6) at (-6, -3.25) {$\TT(\TT Z\otimes X')$};
		\node [style=empty] (7) at (1, -3.25) {$\TT^2 (Z\otimes X')$};
		\node [style=empty] (8) at (-13, 12.5) {};
		\node [style=empty] (9) at (-13, 0.75) {};
		\node [style=empty] (10) at (-13, -0.25) {};
		\node [style=empty] (11) at (-13, -2.75) {};
		\node [style=empty] (12) at (13.75, 7.5) {};
		\node [style=empty] (13) at (13.75, 3.5) {};
		\node [style=empty] (14) at (13.75, 2.5) {};
		\node [style=empty] (15) at (13.75, -2.75) {};
		\node [style=empty] (16) at (-3, 13.75) {$f\otimes\TT X'$};
		\node [style=empty] (17) at (15, 10.75) {$\tau'_{Z,\TT X'}$};
		\node [style=empty] (18) at (15, 5.5) {$\TT\tau_{Z,X'}$};
		\node [style=empty] (19) at (15, 0) {$\mu_{Z\otimes X'}$};
		\node [style=empty] (20) at (-14.5, 7) {$f\otimes\TT X'$};
		\node [style=empty] (21) at (-9.5, -4) {$\tau_{\TT Z,X'}$};
		\node [style=empty] (22) at (-2.25, -4) {$\TT\tau'_{Z,X'}$};
		\node [style=empty] (23) at (7, -4) {$\mu_{Z\otimes X'}$};
		\node [style=empty] (24) at (13.75, 12.5) {};
		\node [style=empty] (25) at (13.75, 8.5) {};
		\node [style=empty] (26) at (13.75, 8) {$\TT (Z\otimes\TT X')$};
		\node [style=empty] (27) at (-13, 0.25) {$\TT Y\otimes\TT X'$};
		\node [style=empty] (28) at (-14.75, -1.5) {$\TT g\otimes\TT X'$};
		\node [style=empty] (29) at (10.25, 13.75) {$\TT g\otimes\TT X'$};
		\node [style=empty] (30) at (6.75, 8) {$\TT (Y\otimes\TT X')$};
		\node [style=empty] (31) at (6.75, 3) {$\TT^2 (Y\otimes X')$};
		\node [style=empty] (32) at (6.75, 12.5) {};
		\node [style=empty] (33) at (6.75, 8.5) {};
		\node [style=empty] (34) at (6.75, 7.5) {};
		\node [style=empty] (35) at (6.75, 3.5) {};
		\node [style=empty] (36) at (6.75, 0.25) {$\TT (Y\otimes X')$};
		\node [style=empty] (37) at (7.75, 0) {};
		\node [style=empty] (38) at (12.75, -2.75) {};
		\node [style=empty] (39) at (6.75, 2.5) {};
		\node [style=empty] (40) at (6.75, 0.75) {};
		\node [style=empty] (41) at (5.5, 10.75) {$\tau'_{Y,\TT X'}$};
		\node [style=empty] (42) at (5.5, 5.5) {$\TT\tau_{Y,X'}$};
		\node [style=empty] (43) at (5.5, 1.75) {$\mu_{Y\otimes X'}$};
		\node [style=empty] (44) at (10.25, 8.75) {$\TT (g\otimes\TT X')$};
		\node [style=empty] (45) at (10.25, 3.75) {$\TT^2( g\otimes X')$};
		\node [style=empty] (46) at (11, -0.25) {$\TT( g\otimes X')$};
		\node [style=empty] (47) at (-6, 0.25) {$\TT(\TT Y\otimes X')$};
		\node [style=empty] (48) at (1, 0.25) {$\TT^2 (Y\otimes X')$};
		\node [style=empty] (49) at (1, -0.25) {};
		\node [style=empty] (50) at (1, -2.75) {};
		\node [style=empty] (51) at (-6, -0.25) {};
		\node [style=empty] (52) at (-6, -2.75) {};
		\node [style=empty] (53) at (-9.5, 1) {$\tau_{\TT Y,X'}$};
		\node [style=empty] (54) at (-2.25, 1) {$\TT\tau'_{Y,X'}$};
		\node [style=empty] (55) at (3.75, 0.75) {$\mu_{Y\otimes X'}$};
		\node [style=empty] (56) at (-8, -1.5) {$\TT (\TT g\otimes X')$};
		\node [style=empty] (57) at (3.25, -1.5) {$\TT^2( g\otimes X')$};
		\node [style=none] (58) at (-4, 7) {\textcolor{blue}{(1)}};
		\node [style=none] (59) at (10.25, 11.25) {\textcolor{blue}{(2)}};
		\node [style=none] (60) at (10.25, 5.75) {\textcolor{blue}{(3)}};
		\node [style=none] (61) at (10.25, 1.5) {\textcolor{blue}{(4)}};
		\node [style=none] (62) at (-11.25, -1.5) {\textcolor{blue}{(5)}};
		\node [style=none] (63) at (-2.5, -1.5) {\textcolor{blue}{(6)}};
		\node [style=none] (64) at (6.75, -1.5) {\textcolor{blue}{(7)}};
	\end{pgfonlayer}
	\begin{pgfonlayer}{edgelayer}
		\draw [style=to] (0) to (1);
		\draw [style=to] (1) to (2);
		\draw [style=to] (6) to (7);
		\draw [style=to] (7) to (4);
		\draw [style=to] (8) to (9);
		\draw [style=to] (10) to (11);
		\draw [style=to] (14) to (15);
		\draw [style=to] (12) to (13);
		\draw (5) to (6);
		\draw [style=to] (24) to (25);
		\draw [style=to] (30) to (26);
		\draw [style=to] (32) to (33);
		\draw [style=to] (34) to (35);
		\draw [style=to] (31) to (3);
		\draw [style=to] (37) to (38);
		\draw [style=to] (39) to (40);
		\draw [style=to] (27) to (47);
		\draw [style=to] (47) to (48);
		\draw [style=to] (48) to (36);
		\draw [style=to] (51) to (52);
		\draw [style=to] (49) to (50);
	\end{pgfonlayer}
\end{tikzpicture}

%% file: figures/z-functor.tikz
\begin{tikzpicture}
	\begin{pgfonlayer}{nodelayer}
		\node [style=empty] (0) at (-2, 1.5) {$\ZZ X$};
		\node [style=empty] (1) at (2, 1.5) {$\ZZ Y$};
		\node [style=empty] (2) at (-2, -1.5) {$\TT X$};
		\node [style=empty] (3) at (2, -1.5) {$\TT Y$};
		\node [style=empty] (4) at (-2, 1) {};
		\node [style=empty] (5) at (2, 1) {};
		\node [style=empty] (6) at (2, -1) {};
		\node [style=empty] (7) at (-2, -1) {};
		\node [style=empty] (8) at (-2.75, 0) {$\iota_X$};
		\node [style=empty] (9) at (2.75, 0) {$\iota_Y$};
		\node [style=empty] (10) at (0, -2.125) {$\TT f$};
		\node [style=empty] (11) at (0, 2.125) {$\ZZ f$};
	\end{pgfonlayer}
	\begin{pgfonlayer}{edgelayer}
		\draw [style=to] (4) to (7);
		\draw [style=to] (5) to (6);
		\draw [style=to] (2) to (3);
		\draw [style=dashed-arrow] (0) to (1);
	\end{pgfonlayer}
\end{tikzpicture}

%% file: figures/z-functor-comp.tikz
\begin{tikzpicture}
	\begin{pgfonlayer}{nodelayer}
		\node [style=empty] (0) at (-1.75, 3) {$\ZZ A$};
		\node [style=empty] (1) at (-1.75, 0) {$\ZZ B$};
		\node [style=empty] (2) at (1.75, 3) {$\TT A$};
		\node [style=empty] (3) at (1.75, 0) {$\TT B$};
		\node [style=empty] (4) at (-1.25, 3) {};
		\node [style=empty] (5) at (-1.25, 0) {};
		\node [style=empty] (6) at (1.25, 0) {};
		\node [style=empty] (7) at (1.25, 3) {};
		\node [style=empty] (8) at (0, 3.375) {$\iota_A$};
		\node [style=empty] (9) at (0, 0.375) {$\iota_B$};
		\node [style=empty] (10) at (2.5, 1.5) {$\TT g$};
		\node [style=empty] (11) at (-2.5, 1.5) {$\ZZ g$};
		\node [style=empty] (12) at (-1.75, -3) {$\ZZ C$};
		\node [style=empty] (13) at (-2.5, -1.5) {$\ZZ f$};
		\node [style=empty] (14) at (-6, 0) {$\ZZ(f\circ g)$};
		\node [style=empty] (15) at (-1.25, -3) {};
		\node [style=empty] (16) at (1.25, -3) {};
		\node [style=empty] (17) at (1.75, -3) {$\TT C$};
		\node [style=empty] (18) at (0, -3.5) {$\iota_C$};
		\node [style=empty] (19) at (2.5, -1.5) {$\TT f$};
		\node [style=empty] (20) at (6, 0) {$\TT(f\circ g)$};
	\end{pgfonlayer}
	\begin{pgfonlayer}{edgelayer}
		\draw [style=to] (4) to (7);
		\draw [style=to] (5) to (6);
		\draw [style=to] (2) to (3);
		\draw [style=dashed-arrow, bend right=75] (0) to (12);
		\draw [style=to] (3) to (17);
		\draw [style=to] (15) to (16);
		\draw [style=to] (0) to (1);
		\draw [style=to] (1) to (12);
		\draw [style=to, bend left=75] (2) to (17);
	\end{pgfonlayer}
\end{tikzpicture}

%% file: figures/central-unit.tikz
\begin{tikzpicture}
	\begin{pgfonlayer}{nodelayer}
		\node [style=empty] (0) at (-1.5, 1) {$X$};
		\node [style=empty] (1) at (0, -1.5) {$\TT X$};
		\node [style=empty] (2) at (1.75, 1) {$\ZZ X$};
		\node [style=empty] (3) at (-1.25, 1) {};
		\node [style=empty] (4) at (1.25, 1) {};
		\node [style=empty] (5) at (0.25, -1.25) {};
		\node [style=empty] (6) at (1.75, 0.75) {};
		\node [style=empty] (7) at (-1.45, -0.425) {$\eta_X$};
		\node [style=empty] (8) at (1.55, -0.4) {$\iota_X$};
		\node [style=empty] (9) at (-0.025, 1.525) {$\eta^\ZZ_X$};
		\node [style=none] (10) at (-0.5, -1) {};
		\node [style=none] (11) at (-1.375, 0.5) {};
	\end{pgfonlayer}
	\begin{pgfonlayer}{edgelayer}
		\draw [style=dashed-arrow] (3) to (4);
		\draw [style=to] (11.center) to (10.center);
		\draw [style=to] (6) to (5);
	\end{pgfonlayer}
\end{tikzpicture}

%% file: figures/central-mult.tikz
\begin{tikzpicture}
	\begin{pgfonlayer}{nodelayer}
		\node [style=empty] (0) at (-3, 2) {$\ZZ^2 X$};
		\node [style=empty] (1) at (-3, -1) {$\TT\ZZ X$};
		\node [style=empty] (2) at (-3, 1.5) {};
		\node [style=empty] (3) at (-3, -0.5) {};
		\node [style=empty] (4) at (0.5, -1) {$\TT^2 X$};
		\node [style=empty] (5) at (4, -1) {$\TT X$};
		\node [style=empty] (6) at (4, 2) {$\ZZ X$};
		\node [style=empty] (7) at (4, 1.5) {};
		\node [style=empty] (8) at (4, -0.5) {};
		\node [style=empty] (9) at (0.5, 2.75) {$\mu^\ZZ_X$};
		\node [style=empty] (10) at (-1, -1.5) {$\TT\iota_X$};
		\node [style=empty] (11) at (2.5, -1.5) {$\mu_X$};
		\node [style=empty] (12) at (-3.75, 0.5) {$\iota_{\ZZ X}$};
		\node [style=empty] (13) at (4.75, 0.5) {$\iota_X$};
	\end{pgfonlayer}
	\begin{pgfonlayer}{edgelayer}
		\draw [style=to] (7) to (8);
		\draw [style=to] (1) to (4);
		\draw [style=to] (4) to (5);
		\draw [style=to] (2) to (3);
		\draw [style=dashed-arrow] (0) to (6);
	\end{pgfonlayer}
\end{tikzpicture}

%% file: figures/central-strength.tikz
\begin{tikzpicture}
	\begin{pgfonlayer}{nodelayer}
		\node [style=empty] (0) at (-3.5, 2) {$A\otimes\ZZ B$};
		\node [style=empty] (1) at (2.5, 2) {$\ZZ(A\otimes B)$};
		\node [style=empty] (2) at (-3.5, -1) {$A\otimes\TT B$};
		\node [style=empty] (3) at (2.5, -1) {$\TT(A\otimes B)$};
		\node [style=empty] (4) at (-3.5, 1.5) {};
		\node [style=empty] (5) at (-3.5, -0.5) {};
		\node [style=empty] (6) at (2.5, 1.5) {};
		\node [style=empty] (7) at (2.5, -0.5) {};
		\node [style=empty] (8) at (-4.75, 0.5) {$A\otimes\iota_B$};
		\node [style=empty] (9) at (3.5, 0.5) {$\iota_{A\otimes B}$};
		\node [style=empty] (10) at (-0.625, 2.625) {$\tau^\ZZ_{A,B}$};
		\node [style=empty] (11) at (-0.625, -1.5) {$\tau_{A,B}$};
	\end{pgfonlayer}
	\begin{pgfonlayer}{edgelayer}
		\draw [style=to] (4) to (5);
		\draw [style=to] (2) to (3);
		\draw [style=to] (6) to (7);
		\draw [style=dashed-arrow] (0) to (1);
	\end{pgfonlayer}
\end{tikzpicture}

%% file: figures/eta-z-natural.tikz
\begin{tikzpicture}
	\begin{pgfonlayer}{nodelayer}
		\node [style=empty] (0) at (-6, 5) {$X$};
		\node [style=empty] (1) at (6, 5) {$\ZZ X$};
		\node [style=empty] (2) at (-6, -3) {$Y$};
		\node [style=empty] (3) at (6, -3) {$\TT Y$};
		\node [style=empty] (4) at (-6, 4.5) {};
		\node [style=empty] (5) at (-6, -2.5) {};
		\node [style=empty] (6) at (6, 4.5) {};
		\node [style=empty] (7) at (6, 1) {$\ZZ Y$};
		\node [style=empty] (8) at (6, 1.5) {};
		\node [style=empty] (9) at (6, 0.5) {};
		\node [style=empty] (10) at (6, -2.5) {};
		\node [style=empty] (11) at (0, -3) {$\ZZ Y$};
		\node [style=empty] (12) at (0, 3) {$\TT X$};
		\node [style=empty] (13) at (-0.5, 3.25) {};
		\node [style=empty] (14) at (0.5, 3.25) {};
		\node [style=empty] (15) at (5.5, 4.5) {};
		\node [style=empty] (16) at (-5.5, 4.5) {};
		\node [style=empty] (17) at (0.5, 2.5) {};
		\node [style=empty] (18) at (5.5, -2.5) {};
		\node [style=empty] (19) at (0, 5.75) {$\eta^\ZZ_X$};
		\node [style=empty] (20) at (-7, 1.25) {$f$};
		\node [style=empty] (21) at (7, 3.25) {$\ZZ f$};
		\node [style=empty] (22) at (1.5, 0.25) {$\TT f$};
		\node [style=empty] (23) at (0, -1.75) {$\eta_y$};
		\node [style=empty] (24) at (-3, -3.75) {$\eta^\ZZ_Y$};
		\node [style=empty] (25) at (3, -3.75) {$\iota_Y$};
		\node [style=empty] (26) at (6.75, -1) {$\iota_Y$};
		\node [style=empty] (27) at (-3.75, 3.25) {$\eta_X$};
		\node [style=empty] (28) at (3.75, 3.25) {$\iota_X$};
		\node [style=none] (29) at (0, 4.25) {\textcolor{blue}{(1)}};
		\node [style=none] (30) at (4, 1.25) {\textcolor{blue}{(2)}};
		\node [style=none] (31) at (-2.5, 0.5) {\textcolor{blue}{(3)}};
		\node [style=none] (32) at (-1.75, -2.25) {\textcolor{blue}{(4)}};
	\end{pgfonlayer}
	\begin{pgfonlayer}{edgelayer}
		\draw [style=to] (0) to (1);
		\draw [style=to] (6) to (8);
		\draw [style=to] (4) to (5);
		\draw [style=to] (2) to (11);
		\draw [style=mono] (11) to (3);
		\draw [style=mono] (9) to (10);
		\draw [style=to, bend left] (2) to (3);
		\draw [style=to] (16) to (13);
		\draw [style=to] (15) to (14);
		\draw [style=to] (17) to (18);
	\end{pgfonlayer}
\end{tikzpicture}

%% file: figures/mu-z-natural.tikz
\begin{tikzpicture}
	\begin{pgfonlayer}{nodelayer}
		\node [style=empty] (0) at (-8, 5) {$\ZZ^2 X$};
		\node [style=empty] (1) at (8, 5) {$\ZZ X$};
		\node [style=empty] (2) at (-8, -7) {$\ZZ^2 Y$};
		\node [style=empty] (3) at (8, -7) {$\TT Y$};
		\node [style=empty] (4) at (-8, 4.5) {};
		\node [style=empty] (5) at (-8, -6.5) {};
		\node [style=empty] (6) at (0, -7) {$\ZZ Y$};
		\node [style=empty] (7) at (8, -1) {$\ZZ Y$};
		\node [style=empty] (8) at (8, 4.5) {};
		\node [style=empty] (9) at (8, -0.5) {};
		\node [style=empty] (10) at (8, -1.5) {};
		\node [style=empty] (11) at (8, -6.5) {};
		\node [style=empty] (12) at (-3, 2) {$\TT^2 X$};
		\node [style=empty] (13) at (3, 2) {$\TT X$};
		\node [style=empty] (14) at (-7.5, 4.5) {};
		\node [style=empty] (15) at (-3.5, 2.25) {};
		\node [style=empty] (16) at (3.5, 2.25) {};
		\node [style=empty] (17) at (7.5, 4.5) {};
		\node [style=empty] (18) at (0, -3) {$\TT^2 Y$};
		\node [style=empty] (19) at (-7.5, -6.5) {};
		\node [style=empty] (20) at (-0.5, -3.25) {};
		\node [style=empty] (21) at (0.5, -3.25) {};
		\node [style=empty] (22) at (7.25, -6.75) {};
		\node [style=empty] (23) at (-2.5, 1.5) {};
		\node [style=empty] (24) at (-0.25, -2.5) {};
		\node [style=empty] (25) at (3.25, 1.5) {};
		\node [style=empty] (26) at (7.5, -6.25) {};
		\node [style=empty] (27) at (-9, -1) {$\ZZ^2 f$};
		\node [style=empty] (28) at (9, 2) {$\ZZ f$};
		\node [style=empty] (29) at (9, -4) {$\iota_Y$};
		\node [style=empty] (30) at (4, -7.75) {$\iota_Y$};
		\node [style=empty] (31) at (0, 5.75) {$\mu^\ZZ_X$};
		\node [style=empty] (32) at (-6, 3) {$\iota\iota$};
		\node [style=empty] (33) at (-4.25, -4) {$\iota\iota$};
		\node [style=empty] (34) at (5.25, -0.25) {$\TT f$};
		\node [style=empty] (35) at (0, 2.5) {$\mu_X$};
		\node [style=empty] (36) at (-2.5, -0.75) {$\TT^2 f$};
		\node [style=empty] (37) at (2.5, -5) {$\mu_Y$};
		\node [style=empty] (38) at (-4, -7.75) {$\mu^\ZZ_Y$};
		\node [style=none] (39) at (0, 4) {\textcolor{blue}{(1)}};
		\node [style=none] (40) at (-5.5, -1) {\textcolor{blue}{(2)}};
		\node [style=none] (41) at (2, -1) {\textcolor{blue}{(3)}};
		\node [style=none] (42) at (6.5, -1.75) {\textcolor{blue}{(4)}};
		\node [style=none] (43) at (0, -5.5) {\textcolor{blue}{(5)}};
		\node [style=none] (44) at (5.75, 3) {$\iota$};
	\end{pgfonlayer}
	\begin{pgfonlayer}{edgelayer}
		\draw [style=to] (4) to (5);
		\draw [style=mono] (10) to (11);
		\draw [style=to] (8) to (9);
		\draw [style=to] (2) to (6);
		\draw [style=mono] (6) to (3);
		\draw [style=to] (0) to (1);
		\draw [style=to] (12) to (13);
		\draw [style=to] (14) to (15);
		\draw [style=to] (17) to (16);
		\draw [style=to] (19) to (20);
		\draw [style=to] (21) to (22);
		\draw [style=to] (23) to (24);
		\draw [style=to] (25) to (26);
	\end{pgfonlayer}
\end{tikzpicture}

%% file: figures/tau-z-natural.tikz
\begin{tikzpicture}
	\begin{pgfonlayer}{nodelayer}
		\node [style=empty] (0) at (-13, 6) {$A\otimes\ZZ C$};
		\node [style=empty] (1) at (5, 6) {$\ZZ(A\otimes C)$};
		\node [style=empty] (2) at (-13, -2) {$B\otimes\ZZ C$};
		\node [style=empty] (3) at (5, -2) {$\TT(B\otimes C)$};
		\node [style=empty] (4) at (-6, -2) {$\ZZ(B\otimes C)$};
		\node [style=empty] (5) at (5, 2) {$\ZZ(B\otimes C)$};
		\node [style=empty] (6) at (5, 5.5) {};
		\node [style=empty] (7) at (5, 2.5) {};
		\node [style=empty] (8) at (5, 1.5) {};
		\node [style=empty] (9) at (5, -1.5) {};
		\node [style=empty] (10) at (-13, 5.5) {};
		\node [style=empty] (11) at (-13, -1.5) {};
		\node [style=empty] (12) at (-7, 0.5) {$B\otimes\TT C$};
		\node [style=empty] (13) at (-7, 4) {$A\otimes\TT C$};
		\node [style=empty] (14) at (-1, 4) {$\TT(A\otimes C)$};
		\node [style=empty] (15) at (-7, 3.5) {};
		\node [style=empty] (16) at (-7, 1) {};
		\node [style=empty] (17) at (-5.5, 0.25) {};
		\node [style=empty] (18) at (3.5, -1.5) {};
		\node [style=empty] (19) at (4.25, -1.25) {};
		\node [style=empty] (20) at (-0.75, 3.5) {};
		\node [style=empty] (21) at (-12.5, 5.5) {};
		\node [style=empty] (22) at (-8.5, 4.25) {};
		\node [style=empty] (23) at (4.5, 5.5) {};
		\node [style=empty] (24) at (0.5, 4.25) {};
		\node [style=empty] (25) at (-12.5, -1.75) {};
		\node [style=empty] (26) at (-8.25, 0) {};
		\node [style=empty] (27) at (-4, 6.75) {$\tau^\ZZ_{A,C}$};
		\node [style=empty] (28) at (-4, 4.5) {$\tau_{A,C}$};
		\node [style=empty] (29) at (-9.5, -2.75) {$\tau^\ZZ_{B,C}$};
		\node [style=empty] (30) at (-1, -2.75) {$\iota_{B\otimes C}$};
		\node [style=empty] (31) at (6, 0.25) {$\iota_{B\otimes C}$};
		\node [style=empty] (32) at (-3, -0.75) {$\tau_{B,C}$};
		\node [style=empty] (33) at (-11.5, 2) {$f\otimes\ZZ C$};
		\node [style=empty] (34) at (-9.25, 5) {$A\otimes\iota$};
		\node [style=empty] (35) at (-9.25, -1) {$B\otimes\iota$};
		\node [style=empty] (36) at (6.5, 4) {$\ZZ(f\otimes C)$};
		\node [style=empty] (37) at (0, 0.75) {$\TT(f\otimes C)$};
		\node [style=empty] (38) at (-5.5, 2.25) {$f\otimes\TT C$};
		\node [style=empty] (39) at (2.75, 4.5) {$\iota$};
		\node [style=none] (40) at (-4, 5.25) {\textcolor{blue}{(1)}};
		\node [style=none] (41) at (-9, 2) {\textcolor{blue}{(2)}};
		\node [style=none] (42) at (-2.5, 2) {\textcolor{blue}{(3)}};
		\node [style=none] (43) at (2.5, 2.25) {\textcolor{blue}{(4)}};
		\node [style=none] (44) at (-6, -1) {\textcolor{blue}{(5)}};
	\end{pgfonlayer}
	\begin{pgfonlayer}{edgelayer}
		\draw [style=to] (0) to (1);
		\draw [style=to] (10) to (11);
		\draw [style=to] (21) to (22);
		\draw [style=to] (23) to (24);
		\draw [style=to] (13) to (14);
		\draw [style=to] (15) to (16);
		\draw [style=to] (25) to (26);
		\draw [style=to] (2) to (4);
		\draw [style=to] (17) to (18);
		\draw [style=to] (20) to (19);
		\draw [style=to] (6) to (7);
		\draw [style=mono] (4) to (3);
		\draw [style=mono] (8) to (9);
	\end{pgfonlayer}
\end{tikzpicture}

%% file: figures/tau-z-natural-1.tikz
\begin{tikzpicture}
	\begin{pgfonlayer}{nodelayer}
		\node [style=empty] (0) at (-9, 4.25) {$A\otimes\ZZ B$};
		\node [style=empty] (1) at (9, 4.25) {$\ZZ(A\otimes B)$};
		\node [style=empty] (2) at (-9, -3.75) {$A\otimes\ZZ C$};
		\node [style=empty] (3) at (9, -3.75) {$\TT(A\otimes C)$};
		\node [style=empty] (4) at (-2, -3.75) {$\ZZ(A\otimes C)$};
		\node [style=empty] (5) at (9, 0.25) {$\ZZ(A\otimes C)$};
		\node [style=empty] (6) at (9, 3.75) {};
		\node [style=empty] (7) at (9, 0.75) {};
		\node [style=empty] (8) at (9, -0.25) {};
		\node [style=empty] (9) at (9, -3.25) {};
		\node [style=empty] (10) at (-9, 3.75) {};
		\node [style=empty] (11) at (-9, -3.25) {};
		\node [style=empty] (12) at (-3, -1.25) {$A\otimes\TT C$};
		\node [style=empty] (13) at (-3, 2.25) {$A\otimes\TT B$};
		\node [style=empty] (14) at (3, 2.25) {$\TT(A\otimes B)$};
		\node [style=empty] (15) at (-3, 1.75) {};
		\node [style=empty] (16) at (-3, -0.75) {};
		\node [style=empty] (17) at (-1.5, -1.5) {};
		\node [style=empty] (18) at (7.5, -3.25) {};
		\node [style=empty] (19) at (8.25, -3) {};
		\node [style=empty] (20) at (3.25, 1.75) {};
		\node [style=empty] (21) at (-8.5, 3.75) {};
		\node [style=empty] (22) at (-4.5, 2.5) {};
		\node [style=empty] (23) at (8.5, 3.75) {};
		\node [style=empty] (24) at (4.5, 2.5) {};
		\node [style=empty] (25) at (-8.5, -3.5) {};
		\node [style=empty] (26) at (-4.25, -1.75) {};
		\node [style=empty] (27) at (0, 5) {$\tau^\ZZ_{A,B}$};
		\node [style=empty] (28) at (0, 2.75) {$\tau_{A,B}$};
		\node [style=empty] (29) at (-5.5, -4.5) {$\tau^\ZZ_{A,C}$};
		\node [style=empty] (30) at (3, -4.5) {$\iota_{A\otimes C}$};
		\node [style=empty] (31) at (10, -1.5) {$\iota_{A\otimes C}$};
		\node [style=empty] (32) at (1, -2.5) {$\tau_{A,C}$};
		\node [style=empty] (33) at (-7.5, 0.25) {$A\otimes\ZZ f$};
		\node [style=empty] (34) at (-5.25, 3.25) {$A\otimes\iota$};
		\node [style=empty] (35) at (-5.25, -2.75) {$A\otimes\iota$};
		\node [style=empty] (36) at (10.5, 2.25) {$\ZZ(A\otimes f)$};
		\node [style=empty] (37) at (4, -1) {$\TT(A\otimes f)$};
		\node [style=empty] (38) at (-1.5, 0.5) {$A\otimes\TT f$};
		\node [style=empty] (39) at (6.75, 2.75) {$\iota$};
		\node [style=none] (40) at (0, 3.5) {\textcolor{blue}{(1)}};
		\node [style=none] (41) at (-5, 0.25) {\textcolor{blue}{(2)}};
		\node [style=none] (42) at (1.5, 0.25) {\textcolor{blue}{(3)}};
		\node [style=none] (43) at (6.25, 1) {\textcolor{blue}{(4)}};
		\node [style=none] (44) at (-1.75, -2.75) {\textcolor{blue}{(5)}};
	\end{pgfonlayer}
	\begin{pgfonlayer}{edgelayer}
		\draw [style=to] (0) to (1);
		\draw [style=to] (10) to (11);
		\draw [style=to] (21) to (22);
		\draw [style=to] (23) to (24);
		\draw [style=to] (13) to (14);
		\draw [style=to] (15) to (16);
		\draw [style=to] (25) to (26);
		\draw [style=to] (2) to (4);
		\draw [style=to] (17) to (18);
		\draw [style=to] (20) to (19);
		\draw [style=to] (6) to (7);
		\draw [style=mono] (4) to (3);
		\draw [style=mono] (8) to (9);
	\end{pgfonlayer}
\end{tikzpicture}

%% file: figures/z-monad-1.tikz
\begin{tikzpicture}
	\begin{pgfonlayer}{nodelayer}
		\node [style=empty] (0) at (-6, 6) {$\ZZ^3 X$};
		\node [style=empty] (1) at (6, 6) {$\ZZ^2 X$};
		\node [style=empty] (2) at (-6, -2) {$\ZZ^2 X$};
		\node [style=empty] (3) at (6, -2) {$\TT X$};
		\node [style=empty] (4) at (6, 2) {$\ZZ X$};
		\node [style=empty] (5) at (0, -2) {$\ZZ X$};
		\node [style=empty] (6) at (6, 1.5) {};
		\node [style=empty] (7) at (6, -1.5) {};
		\node [style=empty] (8) at (6, 5.5) {};
		\node [style=empty] (9) at (6, 2.5) {};
		\node [style=empty] (10) at (-2, 4) {$\TT^3 X$};
		\node [style=empty] (11) at (-2, 1) {$\TT^2 X$};
		\node [style=empty] (12) at (2, 4) {$\TT^2 X$};
		\node [style=empty] (13) at (2.5, 4.25) {};
		\node [style=empty] (14) at (5.5, 5.5) {};
		\node [style=empty] (15) at (-5.5, 5.5) {};
		\node [style=empty] (16) at (-2.5, 4.25) {};
		\node [style=empty] (17) at (2.25, 3.5) {};
		\node [style=empty] (18) at (5.5, -1.25) {};
		\node [style=empty] (19) at (5.25, -1.75) {};
		\node [style=empty] (20) at (-1.5, 0.75) {};
		\node [style=empty] (21) at (-2, 3.5) {};
		\node [style=empty] (22) at (-2, 1.5) {};
		\node [style=empty] (23) at (-6, 5.5) {};
		\node [style=empty] (24) at (-6, -1.5) {};
		\node [style=empty] (25) at (0, 6.75) {$\mu^\ZZ_{\ZZ X}$};
		\node [style=empty] (26) at (-3, -2.75) {$\mu^\ZZ_X$};
		\node [style=empty] (27) at (6.75, 4) {$\mu^\ZZ_X$};
		\node [style=empty] (28) at (3, -2.75) {$\iota_X$};
		\node [style=empty] (29) at (6.75, 0) {$\iota_X$};
		\node [style=empty] (30) at (-7, 2) {$\ZZ\mu^\ZZ_X$};
		\node [style=empty] (31) at (-3, 2.5) {$\TT\mu_X$};
		\node [style=empty] (32) at (0, 3.5) {$\mu_{\TT X}$};
		\node [style=empty] (33) at (2.75, 1.5) {$\mu_X$};
		\node [style=empty] (34) at (1.25, 0.25) {$\mu_X$};
		\node [style=empty] (35) at (-5.5, -1.5) {};
		\node [style=empty] (36) at (-2.5, 0.5) {};
		\node [style=empty] (37) at (-3.25, 5.25) {$\iota^3$};
		\node [style=empty] (38) at (3.5, 5.25) {$\iota^2$};
		\node [style=empty] (39) at (-4, 0) {$\iota^2$};
		\node [style=empty] (40) at (10.25, 6) {$\ZZ X$};
		\node [style=empty] (41) at (22.25, 6) {$\ZZ^2 X$};
		\node [style=empty] (42) at (10.25, -2) {$\ZZ^2 X$};
		\node [style=empty] (43) at (22.25, -2) {$\TT X$};
		\node [style=empty] (44) at (22.25, 2) {$\ZZ X$};
		\node [style=empty] (45) at (16.25, -2) {$\ZZ X$};
		\node [style=empty] (46) at (22.25, 1.5) {};
		\node [style=empty] (47) at (22.25, -1.5) {};
		\node [style=empty] (48) at (22.25, 5.5) {};
		\node [style=empty] (49) at (22.25, 2.5) {};
		\node [style=empty] (50) at (14.25, 4) {$\TT X$};
		\node [style=empty] (51) at (14.25, 1) {$\TT^2 X$};
		\node [style=empty] (52) at (18.25, 4) {$\TT^2 X$};
		\node [style=empty] (53) at (18.75, 4.25) {};
		\node [style=empty] (54) at (21.75, 5.5) {};
		\node [style=empty] (55) at (10.75, 5.5) {};
		\node [style=empty] (56) at (13.75, 4.25) {};
		\node [style=empty] (57) at (18.5, 3.5) {};
		\node [style=empty] (58) at (21.75, -1.25) {};
		\node [style=empty] (59) at (21.5, -1.75) {};
		\node [style=empty] (60) at (14.75, 0.75) {};
		\node [style=empty] (61) at (14.25, 3.5) {};
		\node [style=empty] (62) at (14.25, 1.5) {};
		\node [style=empty] (63) at (10.25, 5.5) {};
		\node [style=empty] (64) at (10.25, -1.5) {};
		\node [style=empty] (65) at (16.25, 6.75) {$\eta^\ZZ_{\ZZ X}$};
		\node [style=empty] (66) at (13.25, -2.75) {$\mu^\ZZ_X$};
		\node [style=empty] (67) at (23, 4) {$\mu^\ZZ_X$};
		\node [style=empty] (68) at (19.25, -2.75) {$\iota_X$};
		\node [style=empty] (69) at (23, 0) {$\iota_X$};
		\node [style=empty] (70) at (9.25, 2) {$\ZZ\eta^\ZZ_X$};
		\node [style=empty] (71) at (13.25, 2.5) {$\TT\eta_X$};
		\node [style=empty] (72) at (16.25, 3.5) {$\eta_{\TT X}$};
		\node [style=empty] (73) at (19, 1.5) {$\mu_X$};
		\node [style=empty] (74) at (17.5, 0.25) {$\mu_X$};
		\node [style=empty] (75) at (10.75, -1.5) {};
		\node [style=empty] (76) at (13.75, 0.5) {};
		\node [style=empty] (77) at (12.75, 5.25) {$\iota$};
		\node [style=empty] (78) at (20, 5.25) {$\iota^2$};
		\node [style=empty] (79) at (12, 0) {$\iota^2$};
		\node [style=none] (80) at (0, 5) {\textcolor{blue}{(1)}};
		\node [style=none] (81) at (-4.5, 2) {\textcolor{blue}{(2)}};
		\node [style=none] (82) at (0.5, 1.75) {\textcolor{blue}{(3)}};
		\node [style=none] (83) at (4.25, 2.75) {\textcolor{blue}{(4)}};
		\node [style=none] (84) at (-1.25, -0.75) {\textcolor{blue}{(5)}};
		\node [style=none] (85) at (16.25, 5) {\textcolor{blue}{(6)}};
		\node [style=none] (86) at (11.75, 2) {\textcolor{blue}{(7)}};
		\node [style=none] (87) at (16.75, 1.75) {\textcolor{blue}{(8)}};
		\node [style=none] (88) at (20.5, 2.75) {\textcolor{blue}{(9)}};
		\node [style=none] (89) at (15.75, -0.75) {\textcolor{blue}{(10)}};
	\end{pgfonlayer}
	\begin{pgfonlayer}{edgelayer}
		\draw [style=to] (23) to (24);
		\draw [style=to] (0) to (1);
		\draw [style=to] (15) to (16);
		\draw [style=to] (14) to (13);
		\draw [style=to] (10) to (12);
		\draw [style=to] (21) to (22);
		\draw [style=to] (17) to (18);
		\draw [style=to] (20) to (19);
		\draw [style=to] (2) to (5);
		\draw [style=to] (8) to (9);
		\draw [style=mono] (5) to (3);
		\draw [style=mono] (6) to (7);
		\draw [style=to] (35) to (36);
		\draw [style=to] (63) to (64);
		\draw [style=to] (40) to (41);
		\draw [style=to] (55) to (56);
		\draw [style=to] (54) to (53);
		\draw [style=to] (50) to (52);
		\draw [style=to] (61) to (62);
		\draw [style=to] (57) to (58);
		\draw [style=to] (60) to (59);
		\draw [style=to] (42) to (45);
		\draw [style=to] (48) to (49);
		\draw [style=mono] (45) to (43);
		\draw [style=mono] (46) to (47);
		\draw [style=to] (75) to (76);
	\end{pgfonlayer}
\end{tikzpicture}

%% file: figures/z-monad-commutative.tikz
\begin{tikzpicture}
	\begin{pgfonlayer}{nodelayer}
		\node [style=empty] (0) at (-10, 6) {$\ZZ A\otimes\ZZ B$};
		\node [style=empty] (1) at (21, 6) {$\ZZ^2(A\otimes B)$};
		\node [style=empty] (2) at (-10, -18) {$\ZZ^2(A\otimes B)$};
		\node [style=empty] (3) at (21, -18) {$\TT(A\otimes B)$};
		\node [style=empty] (4) at (-10, -6) {$\ZZ(\ZZ A\otimes B)$};
		\node [style=empty] (5) at (21, -6) {$\ZZ(A\otimes B)$};
		\node [style=empty] (6) at (5, -18) {$\ZZ(A\otimes B)$};
		\node [style=empty] (7) at (5, 6) {$\ZZ(A\otimes\ZZ B)$};
		\node [style=empty] (8) at (-10, 5.5) {};
		\node [style=empty] (9) at (-10, -5.5) {};
		\node [style=empty] (10) at (-10, -6.5) {};
		\node [style=empty] (11) at (-10, -17.5) {};
		\node [style=empty] (12) at (21, 5.5) {};
		\node [style=empty] (13) at (21, -5.5) {};
		\node [style=empty] (14) at (21, -6.5) {};
		\node [style=empty] (15) at (21, -17.5) {};
		\node [style=empty] (16) at (-3, 6.5) {$\tau'^\ZZ_{A,\ZZ B}$};
		\node [style=empty] (17) at (12, 6.5) {$\ZZ\tau^\ZZ_{A,B}$};
		\node [style=empty] (18) at (22.25, 0) {$\mu^\ZZ_{A\otimes B}$};
		\node [style=empty] (19) at (-3, -18.75) {$\mu^\ZZ_{A\otimes B}$};
		\node [style=empty] (20) at (-11.25, -12) {$\ZZ\tau'^\ZZ_{A,B}$};
		\node [style=empty] (21) at (-11.25, 0) {$\tau^\ZZ_{\ZZ A,B}$};
		\node [style=empty] (22) at (-2, -15) {$\ZZ\TT(A\otimes B)$};
		\node [style=empty] (23) at (11, -15) {$\TT^2(A\otimes B)$};
		\node [style=empty] (24) at (-8.25, -17.5) {};
		\node [style=empty] (25) at (-3.75, -15.25) {};
		\node [style=empty] (26) at (13, -15.5) {};
		\node [style=empty] (27) at (19.5, -17.5) {};
		\node [style=empty] (28) at (16, -1) {$\ZZ\TT(A\otimes B)$};
		\node [style=empty] (29) at (16, -9) {$\TT^2(A\otimes B)$};
		\node [style=empty] (30) at (16, -1.5) {};
		\node [style=empty] (31) at (16, -8.5) {};
		\node [style=empty] (32) at (20.5, 5.5) {};
		\node [style=empty] (33) at (16.25, -0.5) {};
		\node [style=empty] (34) at (20.25, -17) {};
		\node [style=empty] (35) at (16.25, -9.5) {};
		\node [style=empty] (36) at (-2, -9) {$\ZZ(\TT A\otimes B)$};
		\node [style=empty] (37) at (-8, -6.25) {};
		\node [style=empty] (38) at (-3.75, -8.5) {};
		\node [style=empty] (39) at (-2, -9.5) {};
		\node [style=empty] (40) at (-2, -14.5) {};
		\node [style=empty] (41) at (8, -1) {$\ZZ(A\otimes\TT B)$};
		\node [style=empty] (42) at (5.25, 5.5) {};
		\node [style=empty] (43) at (7.75, -0.5) {};
		\node [style=empty] (44) at (5.5, -11.5) {$\TT(\TT A\otimes B)$};
		\node [style=empty] (45) at (-0.25, -9.5) {};
		\node [style=empty] (46) at (3.5, -11.25) {};
		\node [style=empty] (47) at (6, -12) {};
		\node [style=empty] (48) at (10.25, -14.5) {};
		\node [style=empty] (49) at (11.5, -5.5) {$\TT(A\otimes\TT B)$};
		\node [style=empty] (50) at (8.25, -1.5) {};
		\node [style=empty] (51) at (11.5, -5) {};
		\node [style=empty] (52) at (12, -6) {};
		\node [style=empty] (53) at (15, -8.5) {};
		\node [style=empty] (54) at (3.75, -3.25) {$\TT A\otimes\TT B$};
		\node [style=empty] (55) at (2.5, -5.75) {};
		\node [style=empty] (56) at (5.25, -10.75) {};
		\node [style=empty] (57) at (-3, -3) {$\TT A\otimes\ZZ B$};
		\node [style=empty] (58) at (1, 2) {$\ZZ A\otimes\TT B$};
		\node [style=empty] (59) at (-9.5, 5.25) {};
		\node [style=empty] (60) at (-3.25, -2.5) {};
		\node [style=empty] (61) at (-8.25, 5.5) {};
		\node [style=empty] (62) at (-0.75, 2.25) {};
		\node [style=empty] (63) at (-2.75, -3.5) {};
		\node [style=empty] (64) at (-2, -8.5) {};
		\node [style=empty] (65) at (1.25, 1.5) {};
		\node [style=empty] (66) at (3.5, -2.75) {};
		\node [style=empty] (67) at (-3.75, 4) {$\iota$};
		\node [style=empty] (68) at (-7, 1.25) {$\iota$};
		\node [style=empty] (69) at (-6.5, -16) {$\iota$};
		\node [style=empty] (70) at (17.75, 3) {$\iota$};
		\node [style=empty] (71) at (-0.25, -3.75) {$\iota$};
		\node [style=empty] (72) at (3.25, -0.75) {$\iota$};
		\node [style=empty] (73) at (7, 3) {$\iota$};
		\node [style=empty] (74) at (4.5, 1.25) {$\tau'^\ZZ_{A,\TT B}$};
		\node [style=empty] (75) at (-3.5, -6) {$\tau^\ZZ_{\TT A,B}$};
		\node [style=empty] (76) at (-6.25, -7.75) {$\iota$};
		\node [style=empty] (77) at (-3.25, -12) {$\ZZ\tau'_{A,B}$};
		\node [style=empty] (78) at (1.25, -10.75) {$\iota$};
		\node [style=empty] (79) at (5.25, -7.75) {$\tau_{\TT A,B}$};
		\node [style=empty] (80) at (7.5, -5.25) {$\tau'_{A,\TT B}$};
		\node [style=empty] (81) at (8.5, -12.75) {$\TT\tau'_{A,B}$};
		\node [style=empty] (82) at (12.5, -7.5) {$\TT\tau_{A,B}$};
		\node [style=empty] (83) at (3.25, -14.5) {$\iota$};
		\node [style=empty] (84) at (16.5, -12.75) {$\mu_{A\otimes B}$};
		\node [style=empty] (85) at (16.25, -15.75) {$\mu_{A\otimes B}$};
		\node [style=empty] (86) at (16.5, -5) {$\iota$};
		\node [style=empty] (87) at (9.25, -3.5) {$\iota$};
		\node [style=empty] (88) at (12, -0.5) {$\ZZ\tau_{A,B}$};
		\node [style=empty] (89) at (22, -12) {$\iota_{A\otimes B}$};
		\node [style=empty] (90) at (13, -18.75) {$\iota_{A\otimes B}$};
		\node [style=none] (91) at (1, 4) {\textcolor{blue}{(1)}};
		\node [style=none] (92) at (12.75, 3) {\textcolor{blue}{(2)}};
		\node [style=none] (93) at (-7, -3) {\textcolor{blue}{(3)}};
		\node [style=none] (94) at (-2, 0) {\textcolor{blue}{(4)}};
		\node [style=none] (95) at (6, -2.25) {\textcolor{blue}{(5)}};
		\node [style=none] (96) at (13.25, -3.75) {\textcolor{blue}{(6)}};
		\node [style=none] (97) at (18.5, -5) {\textcolor{blue}{(7)}};
		\node [style=none] (98) at (1, -7) {\textcolor{blue}{(8)}};
		\node [style=none] (99) at (11, -10) {\textcolor{blue}{(9)}};
		\node [style=none] (100) at (-6.5, -12) {\textcolor{blue}{(10)}};
		\node [style=none] (101) at (1.25, -12.75) {\textcolor{blue}{(11)}};
		\node [style=none] (102) at (6.25, -16.5) {\textcolor{blue}{(12)}};
		\node [style=empty] (103) at (2, -5.25) {$\TT A\otimes\TT B$};
		\node [style=empty] (104) at (-2, -3.5) {};
		\node [style=empty] (105) at (0.75, -4.75) {};
		\node [style=empty] (106) at (0.75, 1.5) {};
		\node [style=empty] (107) at (1.75, -4.75) {};
		\node [style=empty] (108) at (0.75, -1.25) {$\iota$};
	\end{pgfonlayer}
	\begin{pgfonlayer}{edgelayer}
		\draw [style=to] (0) to (7);
		\draw [style=to] (7) to (1);
		\draw [style=to] (12) to (13);
		\draw [style=to] (2) to (6);
		\draw [style=to] (10) to (11);
		\draw [style=to] (8) to (9);
		\draw [style=to] (22) to (23);
		\draw [style=to] (24) to (25);
		\draw [style=to] (26) to (27);
		\draw [style=to] (30) to (31);
		\draw [style=to] (32) to (33);
		\draw [style=to] (35) to (34);
		\draw [style=to] (37) to (38);
		\draw [style=to] (39) to (40);
		\draw [style=to] (42) to (43);
		\draw [style=to] (41) to (28);
		\draw [style=to] (45) to (46);
		\draw [style=to] (47) to (48);
		\draw [style=to] (50) to (51);
		\draw [style=to] (52) to (53);
		\draw [style=to] (55) to (56);
		\draw [style=to] (54) to (49);
		\draw [style=to] (59) to (60);
		\draw [style=to] (61) to (62);
		\draw [style=to] (63) to (64);
		\draw [style=to] (65) to (66);
		\draw [style=to] (58) to (41);
		\draw [style=mono] (6) to (3);
		\draw [style=mono] (14) to (15);
		\draw [style=to] (104) to (105);
		\draw [style=to] (106) to (107);
	\end{pgfonlayer}
\end{tikzpicture}

%% file: figures/main-theorem-2-to-3.tikz
\begin{tikzpicture}
	\begin{pgfonlayer}{nodelayer}
		\node [style=empty] (0) at (-4, 2) {$\CC$};
		\node [style=empty] (1) at (1, 2) {$\CC_\TT$};
		\node [style=empty] (2) at (-4, -2) {$\CC_\ZZ$};
		\node [style=empty] (3) at (1, -2) {$Z(\CC_\TT)$};
		\node [style=empty] (4) at (-4, 1.5) {};
		\node [style=empty] (5) at (-4, -1.5) {};
		\node [style=empty] (6) at (1, 1.5) {};
		\node [style=empty] (7) at (1, -1.5) {};
		\node [style=empty] (8) at (-3.5, 1.5) {};
		\node [style=empty] (9) at (0.5, -1.5) {};
		\node [style=empty] (10) at (-1.75, -1.75) {$\cong$};
		\node [style=empty] (11) at (-1.75, -2.5) {$\hat \II$};
		\node [style=empty] (12) at (-4.75, 0) {$\JJ^\ZZ$};
		\node [style=empty] (13) at (1.75, 0) {};
		\node [style=empty] (14) at (-1.5, 2.5) {$\JJ$};
		\node [style=empty] (15) at (-1.025, 0.375) {$\hat \JJ$};
	\end{pgfonlayer}
	\begin{pgfonlayer}{edgelayer}
		\draw [style=to] (4) to (5);
		\draw [style=to] (0) to (1);
		\draw [style=to] (2) to (3);
		\draw [style=to] (8) to (9);
		\draw [style=hook] (7) to (6);
	\end{pgfonlayer}
\end{tikzpicture}

%% file: figures/adjoint-diagram-unique-centre.tikz
\begin{tikzpicture}
	\begin{pgfonlayer}{nodelayer}
		\node [style=empty] (0) at (-6.5, 2) {$\hat\JJ Y$};
		\node [style=empty] (1) at (-6.5, -2) {$\hat\JJ \mathcal RX$};
		\node [style=empty] (2) at (-2.5, -2) {$X$};
		\node [style=empty] (3) at (-6, 1.5) {};
		\node [style=empty] (4) at (-3, -1.5) {};
		\node [style=empty] (5) at (-6.5, 1.5) {};
		\node [style=empty] (6) at (-6.5, -1.5) {};
		\node [style=empty] (7) at (-8, 0) {$\hat\JJ\Phi(f)$};
		\node [style=empty] (8) at (-4, 0.75) {$f$};
		\node [style=empty] (9) at (-4.5, -2.5) {$\varepsilon_X$};
		\node [style=empty] (10) at (6.25, 2) {$Y$};
		\node [style=empty] (11) at (6.25, -2) {$\mathcal RX$};
		\node [style=empty] (12) at (10.25, -2) {$\TT X$};
		\node [style=empty] (13) at (6.75, 1.5) {};
		\node [style=empty] (14) at (9.75, -1.5) {};
		\node [style=empty] (15) at (6.25, 1.5) {};
		\node [style=empty] (16) at (6.25, -1.5) {};
		\node [style=empty] (17) at (5.25, 0) {$\Phi(f)$};
		\node [style=empty] (18) at (8.75, 0.75) {$f$};
		\node [style=empty] (19) at (8.25, -2.5) {$\varepsilon_X$};
	\end{pgfonlayer}
	\begin{pgfonlayer}{edgelayer}
		\draw [style=to] (3) to (4);
		\draw [style=to] (1) to (2);
		\draw [style=dashed-arrow] (5) to (6);
		\draw [style=to] (13) to (14);
		\draw [style=to] (11) to (12);
		\draw [style=dashed-arrow] (15) to (16);
	\end{pgfonlayer}
\end{tikzpicture}

%% file: figures/suff-cond-com.tikz
\begin{tikzpicture}
	\begin{pgfonlayer}{nodelayer}
		\node [style=empty] (0) at (-2.5, 1.5) {$c\otimes X$};
		\node [style=empty] (1) at (-2.5, -2.75) {$c\otimes Y$};
		\node [style=empty] (2) at (4.5, 1.5) {$c'\otimes X$};
		\node [style=empty] (3) at (4.5, -2.75) {$c'\otimes Y$};
		\node [style=empty] (4) at (-2.5, 1) {};
		\node [style=empty] (5) at (-2.5, -2.25) {};
		\node [style=empty] (6) at (4.5, 1) {};
		\node [style=empty] (7) at (4.5, -2.25) {};
		\node [style=empty] (16) at (1, 2) {$h\otimes_l X$};
		\node [style=empty] (17) at (-3.75, -0.5) {$c\otimes_r f$};
		\node [style=empty] (18) at (5.75, -0.5) {$c' \otimes_r f$};
		\node [style=empty] (19) at (1, -3.25) {$h\otimes_l Y$};
	\end{pgfonlayer}
	\begin{pgfonlayer}{edgelayer}
		\draw [style=to] (0) to (2);
		\draw [style=to] (6) to (7);
		\draw [style=to] (4) to (5);
		\draw [style=to] (1) to (3);
	\end{pgfonlayer}
\end{tikzpicture}

%% file: figures/suff-cond-ct.tikz
\begin{tikzpicture}
	\begin{pgfonlayer}{nodelayer}
		\node [style=empty] (0) at (-2.5, 1.5) {$c\otimes X$};
		\node [style=empty] (1) at (-2.5, -2.75) {$c\otimes Y$};
		\node [style=empty] (2) at (4.5, 1.5) {$c'\otimes X$};
		\node [style=empty] (3) at (4.5, -2.75) {$c'\otimes Y$};
		\node [style=empty] (4) at (-2.5, 1) {};
		\node [style=empty] (5) at (-2.5, -2.25) {};
		\node [style=empty] (6) at (4.5, 1) {};
		\node [style=empty] (7) at (4.5, -2.25) {};
		\node [style=empty] (16) at (1, 2) {$h\otimes_l X$};
		\node [style=empty] (17) at (-3.75, -0.5) {$c\otimes_r f$};
		\node [style=empty] (18) at (5.75, -0.5) {$c' \otimes_r f$};
		\node [style=empty] (19) at (1, -3.25) {$h\otimes_l Y$};
		\node [style=empty] (20) at (-6, 2) {$\mathcal{J}(\epsilon_A \otimes X)$};
		\node [style=empty] (21) at (-9.5, 1.5) {$A \otimes X$};
	\end{pgfonlayer}
	\begin{pgfonlayer}{edgelayer}
		\draw [style=to] (0) to (2);
		\draw [style=to] (6) to (7);
		\draw [style=to] (4) to (5);
		\draw [style=to] (1) to (3);
		\draw [style=to] (21) to (0);
	\end{pgfonlayer}
\end{tikzpicture}

%% file: figures/suff-cond-c.tikz
\begin{tikzpicture}
	\begin{pgfonlayer}{nodelayer}
		\node [style=empty] (1) at (-14, -3.5) {$\TT (c \otimes Y)$};
		\node [style=empty] (2) at (0, 3.5) {$ \TT  c'  \otimes X $};
		\node [style=empty] (3) at (0, -3.5) {$\TT(\TT c'\otimes Y)$};
		\node [style=empty] (6) at (14, 3) {};
		\node [style=empty] (7) at (14, 0.5) {};
		\node [style=empty] (8) at (-0.5, 6) {$h\otimes X$};
		\node [style=empty] (10) at (10.5, 4) {$\TT (c' \otimes f)$};
		\node [style=empty] (11) at (-7, -6.5) {$\TT (h\otimes Y)$};
		\node [style=empty] (12) at (7, 3.5) {$\TT  (c' \otimes X)$};
		\node [style=empty] (13) at (3.5, 4) {$\tau'_{c',X}$};
		\node [style=empty] (15) at (-14, 3) {};
		\node [style=empty] (16) at (-14, 0.5) {};
		\node [style=empty] (17) at (14, 0) {$\TT^2(c' \otimes Y)$};
		\node [style=empty] (18) at (14, -0.5) {};
		\node [style=empty] (19) at (7, -3.5) {$\TT^2 (c' \otimes Y)$};
		\node [style=empty] (20) at (14, -3) {};
		\node [style=empty] (21) at (15.5, 1.75) {$\TT\tau_{c' ,Y}$};
		\node [style=empty] (22) at (3.5, -3) {$\TT\tau'_{c' ,Y}$};
		\node [style=empty] (23) at (-15.25, 1.75) {$c \otimes f $};
		\node [style=empty] (24) at (14, 3.5) {$\TT (c' \otimes\TT Y)$};
		\node [style=empty] (25) at (15.5, -1.75) {$\mu_{c' \otimes Y}$};
		\node [style=empty] (41) at (-7, 3.5) {$A \otimes X$};
		\node [style=empty] (42) at (14, -3.5) {$\TT(c' \otimes Y)$};
		\node [style=empty] (50) at (-14, -0.5) {};
		\node [style=empty] (51) at (-14, -3) {};
		\node [style=empty] (52) at (-15.25, -1.75) {$\tau_{c ,Y}$};
		\node [style=empty] (54) at (11, -3) {$\mu_{c' \otimes Y}$};
		\node [style=empty] (56) at (-7, 6) {$\epsilon_A \otimes X $};
		\node [style=empty] (57) at (-3.5, 7) {$c \otimes X$};
		\node [style=empty] (61) at (-14, 3.5) {$c \otimes X$};
		\node [style=empty] (63) at (-10.5, 4) {$\epsilon_A \otimes X $};
		\node [style=empty] (64) at (-14, 0) {$c \otimes \TT Y $};
		\node [style=empty] (65) at (-3.5, 4) {$\epsilon'_A \otimes X $};
		\node [style=empty] (66) at (-7, 0) {$A \otimes \TT Y $};
		\node [style=empty] (67) at (-6, 1.75) {$A \otimes f $};
		\node [style=empty] (68) at (-10.5, 0.5) {$\epsilon_A \otimes \TT Y $};
		\node [style=empty] (69) at (-7, -0.5) {};
		\node [style=empty] (70) at (-7, -3) {};
		\node [style=empty] (71) at (-6, -1.75) {$\tau_{A ,Y}$};
		\node [style=empty] (72) at (-7, -3.5) {$\TT (A \otimes Y)$};
		\node [style=empty] (73) at (-10.5, -3) {$\TT (\epsilon_A \otimes Y ) $};
		\node [style=empty] (74) at (-3.5, -3) {$\TT (\epsilon'_A \otimes Y ) $};
		\node [style=empty] (75) at (-7, 3) {};
		\node [style=empty] (76) at (-7, 0.5) {};
		\node [style=none] (77) at (-3.5, 5.25) {\textcolor{blue}{(1)}};
		\node [style=none] (78) at (-10.5, 2) {\textcolor{blue}{(2)}};
		\node [style=none] (79) at (-10.5, -1.5) {\textcolor{blue}{(3)}};
		\node [style=none] (80) at (-7, -4.75) {\textcolor{blue}{(4)}};
		\node [style=none] (81) at (3.5, 0) {\textcolor{blue}{(5)}};
	\end{pgfonlayer}
	\begin{pgfonlayer}{edgelayer}
		\draw [style=to] (6) to (7);
		\draw [style=to, bend right] (1) to (3);
		\draw [style=to] (2) to (12);
		\draw [style=to] (15) to (16);
		\draw [style=to] (18) to (20);
		\draw [style=to] (3) to (19);
		\draw [style=to] (12) to (24);
		\draw [style=to] (50) to (51);
		\draw [style=to] (41) to (57);
		\draw [style=to] (57) to (2);
		\draw [style=to] (19) to (42);
		\draw [style=to] (41) to (2);
		\draw [style=to] (66) to (64);
		\draw [style=to] (69) to (70);
		\draw [style=to] (72) to (1);
		\draw [style=to] (72) to (3);
		\draw [style=to] (75) to (76);
		\draw [style=to] (41) to (61);
	\end{pgfonlayer}
\end{tikzpicture}